\documentclass[a4paper,11pt]{article}
\pdfoutput=1 
\usepackage{jheppub} 
\usepackage[T1]{fontenc} 
\usepackage{lmodern}
\usepackage{multirow}
\usepackage{float}
\usepackage{tikz}
\usepackage{subcaption}
\usepackage{amsmath,amssymb,xfrac,framed,verbatim,amsthm}
\usepackage{mathrsfs,accents,bbm,bbold} 
\usepackage{empheq}

\usepackage{microtype, ctable}
\usepackage{simplewick}
\usepackage{slashed,mathtools} 
\usepackage{sectsty} 
\usepackage{graphicx,enumerate,bm} 
\usepackage{comment}
\usepackage{young}
\usepackage[vcentermath]{youngtab}
\usepackage{ytableau}

\renewcommand{\i}{\text{i}}

\preprint{TIFR/TH/22-41}
\title{\boldmath Crossing Symmetry in Matter Chern-Simons Theories at finite $N$ and $k$}

\author[a,1]{Umang Mehta,\note{umangmehta@uchicago.edu}}
\author[b,2]{Shiraz  Minwalla,\note{minwalla@theory.tifr.res.in}}
\author[b,3]{Chintan Patel,\note{chintan.patel@tifr.res.in}}
\author[c,4]{Shiroman Prakash \note{sprakash@dei.ac.in}}
\author[d,b, 5]{Kartik Sharma \note{kartik.sharma@students.iiserpune.ac.in}}

\affiliation[a]{Kadanoff Center for Theoretical Physics, \\ University of Chicago, Chicago, IL 60637}
\affiliation[b]{Department of Theoretical Physics, \\ Tata Institute
	of Fundamental Research, Homi Bhabha Rd, Mumbai 400005, India}
\affiliation[c]{Department of Physics and Computer Science, \\ Dayalbagh Educational Institute, Dayalbagh Road, Agra 282005, India} \affiliation[d]{Indian Institute of Science Education and Research, Dr Homi Bhabha Rd, Pashan, Pune, Maharashtra 411008 }

\abstract{We present a conjecture for the crossing symmetry rules for Chern-Simons gauge theories interacting with  massive matter in $2+1$ dimensions. Our crossing rules are given in terms of  the expectation values of particular tangles of Wilson lines, and reduce to the standard rules at large Chern-Simons level. We present completely explicit results for the special case of two  fundamental and two antifundamental insertions in $SU(N)_k$ and $U(N)_k$ theories. These formulae are  consistent with the conjectured level-rank,  Bose-Fermi duality between these theories and take the form of a $q=e^{\frac{ 2 \pi i }{\kappa}}$ deformation of their large $k$  counterparts. In the 't Hooft large $N$ limit our results reduce to standard rules with one twist: the $S$-matrix in the singlet channel is reduced by the factor  $\frac{\sin \pi \lambda}{\pi \lambda} $ (where $\lambda$ is the 't Hooft coupling), explaining `anomalous' crossing properties observed in earlier direct large $N$ computations. }

\begin{document}
	\maketitle
	\section{Introduction}

	$S$-matrices are among the best studied and most interesting observables in quantum field theories (and also asymptotically flat backgrounds of string theory.) Of course, the explicit formulae for $S$-matrices differ from QFT to QFT (and from one vacuum of string theory to another.)  However, the $S$-matrices of  all massive theories that are genuinely trivial in the IR  -- so-called trivially-gapped theories --  have been shown to share several universal properties. First,  the $S$-matrix in all such theories takes the form $S=\mathcal S_{ {\rm id}}  +i\tau$, where $\mathcal S_{ {\rm id}}$ is the identity $S$-matrix (a $\delta$-function localized on forward scattering), while $\tau$
	is an analytic function of the scattering momenta (apart, of course, from the overall momentum-conserving delta-function)\footnote{In the case of $2 \rightarrow 2$ and scattering it is believed that the only non-analyticity of the $\tau$ matrix on the principal sheet are those required by Cutkosky's rules. For more complicated scattering processes (e.g. $3 \rightarrow 3$ scattering), unitarity actually also requires additional anomalous thresholds on the principal sheet, see \cite{Hannesdottir:2022bmo}.}. Second, atleast in the case of $2 \rightarrow 2$ scattering,  $S$-matrices obey crossing symmetry \footnote{See \cite{Bros:1965kbd} for $2\rightarrow 2$ scattering, \cite{Williams:1963zz, Cohen-Tannoudji:1968lnm}  for $2 \rightarrow 3$ scattering, and also the relevant more recent papers \cite{DeLacroix:2018arq, Mizera:2021fap}.} : i.e. analytic continuation along a suitable path relates the formula for the $S$-matrix involving an {\it antiparticle} in an {\it initial/final} state to the
	formula for the same $S$-matrix involving the corresponding {\it particle} in the {\it final/initial} state. Finally, of course, these  $S$-matrices obey the unitarity equation $S^\dagger S=I$.

	In contrast to trivially-gapped theories, $S$-matrices in theories with massless particles often display IR divergences, and consequently are less well understood.
	\footnote{The severity of IR issues varies from theory to theory. IR divergences are absent in gravity and  QED in $D\geq 5$; these $S$-matrices may well share the structural properties of their massive cousins. The relative simplicity of IR divergences in $D=4$ dimensional massive QED allows for the definition of the finite Faddeev Kullish $S$-matrix, which may also enjoy good properties. IR divergences are more severe in gravity and massless QED in $D=4$, and the existence of a finite $S$-matrix in such theories is unclear. Finally IR problems  are so severe in theories that flow to interacting CFT's at low energies that standard lore asserts that finite $S$-matrices simply do not exist in such theories (however, in our opinion, the the successful and elegant computations of this quantity in ${\cal N}=4$ Yang Mills require an explanation: perhaps a useful definition of $S$-matrices in CFT's in terms of asymptotic showers will eventually be found). In contrast, gluon $S$-matrices in  `even-more-strongly-coupled' confining gauge theories like pure Yang Mills deal with infinite energy states, and so presumably cannot be made sense of in any manner.}.

	There is a third class of quantum field theories -- the so-called \textit{topologically}-gapped theories --  which, in some sense, lie somewhere in-between the class of gapped theories with a trivial vacuum and theories with massless degrees of freedom. This class consists of those theories that host only massive particles, but are nonetheless nontrivial in the IR because they flow at low energies to topological field theories rather than completely empty theories. The fact that these theories do not possess a continuum of low energy degrees of freedom suggests that  their  $S$-matrices should be well-defined. However, the non-triviality of these theories in the IR raises the possibility that their $S$-matrices will have novel structural properties.

	In this paper, we study the structural properties of $S$-matrices in a well-studied class of examples of topologically-gapped theories: Chern-Simons theories minimally coupled to massive matter fields in 2+1 dimensions. At low energies these theories reduce to extremely familiar Topological Field theories (TFT's) -- pure Chern-Simons theory in 2+1 dimensions. In this paper we use well-studied properties of these exactly solvable TFT's to make sharp predictions for the structural properties of the $S$-matrices of the QFTs -- i.e., Chern-Simons matter theories --  that flow to them in the IR.

	While the analysis presented in this paper applies to all Chern-Simons theories with massive matter, it is motivated by explicit results in a particular class of examples:  $SU(N)_k$ Chern-Simons interacting with fundamental matter fields. It was noted in \cite{Giombi:2011kc} that these theories are exactly solvable in the large $N$ limit. The exact large $N$ solution has been explored in several directions in \cite{Sezgin:2002rt, Klebanov:2002ja, Giombi:2009wh,
		Benini:2011mf, Giombi:2011kc, Aharony:2011jz, Maldacena:2011jn,
		Maldacena:2012sf, Chang:2012kt, Jain:2012qi, Aharony:2012nh,
		Yokoyama:2012fa, GurAri:2012is, Aharony:2012ns, Jain:2013py,
		Takimi:2013zca, Jain:2013gza, Yokoyama:2013pxa, Bardeen:2014paa,
		Jain:2014nza, Bardeen:2014qua, Gurucharan:2014cva, Dandekar:2014era,
		Frishman:2014cma, Moshe:2014bja, Aharony:2015pla, Inbasekar:2015tsa,
		Bedhotiya:2015uga, Gur-Ari:2015pca, Minwalla:2015sca,
		Radicevic:2015yla, Geracie:2015drf, Aharony:2015mjs,
		Yokoyama:2016sbx, Gur-Ari:2016xff, Karch:2016sxi, Murugan:2016zal,
		Seiberg:2016gmd, Giombi:2016ejx, Hsin:2016blu, Radicevic:2016wqn,
		Karch:2016aux, Giombi:2016zwa, Wadia:2016zpd, Aharony:2016jvv,
		Giombi:2017rhm, Benini:2017dus, Sezgin:2017jgm, Nosaka:2017ohr,
		Komargodski:2017keh, Giombi:2017txg, Gaiotto:2017tne,
		Jensen:2017dso, Jensen:2017xbs, Gomis:2017ixy, Inbasekar:2017ieo,
		Inbasekar:2017sqp, Cordova:2017vab, Charan:2017jyc, Benini:2017aed,
		Aitken:2017nfd, Argurio:2018uup, Jensen:2017bjo, Chattopadhyay:2018wkp,
		Turiaci:2018nua, Choudhury:2018iwf, Karch:2018mer, Aharony:2018npf,
		Yacoby:2018yvy, Aitken:2018cvh, Aharony:2018pjn, Dey:2018ykx, Skvortsov:2018uru, Argurio:2019tvw, Armoni:2019lgb,
		Chattopadhyay:2019lpr, Dey:2019ihe, Halder:2019foo, Aharony:2019mbc,
		Li:2019twz, Jain:2019fja, Inbasekar:2019wdw, Inbasekar:2019azv,
		Jensen:2019mga, Kalloor:2019xjb, Ghosh:2019sqf, Argurio:2020her, Inbasekar:2020hla,
		Jain:2020rmw, Minwalla:2020ysu, Jain:2020puw, Mishra:2020wos,
		Jain:2021wyn, Jain:2021vrv, Gandhi:2021gwn, Gabai:2022snc,Gabai:2022vri}.
	In particular, the $2 \times 2$ $S$-matrices of fundamental and anitfundamental matter fields in these  theories were computed to all orders in the 't Hooft coupling, in \cite{Jain:2014nza} \cite{Inbasekar:2015tsa}, \cite{Gabai:2022snc}. The authors of \cite{Jain:2014nza} noted that the exact large-$N$ results for $S$-matrices display some structural surprises
	\footnote{These observations were confirmed and strengthened in \cite{Inbasekar:2015tsa}, \cite{Gabai:2022snc}.} and conjectured that the rules of crossing symmetry are modified in Chern-Simons matter theories in a manner we will detail below. The results presented in this paper verify the conjectures of \cite{Jain:2014nza} and also supply generalizations of the conjectures to finite $N$ and $k$ and other matter representations \footnote{Alternatively, the results of the computations of \cite{Jain:2014nza} \cite{Inbasekar:2015tsa} and  \cite{Gabai:2022snc} may be viewed as a consistency test of the analysis presented in this paper.}.

	\subsection{Modification of $S=\mathcal S_{ {\rm id}}+i\tau$ to accommodate Aharonov-Bohm phases}  \label{moi}

	As first noted by Ruijsenaars \cite{Ruijsenaars:1983aa} in the context of the non relativistic theory (see also \cite{Jackiw:1989qp,Bak:1994zz,Amelino-Camelia:1994xrl}), and physically explained 
	by the authors of \cite{Jain:2014nza}, it is easy to see that one key structural property of
	$S$-matrices in topologically trivial gapped theories -- namely, that $S=\mathcal S_{ {\rm id}}+i \tau$ where
	$\tau$ is an analytic function --  must be modified in topologically-nontrivial
	massive theories in 2+1 dimensions in order to accommodate the possibility of (non-abelian) Aharonov-Bohm phases. A conjecture, slightly generalizing the conjecture of \cite{Jain:2014nza}, for how this property is modified can be stated as follows.

	Consider a $2$-to-$2$ scattering process $A B \rightarrow AB$. In the low energy TFT, the world lines of particles $A$ and $B$ are represented by Wilson lines in representations $R_A$ and $R_B$. Let $O_A$ and $O_B$ be the corresponding primary operators in the rational CFT dual to this low energy TFT. Let the fusion rule of these operators be given by
	\begin{equation}\label{oaobintro}
	O_A O_B = \sum_M  N_{ABM} O_M,
	\end{equation}
	and let $h_A$, $h_B$ and $h_M$ denote the holomorphic dimensions of the operators
	$O_A$, $O_B$ and $O_M$ respectively.

	The monodromy (or Aharonov-Bohm phase) operator has eigenvalues
	$e^{2 \pi \i \nu_{AB}^M}= e^{2 \pi i (h_M-h_A-h_B)}$
	\footnote{This is the monodromy corresponding to taking $A$ once around $B$ in an anticlockwise manner.}. In this eigenbasis, the $AB \rightarrow AB$ $S$-matrix takes the form
	\begin{equation}\label{smatrixformintro}
	S_M= \cos \left( \pi \nu_{AB}^{M}\right) \mathcal S_{ {\rm id}} + i\tau_M,
	\end{equation}
	where $\tau_M$ is an analytic function\footnote{ Of course, there is also a momentum-conserving delta function. Throughout this paper $\tau$ refers to the analytic coefficient of this $\delta$ function, which we avoid explicitly displaying in equations in order to avoid clutter.}

	In Appendix \ref{mstn} we present, for completeness,  a detailed
	review of the reasoning that suggests the structure \eqref{smatrixformintro} and also examine some of its consequences. In that Appendix we also demonstrate that 
	the analytic part of the S matrix, $\tau_M$, necessarily has a singularity at 
	$\theta=0$, and demonstrate that both the form of this singularity and its coefficient are precisely determined by $\nu_{AB}^M$, independent of all other 
	dynamical details (see Appendix \ref{upw}).

  The focus of this paper is on the crossing properties of the analytic part of the $S$-matrix, $\tau_M$, which must also be modified in topologically-nontrivial theories.

\subsection{Crossing symmetry}

\subsubsection{Crossing in theories with a global symmetry}

In the limit $k \to \infty$,  Chern-Simons theories reduce (for many purposes) to ungauged theories, with the gauge group turning into an effective global symmetry group. It is thus useful to first recall how crossing symmetry works in topologically-trivial theories with global symmetries $G$ (see section \ref{cugs}).

An $S$-matrix in a topologically trivial theory with a global symmetry is a map from the tensor product of initial, or incoming, $G$-representations, $H_{ {\rm in}}$, to the tensor product of final, or outgoing, $G$-representations, $H_{ {\rm out}}$ \footnote{More precisely, the $S$-matrix is a linear combination of such maps, with coefficients that are functions of initial and final momenta. The momenta play no role in this discussion, and so will be ignored.}. Crossing symmetry relates one $S$-matrix to another in which some particles are replaced by anti-particles, and therefore, alters these representations spaces (e.g., by deleting a final representation space factor $R$ but adding $R^*$ as a factor for the initial representation space). Nonetheless, crossing-symmetry is meaningful because there exists a one-to-one correspondence between $G$-invariant maps from $H_{ {\rm in}}$ to $H_{ {\rm out}}$ and the space of $G$-invariant tensors on
$$H_{ {\rm in}} \otimes H_{ {\rm out}}^* \equiv H,$$
Even though crossing modifies $H_{ {\rm in}}$ and $H_{ {\rm out}}$ individually, it leaves $H$ invariant.
In other words, crossing relates two $S$-matrices that can both be thought of as invariant tensors on $H$, even though they are maps from different initial to different final representation spaces.
From the viewpoint of crossing, therefore, it is more convenient to think of an $S$-matrix as an invariant tensor on $H$ rather than a
map between $H_{ {\rm in}}$ and $H_{ {\rm out}}$.

In order to evaluate the unitarity equation, we are required to multiply the $S$-matrix with its dagger. The rule for multiplication, in invariant tensor language, is given as follows. Let $M$ and $M^*$ respectively be the invariant tensor corresponding to $S$ and $S^\dagger$. In order to compute the invariant tensor corresponding to $S^\dagger S$, we  take the outer product of  $M$ and $M^*$ and then contract all indices associated with $H_{ {\rm out}}$ in $M$  with the corresponding indices associated with $H_{ {\rm out}}$ in $M^*$. This leaves us with an invariant tensor on $H_{ {\rm in}}\otimes H_{ {\rm in}}^*$, which we then equate to the RHS, i.e. to the invariant identity tensor on this space.

Note that while the invariant tensor representation of the $S$-matrix is crossing invariant, the rule for multiplying $S^\dagger$ with $S$  depends on \textit{the crossing frame} (which we define to be the particular decomposition of $H$ into $H_{ {\rm in}}$ and $H_{ {\rm out}}^*$).

In practical computations, it is useful to employ a convenient basis in the space of index structures. As we explain in subsection \ref{pb}  turns out to be possible to find a basis on invariant tensors $T_i$, whose multiplication rules are `orthonormal'. This basis is obtained by Clebsch-Gordon coupling the states in $H_{ {\rm in}}$ -- and separately those of $H_{ {\rm out}}$ -- into product states of definite $G$ representations. Working with this basis, the $S$-matrix may be expanded as
\begin{equation}\label{smatriexpintro}
	S= \sum_i \mathcal{S}^i T_i
\end{equation}
and the unitarity relation, $S^\dagger S=1$, for the $S$-matrices, takes the simple canonical form listed in \eqref{smatrixsuni} below. Note that $\mathcal{S}_i$ denotes the momentum dependent part of the $S$-matrix in the $i^{th}$ channel. Like the multiplication rule itself, the canonical basis for index structures depends on the choice of crossing frame. Let the canonical choice of basis for index structures in crossing frame $\mathcal{F}$ be denoted by $\{ T_i \}$, and the canonical choice in crossing frame $\tilde{\mathcal{F}}$ be denoted by $\{ {\tilde T}_j \}$. As $\{ T_i \}$ and $\{ {\tilde T}_j \}$ span the same space, it follows that
\begin{equation}\label{basischangeintro}
	T_i= M_i^{~j} {\tilde T}_j.
\end{equation}
Consequently, the canonically normalized $S$-matrix coefficients $\tilde {\mathcal{S}}^j$ in the frame ${\tilde{\mathcal{F}}}$ are related to the canonically
normalized $S$-matrix coefficients $\mathcal{S}^i$ in the frame ${\mathcal F}$ via the relationship
\begin{equation}\label{relsmat}
	\tilde{\mathcal{S}}^j= {\mathcal{S}}^i M_i^{~j}.
\end{equation}
(Note that, as usual, \eqref{relsmat} applies after we perform the appropriate analytic continuations of energies.)

Equation \eqref{relsmat}
is the final result for the crossing of canonically normalized $S$-matrices in
topologically trivial massive theories with a global symmetry. As we explain in detail in section \ref{cugs}, the matrices $M_i^{~j}$ -- which are a sort of multi-representation generalizations of the standard $6j$ symbols of classical group theory - 
can be computed using standard group theory techniques.

\subsubsection{Modified crossing rules in topologically-nontrivial gapped theories}

We now turn to the study of crossing in topologically-nontrivial gapped theories.
Throughout this paper, we work only with the example of matter Chern-Simons theories. We suspect our constructions can be generalized to all topologically-nontrivial theories in 3 spacetime dimensions, but we leave the careful verification of this suspicion to future work.

\begin{figure}[h!]
	\centering
	\includegraphics[scale=.25]{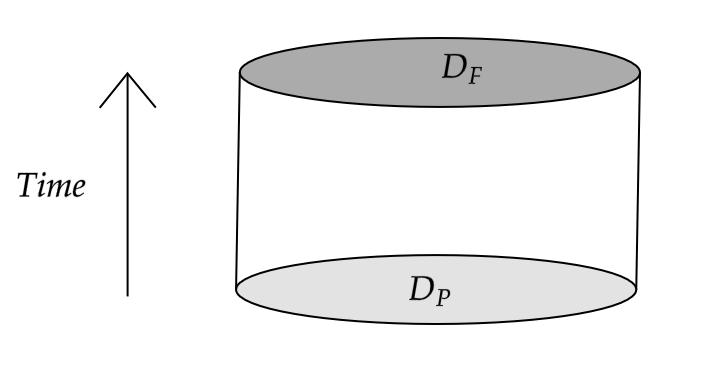}
	\caption{Pillbox depicting regulated version of spacetime}
	\label{pillbox}
\end{figure}

An $S$-matrix is a transition amplitude from early to late times, and so can be extracted from a path integral computed on the spacetime cylinder depicted in Fig \ref{pillbox}. The transition amplitude is the path integral computed as a functional of initial and final data. At every order in perturbation theory, the path  integral over matter fields can be rewritten as a sum over world lines. For any fixed world line configuration, the integral over gauge fields is the expectation value of Chern-Simons Wilson lines with specified
endpoints on the initial and final boundary.

After the usual continuation to  Euclidean space,  the $A_\mu$ path integral is performed  over a solid ball, in the presence of Wilson lines that begin and end at specified points on the top and bottom halves of the $S^2$ that makes up the boundary of this solid ball. \footnote{ We choose the spacetime cylinder in Fig. \ref{pillbox} to be much flatter than it is tall. As a consequence, no massive particle reaches the curved boundary of the cylinder and the Hamiltonian of the boundary WZW modes (which live on  curved boundary of the cylinder) vanishes. After continuation to Euclidean space, therefore, the time interval on this cylinder can be effectively shrunk to zero size and the  boundary cylinder reduces to a  circle. }.
As is well-known, however, such Wilson line expectation values are topologically invariant, and evaluate to conformal blocks (of primary operators in the representations corresponding to the starting-points and endpoints of the Wilson lines) on $S^2$. After performing the sum over all particle trajectories, we thus conclude that the $S$-matrix can be cast into the schematic form
\begin{equation}\label{sumtopintro}
	S= \sum_i \mathcal{S}^i G_i,
\end{equation}
where $G_i$ are a basis of $S^2$ conformal blocks with the given insertions, and $\mathcal{S}^i$ are the momentum dependent coefficient $S$-matrices multiplying these blocks. Comparing \eqref{sumtopintro} with \eqref{smatriexpintro}, we see that the conformal blocks $G_i$ play the same role for $S$-matrices of matter Chern-Simons theories, that the invariant tensors $T_i$ played
for topologically-trivial massive theories with a global symmetry. Like invariant tensors, conformal blocks are also crossing invariant: they do not depend, for their definition, on how one chooses to split the insertions into past and future. Moreover in the limit $k \to \infty$, conformal blocks simply reduce to invariant tensors.

As in the previous subsection, in order to make sense of the unitarity equation, we need to understand the rule for multiplying the blocks corresponding to $S$ and $S^\dagger$. Clearly the physically correct thing to do is to `contract' all
relevant indices \footnote{That is, contract
 the final state indices in
the block corresponding to $S$ with the corresponding complex-conjugated indices in the block corresponding to $S^\dagger$.}  by simply
gluing and continuing the corresponding particle trajectories, i.e.,  gluing and continuing the Wilson lines.
In the limit $k \to \infty$ all gauge fields tend to zero like $\frac{1}{\sqrt{k}}$, so the continuation of Wilson lines reduces to the simple contraction of indices described in the previous subsection on theories with a global symmetry.

In section \ref{ucmcs}, we demonstrate that it is possible to choose a basis in the space of conformal blocks so that the unitarity equation takes exactly the same form, i.e., (\eqref{smatrixsuni} ), when expressed in terms of $\mathcal{S}^i$ in \eqref{sumtopintro} as it did in terms of $\mathcal{S}^i$ in \eqref{smatriexpintro}.
We present an explicit Wilson line construction of the distinguished basis of conformal blocks in section \ref{pbs}. Our construction involves a configuration of bulk Wilson lines that  bifurcate at two or more `three-point bulk vertices' (see Fig. \ref{symasym} for an
example of such a Wilson line configuration).
Our distinguished basis is essentially identical to Witten's construction of an orthogonal basis of conformal blocks given in \cite{Witten:1989wf}, and is obtained  by taking suitable linear combinations of tangles that involve the exchange of an operator in definite representations  between the insertions in $H_{ {\rm in}}$ and $H_{ {\rm out}}$.

As in the previous subsubsection, this canonical basis of blocks depends on the choice of crossing frame. If $G_i$ denotes the canonical choice of basis appropriate to the frame $\mathcal{F}$, and ${\tilde G}_j$ denotes the canonical choice of basis appropriate to the frame $\tilde{\mathcal{F}}$, then
there exist some matrices $N_i^{~j}$ such that
\begin{equation}\label{blockbasechan}
	G_i= N_i^{~j} {\tilde G}_j~~~~
	\tilde {\mathcal{S}}^j= \mathcal{S}^i N_i^{~j}
\end{equation}
\footnote{Even in the case of trivially gapped theories, the rules for crossing symmetry for particles with spin are sometimes more involved than those for their scalar counterparts. In such theories 
S matrices admit an expansion of the form 
$S= \sum_{i} a_i T_i(s, t)$ where $a_i$ are `polarization structures' (see \cite{Chowdhury:2019kaq} for an example of such an expansion). In this situation, while the $T_i(s, t)$ 
enjoy standard crossing properties, the crossing of the polrization structures $a_i$ might involve 
`anomalous' phases, see \cite{Hara:1971kj}. S. Mizera has suggested to us that the modified crossing relations 
proposed in this paper should be viewed as the analogues of 
these anomalous phases, with conformal blocks playing the role 
of polarization tensors. We thank S. Mizera for this interesting comment.}

At finite $k$ the rules for compounding conformal blocks differ from the rules for multiplying two index structures. At finite $k$, consequently, $N_i^{~j}$ in \eqref{blockbasechan} differ from $M_i^{~j}$ in \eqref{basischange}, although, in the limit $k\to \infty$ both are equal. Therefore, at finite $k$, the canonically normalized $S$-matrices $\mathcal{S}_i$ enjoy \textit{different} crossing transformations than  their global symmetry counterparts.

Our construction of the distinguished basis of blocks $G_i$ in terms of Wilson lines tangles is explicit, and allows us to perform practical computations. In section \ref{ucmcs} we use the analysis of \cite{Witten:1989wf} to evaluate the matrices $N$ in terms of expectation values of certain closed tangles of Wilson lines including three-point interaction vertices.
These relevant ratios of tangles can be evaluated explicitly. We demonstrate this fact in the special case $S$-matrices involving the insertion of two fundamentals and two antifundamentals in $SU(N)_k$ or
$U(N)_k$ Chern-Simons theory. Using manipulations explained in \cite{Witten:1989wf}, we are able to explicitly evaluate the matrices $N_i^{~j}$ for this case. We now turn to a study of this special case. \footnote{It is possible that the matrices $N_i^{j}$ will turn out to be standard structures (generalizations of $6j$ symbols?) for the quantum group. We leave an investigation of this suggestion to future work.}
\subsection{Resolution of a puzzle relating to the  scattering of fundamentals at large $N$}

As we have mentioned above, the authors of \cite{Jain:2014nza} ,\cite{Inbasekar:2015tsa} and \cite{Gabai:2022snc} have computed all $2 \rightarrow 2$ $S$-matrices to all orders in the 't Hooft limit in large $N$ fundamental matter Chern-Simons theories. The authors of \cite{Jain:2014nza} \cite{Inbasekar:2015tsa} and \cite{Gabai:2022snc} noted that their results raise a puzzle. The $S$-matrices computed and conjectured in these papers fail to obey standard crossing relations. In order to make sense of their results, the authors of \cite{Jain:2014nza} conjecture that the crossing rules of $S$-matrices in matter Chern-Simons theories differ from those
in trivial theories with a global symmetry. In particular, they conjectured that the modified  crossing rules differ from standard rules by an extra factor of $1/N$ times circular Wilson loop in $S^3$ -- a factor of $ \frac{\sin \pi \lambda}{\pi \lambda}$ -- in the so-called singlet channel.
We will now explain that the analysis of the current paper confirms (and explains) the conjectures of \cite{Jain:2014nza}. More generally, the current paper may be thought of as generalizing the results crossing, conjectured in \cite{Jain:2014nza}, to finite values $N$ and $k$, and also to arbitrary gauge groups.

The general formalism of sections \ref{cugs} and \ref{ucmcs}, reviewed above, instructs us to  choose our canonical basis of  conformal blocks with two fundamentals and two anti-fundamentals as
follows. For fundamental-fundamental scattering we work with the two-dimensional basis of blocks in which the fundamentals fuse to the symmetric or antisymmetric representations, and denote the $S$-matrices that multiply these blocks as $\mathcal{S}_{s}$ and $\mathcal{S}_{a}$ respectively. For fundamental-antifundamental scattering we work with an alternate two dimensional basis of blocks; one in which the initial insertions fuse to either the singlet or the adjoint. The coefficient $S$-matrices are denoted by $\mathcal{S}_{I}$ and $\mathcal{S}_{ {\rm Adj}}$. Following our  general formalism,  we are able to normalize our basis blocks to ensure that their coefficients -- i.e., the $S$-matrices  $\mathcal{S}_{s}$, $\mathcal{S}_{a}$, $\mathcal{S}_{I}$ and $\mathcal{S}_{ {\rm Adj}}$ -- obey the `standard' unitarity relation
\begin{equation}\label{canonunit}
	\mathcal{S}^* \star \mathcal{S}=\mathcal{S}_{ {\rm id}}
\end{equation}
where the operation $\star$ denotes convolution over final state momenta, and
$\mathcal{S}_{ {\rm id}}$ is the delta function at forward scattering.

The discussion of subsection \ref{moi} tells us that the full $S$-matrix in each of these channels must be expanded as
\begin{equation}\begin{split} \label{cosinid}
		& \mathcal S_I=  \mathcal S_{ {\rm id}} \cos \pi \nu_I + i
		\tau_I \\
		& \mathcal S_{ {\rm Adj}}= \mathcal S_{ {\rm id}} \cos \pi\nu_{ {\rm Adj}}  + i \tau_{ {\rm Adj}} \\
		& \mathcal S_s= \mathcal S_{ {\rm id}} \cos \pi\nu_s  + i \tau_s \\
		& \mathcal S_a= \mathcal S_{ {\rm id}} \cos \pi\nu_a  + i \tau_a \\
	\end{split}
\end{equation}
where $\nu$ is defined in \eqref{fluxnu}.\footnote{See \eqref{nuone}, \eqref{nutwo} and \eqref{nusu} for an explicit listing of $\nu$ in the case  Type I, Type II and $SU(N)$ theories.}

As we have mentioned above, it is possible to evaluate the crossing matrices completely explicitly in this very simple case. In section \ref{fundnew} and Appendix \ref{quantexamp}, we argue that the $\tau$ matrices that appear in \eqref{cosinid}
are related by the crossing relations
\begin{equation}\label{Smattrqintro}
	\begin{split}
		&\tau_I=\tau_s \left( \frac{ {\lfloor N+1 \rfloor}_{q}}{{\lfloor 2\rfloor}_{q}} \right) + \tau_a  \left( \frac{ {\lfloor N-1 \rfloor}_{q}}{{\lfloor 2\rfloor}_{q}} \right)  \\
		&\tau_{ {\rm Adj}}= \frac{\tau_s - \tau_a}{{ \lfloor 2 \rfloor}_{q}}\\
		&q=e^{\frac{2\pi i}{\kappa}}, \\
		&\lfloor m\rfloor_q=\frac{q^{m/2}-q^{-m/2}}{q^{1/2}-q^{-1/2}}
	\end{split}
\end{equation}
where the analytic continuation in energies is understood in every equation. Here, $\kappa$ is the renormalized level
$\kappa = {\rm sgn }(k)(|k| + N)$ of the Chern-Simons theory, $k$ is the Chern Simons level (the level of the dual WZW theory). $q=e^{2\pi i/\kappa}$ is the  $\kappa$th root of unity, and $\lfloor m\rfloor_q$ is the `$q$-number' $m$ defined in \eqref{Smattrqintro}. The relations \eqref{Smattrqintro} apply to all of the $SU(N)_k$, Type I $U(N)_k$ theories, and the Type II  $U(N)_k$ theories.

In the limit  $k \to \infty$ with $N$ held fixed, the coefficients of the identity matrices, in \eqref{cosinid} all tend to unity: $\cos \pi\nu_M \to 1$. Moreover, as $q \to 1$, $\lfloor m\rfloor_q \to m$ and
\eqref{Smattrq} reduces to the crossing symmetry for an ungauged theory (i.e., a trivially-gapped theory) with an $SU(N)$ (or $U(N)$) global symmetry
\begin{align}\label{crossclass}
	&\tau_I=\frac{\tau_s(N+1)+\tau_a(N-1)}{2} \nonumber \\
	&\tau_{ {\rm Adj}}=\frac{\tau_s-\tau_a}{2}.
\end{align}

Remarkably enough, the final result for the  `quantum' (i.e., finite $k$) crossing relations are obtained from their `classical'
(i.e., $k \to \infty$) counterparts by simply replacing the numbers $(N+1)$,  $(N-1)$ and $2$ by their $q$ number versions, namely  $\lfloor N-1 \rfloor_q$, $\lfloor N-1 \rfloor_q$, $\lfloor 2 \rfloor_q$.

In the 't Hooft large $N$ limit ($N \to \infty$, $k \to \infty$ with $N/\kappa=\lambda$ held fixed), on the other hand, the coefficients of identity in the second, third and fourth lines of \eqref{cosinid} all tend to unity.  However, the corresponding coefficient of identity in the first line of \eqref{cosinid} becomes $\cos \pi \lambda$. Moreover, the crossing relations \eqref{Smattrqintro} become
\begin{equation}\label{crossthoft}
	\begin{split}
		&\tau_I=N  \left( \frac{\sin\pi\lambda} {\pi \lambda} \right)
		\left( \frac{\tau_s+\tau_a}{2} \right) \\
		&\tau_{ {\rm Adj}}= \left( \frac{\tau_s-\tau_a}{2} \right).
	\end{split}
\end{equation}

The extra factor of $\frac{\sin \pi \lambda}{\pi \lambda}$  in \eqref{crossthoft} -- compared to \eqref{crossclass} --  was precisely the crossing symmetry puzzle flagged by
\cite{Jain:2014nza}, discussed at the beginning of this subsubsection. We see this initially-puzzling extra factor is a simple and automatic consequence of the  modified crossing rules developed in this paper.

\section{Crossing and unitarity in theories with a global symmetry}\label{cugs}

In this section we review crossing symmetry, unitarity, and their interplay in massive, topologically-trivial theories with a continuous global symmetry.

\subsection{Invariant tensors} \label{itcl}

Consider the space
\begin{equation}\label{groupinvariantsmt}
	H\equiv R_1 \otimes R_2 \otimes \ldots \otimes R_{n+m}.
\end{equation}
where $R_1,~R_2,~ \ldots R_{n+m}$ are irreducible representations of the continuous global symmetry $G$.
Consider a tensor
\begin{equation} \label{whatisT}
	T^{{\vec m}_{a_1}  \ldots {\vec m}_{a_{n}}  {\vec n}_{a_{n+1}} \ldots {\vec n}_{a_{n+m}}}
\end{equation}
that lives in the product space \eqref{groupinvariantsmt}. Here ${\vec m}_{a_i}$ are representation indices in the $R_i^{th}$
representation. By definition, every such tensor transforms under group rotations so that the rotated tensor, $T_G$, is given by
\begin{align}\label{grouprotten}
	T_G^{{\vec m}_{a_1}  \ldots {\vec m}_{a_{n}}  {\vec n}_{a_{n+1}}  \ldots {\vec n}_{a_{n+m}}} = (G^{a_1})^{\vec{m}_{a_1}}_{\vec{m}'_{a_1}}  \ldots (G^{a_{n+m}})^{\vec{n}_{a_{n+m}}}_{\vec{n}'_{a_{n+m}}}T^{{\vec m'}_{a_1}  \ldots {\vec m'}_{a_{n}}  {\vec n'}_{a_1} \ldots {\vec n'}_{a_{n+m}}}  \nonumber\\
\end{align}
where $G^{a_i}$ are the group rotation matrices in the $a_i^{th}$ representation.

A tensor is defined to be group invariant if the rotated and unrotated tensors are equal, i.e., if
$T_G=T$.

\subsection{Invariant maps}

Consider a linear operator $O$ that maps the `initial' Hilbert space
\begin{equation} \label{inspacemt}
	H_{ {\rm in}} \equiv R_{1} \otimes R_{2} \otimes \ldots \otimes R_{n}
\end{equation}
to the `final' Hilbert space
\begin{equation}\label{findpacemt}
	H_{ {\rm out}} \equiv R_{{n+1}}^* \otimes R_{{n+2}}^* \otimes \ldots \otimes R_{{n+m}}^*
\end{equation}
Note that $H= H_{ {\rm in}} \otimes H_{ {\rm out}}^*$.

By definition, $O$ is group invariant if
\begin{equation}\label{gio}
	O|\psi \rangle = |\chi \rangle ~~~\implies ~~~O G |\psi \rangle = G|\chi \rangle
	~~~{\rm i.e.,} ~~~{\rm if} ~~~ G^\dagger O G  = O
\end{equation}
for every group element $G$. Here, and throughout this section, we assume that the global symmetry group is unitarily represented on all relevant representations, so that $G^{-1}=G^\dagger$.

Let $|{\vec m_a} \rangle$
constitute an orthogonal basis for the Hilbert space $R_a$ and let the action of the group rotation generators be given by
\begin{equation}\label{grosp}
	G |{\vec m}_a \rangle = \left( G^a \right) ^{\vec m_a'}_{\vec m_a}|{\vec m_a'} \rangle
\end{equation}
Every map from \eqref{inspacemt} to \eqref{findpacemt} can be written in the
form
\begin{equation} \label{optens}
	M_T
	=\left( T^{{\vec m}_{a_{n+1}}  \ldots {\vec m}_{a_{n+m}}  {\vec n}_{a_1} \ldots  {\vec n}_{a_n}} \right) |m_{a_{n+1}}\rangle \ldots |m_{a_{n+m}} \rangle  \langle n_{a_1}|\ldots \langle n_{a_{n}} |.
\end{equation}
It is easily verified that
\begin{equation} \label{mrot}
G M_{T} G^\dagger = M_{T_G}.
\end{equation}
In other words, the $G$ rotation of the operator $M_T$ is  the operator  $M_{T_G}$, where $T_G$ is given in  \eqref{grouprotten}. We have thus established a one to one correspondence  between  $G$ invariant tensors on the space \eqref{groupinvariantsmt} and $G$ invariant maps from  \eqref{inspacemt} to \eqref{findpacemt}.

It is easy to convince oneself that the tensor
corresponding to the operator $M_T^\dagger$
(which maps $H^*_{ {\rm out}}$ to $H_{ {\rm in}}$) is the tensor $T^*$, (here $*$ represents simple complex conjugation). This tensor lives in the space $H^*$.

\subsection{Crossing}  \label{crossing}

Consider a particular invariant tensor $T$ on the space $H$. Two separate divisions of the factor representations of $H$ into initial and final associate the same invariant tensor $T$ with the operators $O$ and $O'$. In general, $O$ and $O'$ are maps between distinct initial and final spaces.  Crossing invariance is the claim that the $S$-matrices (functions of momenta) that multiply $O$ and $O'$ are related by analytic continuation.

The discussion of the previous paragraph can be repeated in equations. Let $M_{T}$ and $\tilde M_{T}$ denote the operators -- corresponding to the same invariant tensor $T$ --  for two different divisions of $H$ into $H_{ {\rm in}}$ and $H_{ {\rm out}}^*$. Let $T_i$ be a for the space of of invariant tensors, and let the $S$-matrix associated with the two different divisions of
	$H$ into $H_{ {\rm in}}$ and $H_{ {\rm out}}^*$ take the form
	\begin{equation} \label{scheform}
		S= \sum_i  \mathcal S_i M_{T_i}
	\end{equation}
	and
	\begin{equation} \label{scheform}
		S= \sum_i\tilde {\mathcal S}_i \tilde M_{T_i}
	\end{equation}
	Then the functions of momenta,  $\mathcal S_i$ and $\tilde{\mathcal S}_i$ are analytically related via crossing.

\subsection{Compounding invariant tensors} \label{csm}

Consider a map $M_{T}$ from \eqref{inspacemt}  to \eqref{findpacemt} followed by map $M'_{T'}$ from \eqref{findpacemt} to the space
\begin{equation} \label{yanmt}
	R^{'*}_{a_1} \otimes \ldots \otimes R^{'*}_{a_p}
\end{equation}
Here $T$ is an  invariant tensor on the space $H$ listed in \eqref{groupinvariantsmt}  and $T'$ an invariant tensor on the space
\begin{equation}\label{spacefnsc}
	R^*_{a_{n+1}} \otimes R^*_{a_{n+2}} \otimes \ldots \otimes R^*_{a_{m+n}} \otimes R'_{a_1}
	\otimes \ldots \otimes R'_{a_p}
\end{equation}
\footnote{Explicitly, $T'$ takes the form $$(T')^{  {\vec n}^*_{a_{n+1}}  \ldots {\vec n}^*_{a_{n+m}}{\vec m}_{a'_1}  \ldots {\vec m}_{a'_{p}}}$$.}
The tensor $TT'$ obtained by compounding the tensors $T$ and $T'$, can be found by multiplying their corresponding operators $M_{T}$ and $M'_{T'}$ as follows:
\begin{equation}
	\label{multpo}
	M'_{T'}M_T=M''_{TT'}.
\end{equation}
By explicitly evaluating the
LHS of \eqref{multpo}, the reader may easily verify that $TT'$ is simply given by multiplying $T$ and $T'$ and contracting the indices associated with the representations $R_{a_{n+1} }
\ldots R_{a_{n+m} }$. \footnote{ The result of this operation is a tensor on the space
	\begin{equation}\label{spacefnsc}
		R_{a_{1}} \otimes R_{a_{2}} \otimes \ldots \otimes R_{a_{n}} \otimes R'_{a_1}
		\otimes \ldots \otimes R'_{a_p}
	\end{equation}
	as expected.} Since the indices we contract transform in mutually complex conjugate
representations, the gluing procedure preserves the group invariance. Assuming that $T$ and $T'$ are $G$ invariant, it follows that the same is true of $TT'$.

In this paper we will use this procedure to multiply $S$ with $S^\dagger$. In this case the tensor corresponding to $S^\dagger$ is the complex conjugate of the tensor corresponding to $S$, and indices we contract correspond to $H_{ {\rm out}}^*$ in $S$ and $H_{ {\rm out}}$ is $S^\dagger$. The result is an invariant tensor on the space
$H_{ {\rm in}} \otimes H_{ {\rm in}}^*$.

\subsection{Unitarity}

Consider the $S$-matrix given in \eqref{scheform}.
The unitarity condition $S^\dagger S=I$ tells us that
\begin{equation} \label{scheform}
	\sum_{{\rm final~states}} \sum_{i, j}  \left( \mathcal S_j^* \star \mathcal S_i \right) ~M^\dagger_{T_j} M_{T_i} = \mathcal S_{ {\rm id}} M_{ {\rm id}}
\end{equation}
where $\star$ denotes convolution in momentum space
\footnote{In more detail, convolution is defined by identifying set of final momenta of $S_1$  matrix with the initial momenta of $S_2$ and then integrating over identified momenta with the measure
	\begin{equation}\label{measure}
		\prod_i \frac{d^3 p_i}{(2 \pi)^3} (2 \pi) \delta(p_i^2+m^2)
\end{equation} }
w.r.t. the measure \eqref{measure}, $\mathcal S_{ {\rm id}}$ is the momentum space representation of identity (no scattering) and $M_{ {\rm id}}$ is the identity matrix on the space \eqref{inspacemt}. The summation over final states in \eqref{scheform} accounts for the fact that a given set of initial representations could scatter into many distinct collections of final representations.

\subsection{The `projector' basis} \label{pb}

In this subsection we  make a convenient choice of basis for the space of invariant tensors $T_i$. As we will see below, this  choice simplifies the unitarity formula \eqref{scheform} and recasts it into a canonical form. As explained in the introduction, the natural basis for a particular scattering process differs from the natural basis for its crossing related counterpart. In this section we discuss these various choice of bases, and the linear transformations between them.
\subsubsection{Basis of projectors}

Let us suppose that the Clebsch-Gordan decomposition of the classical tensor product of the space \eqref{inspacemt}
takes the form
\begin{equation}\label{classdecomp}
	R_{1} \otimes R_{2}\otimes  \ldots \otimes  R_{n}
	= \sum_{a} Q_a \tilde R_a
\end{equation}
where the index $a$ runs over all the unitary irreducible representations of the global symmetry group, and the positive integers $Q_a$ denote the number of times $R_a$ appears in the fusion.

When $Q^a \neq 1$, it is convenient to choose an orthogonal basis in the space of tensor product states that transform in representation $R_a$.
In other words we work with a collection of states  $|{\vec m} \rangle_{a, r}$ ($r=1 \ldots Q_a$) that obey the following properties.
\begin{itemize}
	\item Under the action of a global symmetry generator on the LHS of \eqref{classdecomp}, the states transform according to some standard representation matrices $(G^a)^{{\vec m}}_{{\vec m}'}$ of the $a^{th}$ irrep of the global symmetry algebra, such that the matrices $G^a$ are unitary, i.e., $G^a (G^a)^\dagger =1$.
	More precisely, the transformation rule for the basis states is given by
	\begin{equation}\label{grosp}
	G |{\vec m}_a \rangle = \left( G^a \right) ^{\vec m_a'}_{\vec m_a}|{\vec m_a}' \rangle
	\end{equation}
	\item The states $|{\vec m} \rangle_{a, r}$ are orthonormal, i.e.
	\begin{equation}\label{orthog}
		_{a, r}\langle {\vec m} | {\vec m}' \rangle_{a',r'} = \delta_{{\vec m} {\vec m}'} \delta_{a a'} \delta_{r, r'}
	\end{equation}
\end{itemize}

In subsection \ref{pis} below, we provide one explicit construction of the states $|{\vec m} \rangle_{a, r}$. For the purposes of this section we do not need this explicit construction: all results we need will follow on general grounds.

In a similar manner, suppose
\begin{equation}\label{classdecomptilde}
	{R}^*_{{n+1}} \otimes {R}^*_{{n+2} } \otimes  \ldots \otimes  { R}^*_{m+n}
	= \sum_{a} { Q'}_a R'_a.
\end{equation}
Once again we can define the states $ { |{\tilde{\vec m}} \rangle}_{a, r}$
$r=1 \ldots {Q'}_a$ by the conditions analogous to those above. In particular, the matrices  $(G^{a})^{{\vec m}}_{{\vec m}'}$ govern the symmetry transformations of the states  $ {|{\tilde{\vec m}} \rangle}_{a, r}$ exactly as in \eqref{grosp}.

Let us now define the operators
\begin{equation} \label{projops}
	P_a^{r r'} = \sum_{{\vec m}}   {|{{\vec m}} \rangle}_{a, r'} ~ _{a, r}\langle {\vec m} |.
\end{equation}
By slight misuse of terminology we will sometimes refer to the operators $P_a^{r r'}$
as projectors onto the space $R_a$. \footnote{
The invariance of $P_a^{r r'}$ under group transformations may be verified as follows:
\begin{align} \label{pinvar}
	U P_a^{r r'} U^\dagger &= \sum_{n,l}\sum_m (G^{a})^n_m (G^{\dagger a})^m_l |\vec{ n}\rangle_{a,r'}~ _{a,r}\langle\vec{l}| \nonumber \\
	&=\sum_{n,l}\delta^n_l |\vec{n}\rangle_{a,r'}  ~_{a,r}\langle\vec{l}| \nonumber \\
	&=\sum_n |\vec{ n}\rangle_{a,r'} ~_{a,r} \langle\vec{n}|=P^{rr'}_a.
\end{align}
}

Clearly $P_a^{r r'}$ and $(P^\dagger)_a^{r r'}$, respectively, constitute a basis of invariant maps from $H_{ {\rm in}} \rightarrow H_{ {\rm out}}$ and
from $H_{ {\rm out}} \rightarrow H_{ {\rm in}}$ respectively.
The utility of this basis lies in the fact that the compounding or multiplication rules of
$O^\dagger$ and $O$ are `orthogonal' in this basis:
\begin{equation} \label{projopsnn}
	(P_a^{r_1 r_2})^\dagger  P_{a'}^{r_3 r_4} = \delta_{a a'} \delta_{r_2, r_4}
	\sum_{{\vec m}}  |{\vec m} \rangle_{a, r_1} ~ _{a, r_3} \langle {\vec m} | = \delta_{a a'} \delta_{r_2, r_4}  {\hat P} _a^{r_3 r_1}
\end{equation}
In \eqref{projopsnn}, the operators ${\hat P} _a^{r_3 r_1}$  map the space $H_{ {\rm in}}$ onto itself. These operators themselves obey multiplication rules that are closely analogous to \eqref{projopsnn}
\begin{equation} \label{projopsn}
	{\hat P}_a^{r_1 r_2} ({\hat P}_{a'}^{r_3 r_4})
	=  \delta_{a a'} \delta_{r_1, r_4}  {{\hat P}}_a^{r_3 r_2}, ~~~~~\left( {\hat P}_a^{r_1 r_2} \right)^\dagger
	= {\hat P}_a^{r_2 r_1}
\end{equation}
and also obey the additional identity
\begin{equation}\label{projcomplete}
	\sum_{a, r} {\hat P}_a^{rr} = M_{ {\rm id}};
\end{equation}
where $M_{ {\rm id}}$ is the identity operator
on $H_{ {\rm in}}$.

While we have worded the discussion of this section in the language of operators from $H_{ {\rm in}} \rightarrow H_{ {\rm out}}$, we can
also work in terms of invariant tensors on $H$. We will use the symbol
$$T_a^{r r'}$$
to denote the index structure associated with the projector ${P}_a^{r r'}$ . The identity index structure on $H_{ {\rm in}} \otimes H_{ {\rm in}}^*$ will be denoted by $T_{ {\rm id}}$.
\subsection{Unitarity in the projector basis}
\label{upbc}

The most general $S$-matrix,  for the  scattering of particles in $H_{ {\rm in}}$ to $H_{ {\rm out}}$ can be taken to be
\begin{equation}\label{expofsmatrixmt}
	S = \sum_{a, r_1, r_2} \mathcal{S}^{r_1, r_2}_a P_a^{r_1 r_2}
\end{equation}
Using \eqref{projopsn} and \eqref{projcomplete}, it follows that the unitarity equation
\eqref{scheform} becomes
\begin{equation}\label{smatrixsuni}
	\sum_{\rm final ~states} \sum_{r'} ( \mathcal S_a^{r_1r'})^* \star  \mathcal S_a^{r_2r'} = \mathcal S_{ {\rm id}} \delta_{r_1, r_2}
\end{equation}
where $\star$ denotes convolution in scattering momentum space according to the rule \eqref{measure} and $\mathcal S_{ {\rm id}}$ denotes the identity $S$-matrix in momentum space.

\eqref{smatrixsuni} can be rewritten as
\begin{equation}\label{smuimf}
	\begin{split}
		& \sum_{\rm final ~states} \mathcal S_a \star \mathcal S_a^\dagger =\mathcal{S}_{ {\rm id}} \mathcal{I}_a\\
	\end{split}
\end{equation}
where we have defined the matrices $S_a$
\begin{equation}\label{sasadagcl}
	\left( \mathcal S_a \right)_m^{~n}= \mathcal S_a^{mn}, ~~~~
	\left( \mathcal S_a^\dagger\right)_m^{~n}= \mathcal S_a^{*nm}, ~~~~\left( \mathcal I_a \right)_m^{~n} = \delta^n_m
\end{equation}

We see from \eqref{sasadagcl} that, in the projector basis $S$ and $S^\dagger$ are block diagonal matrices, with one block for every representation $a$.  In the special case $Q_a=Q_a'=1$, the $a^{th}$ block is $1 \times 1$ and the unitarity equation in the block $a$ takes exactly the same form as in a theory with no global symmetry.

As a slight aside, we note that in the special case of scattering in which the initial particles are identical to the final particles, the $S$-matrix is expanded as $\mathcal S_{ {\rm id}}+i\tau$
(see Appendix \ref{mstn}). In this case $P_{a}^{rr'}={\hat P}_a^{rr'}$, and so it follows from  \eqref{projcomplete} that the component $S$-matrices admit the expansion
\begin{equation}\label{compsmat}
{\mathcal S}^{rr'}_a= \delta^{rr'} \mathcal S_{ {\rm id}}
+ i \tau^{rr'}_a
\end{equation}

\subsection{Crossing in the projector basis}

Let $T_a^{r r'}$ denote the index structure dual to the projector $P_a^{r r'}$ in a particular crossing frame, and
let ${\tilde T}_b^{s s'}$ denote the index structure dual to the projector ${\tilde P}_b^{s s'}$ in a different crossing frame.
Since $T_a^{r r'}$ and ${\tilde T}_b^{s s'}$ individually constitute a basis of the space of invariant index structures on
$H$, it follows that
\begin{equation}\label{basisinbolind}
	T_a^{r r'} = M_{a r r'}^{b s s'} {\tilde T}_b^{s s'}.
\end{equation}
The matrix $M_{a r r'}^{b s s'}$ is a purely group-theoretical object. In the case of $2 \times 2$ scattering, $M_{a r r'}^{b s s'}$ are called 6j symbols, and are well studied. In general, $M_{a r r'}^{b s s'}$ are generalizations of 6j symbols to
higher fusions of representations.

Crossing symmetry is the claim that
\begin{equation}\label{claimcrossind}
	\sum_{a r r'}  M_{a r r'}^{b s s'}\left( \mathcal S\right)_a^{r r'} =\left( \tilde{\mathcal S}\right)_b^{s s'}
\end{equation}
where the equality in \eqref{claimcrossind} holds after the appropriate analytic continuation in cross ratios.
As we have emphasized above, the quantities $M_{a r r'}^{b s s'}$ that appear in \eqref{claimcrossind} are purely group-theoretical objects.

The discussion presented so far in this subsection is modified somewhat in the
case when some of the inserted fields are identical. We explore the nature of this
modification -- which however plays no role in the study of scattering -- in
Appendix \ref{ipg}.

\subsection{Basis dependence of the crossing rules} \label{bdcr}

In this section we have discussed how the coefficient $S$ matrices,
$\mathcal{S}^{r_1, r_2}_a$ transform under crossing. By definition, $\mathcal{S}^{r_1, r_2}_a$ are the coefficients, in the general expansion of the $S$-matrix, of
the `projector' invariant tensors ${P}^{r_1, r_2}_a$ (see \eqref{projops}.) These `projectors' were defined in terms of the states $|{\vec m} \rangle_{a, r}$, which, in turn were defined to obey several properties. While the list of requirements for the states $|{\vec m} \rangle_{a, r}$ constrains their form, it does not determine them uniquely.
For instance, a phase rotation $|{\vec m} \rangle_{a, r}
\rightarrow e^{i \phi(a, r)} |{\vec m} \rangle_{a, r}$ preserves all the properties demanded of these states in subsection \ref{pb}.

Since the ${P}^{r_1, r_2}_a$ are only well defined up certain ambiguities, the crossing relations between their coefficients can also be definitely determined upto certain (generalized) phase ambiguities. We discuss this point in greater detail in section \ref{pis} below, after presenting  a more detailed definition of the states $|{\vec m} \rangle_{a, r}$
than we have provided so far.

\subsection{Scattering of two fundamentals and two antifundamentals in $SU(N)$} \label{Classexamp}

Consider a special case of the symmetry group $SU(N)$ or $U(N)$ and tensor product space (special case of \eqref{groupinvariantsmt})
\begin{equation}\label{tpgi}
	R_F \otimes R_F \otimes R_F^* \otimes R_F^*
\end{equation}
where $R_F$ is the fundamental representation and $R_F^*=R_{{A}}$ is the antifundamental representation.

Let $i$ and $i'$ be the indices for the first and second fundamental representations respectively
and let $j$ and $j'$ be the indices for the first and second antifundamental indices. Our index conventions are as follows: while the fundamental state $|i\rangle$ transforms like a lower $i$ index,  the antifundamental state $|j\rangle$ carries an upper $j$ index. Complex conjugation (or changing a ket to a bra) raises/lowers
indices. From \eqref{optens} we see that indices of invariant tensors contract with bras:  consequently `fundamental' indices in invariant tensors are lower while antifundamental indices are upper.

\begin{align}\label{ginve}
	\left( T_d \right)^{~~jj'}_{ii'} =\delta^j_i \delta^{j'}_{i'}  \nonumber \\
	\left( T_e \right)^{~~jj'}_{ii'} =\delta^{j'}_i \delta^j_{i'}
\end{align}
constitute a basis for the two dimensional vector space of invariant tensors.
The Hermitian conjugates of this basis are given by
\begin{align}\label{ginve}
	\left( T_d^\dagger \right)_{~~jj'}^{ii'} =\delta_j^i \delta_{j'}^{i'}  \nonumber \\
	\left( T_e^\dagger \right)_{~~jj'}^{ii'} =\delta_{j'}^i \delta_j^{i'}
\end{align}
The indices $i$ $i'$, $j$, $j'$ respectively are associated with the particles $1$, $2$, $3$ and $4$ (the particles that carry momentum $p_1$, $p_2$, $p_3$ and $p_4$ respectively). \footnote{Through this subsection present all tensors so that  first, second, third and fourth indices (when read from left to right) pertain to the first, second, third and fourth particle.}

We now turn to the construction of the projector basis.
\subsubsection{Projectors for fundamental-fundamental scattering } \label{ffs}

Consider the scattering process
\begin{equation} \label{scpro}
	|i\rangle + |i' \rangle  \rightarrow |j\rangle + |j^{'} \rangle
\end{equation}
Note that in this case $H_{ {\rm in}}=H_{ {\rm out}}= R_F\otimes R_F$ ($R_F$ is the fundamental representation). In particular, all four indices that appear in \eqref{scpro} are fundamental indices. \footnote{The RHS has complex conjugate of antifundamental, and so fundamental indices.}

We define the projector index structures
\begin{align}\label{ppscatff}
	&T_{s}= \frac{T_d + T_e}{2} ~~~~{\rm so ~that} ~~~~~ \left( T_s \right)^{~~jj'}_{ii'}=\frac{\delta^j_i\delta^{j'}_{i'}+\delta^{j'}_i\delta^{j}_{i'}}{2} ~~~{\rm and} ~~~\left( T_s^\dagger \right)_{~~jj'}^{ii'}=\frac{\delta_j^i\delta_{j'}^{i'}+\delta_{j'}^i\delta_{j}^{i'}}{2}\\
	&T_{a}=\frac{T_d - T_e}{2} ~~~~{\rm so ~that} ~~~~~ \left( T_a \right)^{~~jj'}_{ii'}= \frac{\delta^j_i\delta^{j'}_{i'}-\delta^{j'}_i\delta^{j}_{i'}}{2}  ~~~{\rm and} ~~~\left( T_a^\dagger \right)_{~~jj'}^{ii'}=\frac{\delta_j^i\delta_{j'}^{i'}-\delta_{j'}^i\delta_{j}^{i'}}{2}
\end{align}
Multiplication of $T^\dagger$ with $T$ involves contraction of the $j$ and $j'$ indices. It is easy to verify that
\begin{equation}\label{multiplicationrule}  \begin{split}
		& T_s^\dagger T_s=  \left( T_s^\dagger \right)_{~~jj'}^{i_2i'_2}\left( T_s \right)^{~~jj'}_{i_1 i'_1}=\frac{\delta^{i_2}_{i_1}\delta^{i'_2}_{i'_1}+\delta^{i_1'}_{i_1}\delta^{i_2}_
			{i'_1}}{2} = {\hat T}_s\\
		& T_a^\dagger T_a=  \left( T_a^\dagger \right)_{~~jj'}^{i_2i'_2}\left( T_a \right)^{~~jj'}_{i_1 i'_1}=\frac{\delta^{i_2}_{i_1}\delta^{i'_2}_{i'_1}-\delta^{i_1'}_{i_1}\delta^{i_2}_
			{i'_1}}{2}  = {\hat T}_a\\
		& T_s^\dagger T_a=T_a^\dagger T_s=0 \\
	\end{split}
\end{equation}
Of course ${\hat T}_s$ and ${\hat T}_a$ that appear in \eqref{multiplicationrule} obey \eqref{projopsnn}. Note that in this special case, as $H_{ {\rm in}}=H_{ {\rm out}}$, $T_s$ and $T_a$
live in the same space as ${\hat T}_s$ and ${\hat T}_a$. We have chosen the phases of projectors to ensure that $T_s={\hat T}_s$ and $T_a={\hat T}_a$. This choice ensures that $T_s$ and $T_a$ also obey \eqref{projopsnn}.
Moreover $T_{s/a}^\dagger=T_{s/a}$.

\subsubsection{Projectors for fundamental-antifundamental scattering} \label{fas}
Now consider the following scattering process:
\begin{equation}\label{papscat}
	|i \rangle + |j^*\rangle \rightarrow |i'\rangle + |j^{'*}\rangle.
\end{equation}
In this case, $H_{ {\rm in}}=H_{ {\rm out}}= R_F \times R_{{A}}$.
Let us define the projector index structures
\begin{align}\label{papscatfa}
	&T_{I}= \frac{T_d}{N} ~~~~{\rm so ~that} ~~~~~ \left( T_I \right)^{~j~j'}_{i~i'}=\frac{\delta^j_i\delta^{j'}_{i'}}{N} ~~~{\rm and} ~~~\left( T_I^\dagger \right)_{~j~j'}^{i~i'}=\frac{\delta^i_j\delta^{i'}_{j'}}{N}
	\\
	&T_{ {\rm Adj}}=T_e - \frac{T_d}{N} ~~~~{\rm so ~that} ~~~~~ \left( T_{ {\rm Adj}} \right)^{~j~j'}_{i~i'}= \delta^{j'}_i\delta^j_{i'}-\frac{\delta^{j}_i\delta^{j'}_{i'}}{N}  ~~~{\rm and} ~~~\left( T_{ {\rm Adj}}^\dagger \right)_{~j~j'}^{i~i'}=\delta_{j'}^i\delta_j^{i'}-\frac{\delta_{j}^i\delta_{j'}^{i'}}{N}
\end{align}
Multiplication involves the contraction of $i'$ and $j'$ indices. We find
\begin{equation}\label{multiplicationrule2}  \begin{split}
		& T_I^\dagger T_I= \left( T_I^\dagger \right)_{~j~j_2}^{i~i_2}\left( T_I \right)^{~j_1~j}_{i_1~i}=\frac{\delta^{i_2}_{j_2}\delta^{j_1}_{i_1}}{N}
		= {\hat T}_{I}
		\\
		& T_{ {\rm Adj}}^\dagger T_{ {\rm Adj}}=  \left( T_{ {\rm Adj}}^\dagger \right)_{~j~j_2}^{i~i_2}\left( T_{ {\rm Adj}} \right)^{~j_1~j}_{i_1~i}=\delta_{j_2}^{i_1}\delta_{j_1}^{i_2}-\frac{\delta^{i_2}_{j_2}\delta^{j_1}_{i_1}}{N} = {\hat T}_{ {\rm Adj}}\\
		& T_I^\dagger T_{ {\rm Adj}}=T_{ {\rm Adj}}^\dagger T_I=0. \\
	\end{split}
\end{equation}
As in the previous subsubsection, we  have chosen the phases of projectors to ensure that $T_I={\hat T}_I$ and $T_{ {\rm Adj}}={\hat T}_{ {\rm Adj}}$, and choice ensures that $T_I$ and $T_{ {\rm Adj}}$ also obey \eqref{projopsnn}.
Moreover $T_{I}^\dagger=T_{I}$ and $T_{{ \rm Adj}}^\dagger=T_{{ \rm Adj}}$.

\subsubsection{Crossing and unitarity}

Using \eqref{ppscatff} and \eqref{papscatfa}, it is easy to verify that
	\begin{align}\label{pptoap}
		&T_s=\frac{(N+1)T_I+ T_{ {\rm Adj}}}{2} \nonumber \\
		&T_a=\frac{(N-1)T_I - T_{ {\rm Adj}}}{2},
	\end{align}
Let the most general fundamental - fundamental  $S$-matrix be given -- in invariant tensor notation -- by
\begin{align} \label{smatff}
	\left(S\right)^{~~jj'}_{ii'}=\mathcal S_s \left(T_s\right)^{~~jj'}_{ii'} + \mathcal S_a \left(T_a\right)^{~~jj'}_{ii'},
\end{align}

Similarly let the most general fundamental- antifundamental $S$-matrix be given by
\begin{align}
	\left(S \right)^{~j~j'}_{i~i'}=\mathcal S_{ {\rm Adj}} \left(T_{ {\rm Adj}}\right)^{~j~j'}_{i~i'} + \mathcal S_I  \left(T_I\right)^{~j~j'}_{i~i'}.
\end{align}
Crossing is the claim that $S$-matrices that correspond to the
same invariant tensor are related by analytic continuation. It follows that
\begin{align}\label{StSs}
	&\mathcal \tau_I=\frac{\mathcal \tau_s(N+1)+\mathcal \tau_a(N-1)}{2} \nonumber \\
	&\mathcal \tau_{ {\rm Adj}}=\frac{\mathcal \tau_s-\mathcal \tau_a}{2},
\end{align}
where the analytic continuation between the two sides of \eqref{StSs} is understood.

All the $S$-matrices that appear in this section are normalized so their contribution to the unitarity equation is
\begin{equation}\label{unitarityexample} \begin{split}
		&(\mathcal S_I)^*\star \mathcal S_I + \ldots = \mathcal S_{ {\rm id}} \\
		&(\mathcal S_{ {\rm Adj}})^*\star \mathcal S_{ {\rm Adj}} + \ldots = \mathcal S_{ {\rm id}} \\
		&(\mathcal S_s)^*\star \mathcal S_s + \ldots  = \mathcal S_{ {\rm id}} \\
		& (\mathcal S_a)^*\star \mathcal S_a + \ldots = \mathcal S_{ {\rm id}} \\
	\end{split}
\end{equation}
where $\ldots$ denotes the contribution of all other processes (e.g. $2 \rightarrow 3$, $2 \rightarrow 4$ etc) to the unitarity equation.

\subsection{An explicit construction for `projector' index structures} \label{pis}

We end this section by clearing up a loose end.

The projector invariant tensors, $T_a^{r r'}$, were constructed  in subsection \ref{pb} with the aid of the states $|{\vec m} \rangle_{a, r}$ that coupled all initial insertions (and final insertions) into states that transform in the
$R_a$ representation. In subsection \ref{pb} we
defined these states somewhat abstractly. In preparation for the
generalization of the construction of subsection \ref{pb} to the Chern-Simons case, it will be useful to present a more
concrete algorithm for actually constructing the states $|{\vec m} \rangle_{a, r}$.

To start our construction, let  us suppose that the (Lie group) fusion of representations $R_a$ with $R_b$ produces the representation $R_c$ $N^{\rm cl}_{abc}$ times \footnote{The superscript ${\rm cl}$ in $N^{\rm cl}_{abc}$ reminds us that these integers are fusion coefficients of the classical group theory, in contrast with the WZW fusion coefficients we will deal with in the next section.} . For each choice of $a, b, c$ we will  pick a basis in the $N^{\rm cl}_{abc}$ dimensional space of Clebsch-Gordan (CG) coefficients,
$(C^r)^{{\vec m}_a \vec m_b}_{{\vec m}_c}$ ($r=1 \ldots N^{\rm cl}_{abc}$) that ensures that the states
\begin{equation} \label{stateorth}
	|{\vec m_c}\rangle_{r}=  (C^r)_{{\vec m}_c}^{{\vec m}_a{\vec m}_b} | {\vec m}_{a} \rangle 	| {\vec m}_{b} \rangle
\end{equation}
obey the following orthonormality relations
\begin{equation}\label{orthonorm}
	_{r'}\langle {\vec m}'_{c'} | {\vec m_c}\rangle_{r} = \delta_{ {\vec m}, {\vec m}'}  \delta_{c, c'} \delta_{r, r'}
\end{equation}
The fact that the RHS of \eqref{orthonorm} is proportional to  $\delta_{ {\vec m}, {\vec m}'}  \delta_{c, c'}$ is an automatic consequence of group invariance. The requirement that the coefficient $\delta_{ {\vec m}, {\vec m}'}  \delta_{c, c'}$ equals $\delta_{r, r'}$ is a constraint on the choice of basis CG coefficients.
Plugging \eqref{stateorth}
into \eqref{orthonorm}, we find that \eqref{orthonorm} is obeyed if and only if
\begin{equation}\label{cgorthog}
	(C^r)_{{\vec m}_c}^{{\vec m}_a{\vec m}_b}
	(C^{r'})^{*~{\vec m'}_{c'}} _{{\vec m}_a{\vec m}_b}
	=\delta_{\vec m_c}^{\vec m'_{c'}}  \delta_{c, c'} \delta_{r, r'}
\end{equation}
(repeated indices are summed over in this equation).  Because the dependence of the RHS on $\vec m_c$ and $\vec m'_{c'}$ is determined by group invariance, we lose no information by contracting these indices away. Performing this operation we obtain the equation
\begin{equation}\label{cgorthognorm}
	(C^r)_{{\vec m}_c}^{{\vec m}_a{\vec m}_b}
	(C^{r'})^{*~{\vec m}_{c}} _{{\vec m}_a{\vec m}_b}
	=\delta_{c, c'} \delta_{r, r'}~ {D^{\rm cl}_c}
\end{equation}
where $D^{\rm cl}_c$ is the dimension of the Lie algebra representation $R_c$.

The LHS of \eqref{cgorthognorm} defines a positive definite inner product
on the $N^{\rm cl}_{abc}$ dimensional vector space of classical Clebsch-Gordan coefficients. \eqref{cgorthog} is satisfied once we choose an orthonormal basis in this space, and then rescale all basis vectors by
the factor  $\sqrt{D^{\rm cl}_c }$. \footnote{While the LHS of \eqref{cgorthognorm} is symmetric under permutations of $a, b, c$, the
	RHS singles out $c$ and so breaks this invariance. This is a consequence of the fact that the product rule for invariant tensors, like CG coefficients, is not permutation symmetric, but depends on a split of
	operators into initial and final.}  It is clearly always possible to find such a basis, and so it is always possible to solve  \eqref{cgorthognorm}.

Of course the choice of orthonormal basis is not unique: Given any orthonormal choice of CG coefficients $C^r$, as usual the basis change
\begin{equation}\label{basischangeh}
	U^r_{r'} C^{r'}
\end{equation}
where $U$ is a unitary $N^{\rm cl}_{abc} \times N^{\rm cl}_{abc}$  matrix) yields another orthonormal basis. In what follows we proceed making an arbitrary choice of basis.
Since we could have as well made any other choice, clearly every physical result we obtain using this basis choice will be invariant under \eqref{basischangeh}.

With this preparation in hand, it is now easy to obtain an orthonormal basis for those states in  $R_1 \ldots R_m$ that transform in the representation $R_a$. We first fuse $R_1$ with $R_2$ to obtain all possible representations $R^{(1)}$. It follows from the discussion earlier in this subsection that an orthonormal basis for $R^1\otimes R^2$ is labeled by the choice of representation $R^{(1)}$ together with $r_1$, the choice of $`r'$ coefficient labelling CG coefficients for the fusion, see \eqref{cgorthognorm}.
\footnote{In addition, of course, states are also labeled by  internal symmetry labels ${\vec m}$. As these labels `go along for the ride' in the discussion below, we do not make any explicit reference to them.}

\begin{figure}[h]
	\centering
	\includegraphics[scale=.25]{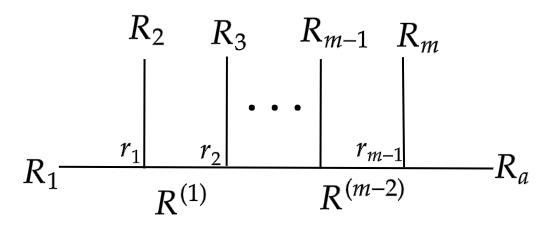}
	\caption{Fusion of $R_1\otimes R_2 \ldots R_m$ to $R_a$}
	\label{classfusion}
\end{figure}

Next we fuse each of these $R^{(1)}$ representations with $R_3$ to obtain every possible representation $R^{(2)}$. The additional labels carried by states arising from this fusion process are the choice of the representation $R^{(2)}$ and $r_2$, the choice of  $`r'$ label for the  CG coefficients that govern this fusion.  Continuing in this manner, the final  fusion is of  $R^{(m-2)}$ with $R_m$ yields $R_a$. In net, the states obtained by this fusion process are labeled by the $m-2$ intermediate representations $R^{(1)} \ldots R^{(m-2)}$, as well as the $m-1$ $r$ values $r_1 \ldots r_{m-1}$ of each of the intermediate fusion processes, as depicted in Fig.
\ref{classfusion}.
It is not difficult to convince oneself  that the basis states obtained through this process are orthonormal, i.e., obey \eqref{orthog}.
\footnote{The argument goes as follows.
	Consider two fusion products that differ in the fusion of $R_1$ with $R_2$ (either because the representations  $R^{(1)}s$ or the $r$ values (labeling the choice of CG coefficients) are different. It follows that the resultant states are orthogonal in the $R_1 \otimes R_2$ subspace, and hence are orthogonal. Now consider two fusion products that are identical in the fusion of $R_1$ and $R_2$, but differ in the fusion of $R^{(1)}$ with $R_3$. The resultant states are, then,  orthogonal in the subspace $R_1 \times R_2 \times R_3$, and so are orthogonal. Proceeding in this manner we conclude that all fusion products are orthogonal to all other distinct fusion products, but  have unit inner product with themselves.}

The states constructed in  this subsection may now be identified with $|{\vec m} \rangle_{a, r}$ of (see under \eqref{classdecomp}), where the index  $r$ is actually the composite index
\begin{equation}\label{compind}
	r = (R^{(1)}, \ldots, R^{(m-2)},
	r_1, \ldots r_{m-1} )
\end{equation}
Similarly, the symbols $r$ and $r'$ that label the projection operators $P_a^{r r'}$ in \eqref{optens}, are also composite indices of the sort \eqref{compind}.

It follows that the orthonormal rotation of CG coefficients, \eqref{basischangeh}, acts on the states ${|{{\vec m}} \rangle}_{a, r'}$ like
\begin{equation}\label{acton}
	{|{{\vec m}} \rangle}_{a, r'} \rightarrow U_1 U_2 \ldots U_{m-1} {|{{\vec m}} \rangle}_{a, r'}
\end{equation}
where $U_1$ is a rotation on the space of CG coefficients that couple $R_1$ and $R_2$ to $R^{(1)}$, $U_2$ is a rotation in the space of CG coefficients that couple $R^{(1)}$ and $R_2$ to
$R^{(2)}$ etc. In the rest of this section we use the shorthand notation
$U$ to denote the product of unitaries $U_1 U_2 \ldots U_{m-1}$.
With this understanding, the states ${|{{\vec m}} \rangle}_{a, r'}$ transform,  under a change of basis of CG coefficients as
\begin{equation}\label{basischangeonm}
	{|{{\vec m}} \rangle}_{a, r'}
	\rightarrow U {|{{\vec m}} \rangle}_{a, r'}
\end{equation}
It follows that the projectors
$P_a^{r r'} = \sum_{{\vec m}}   {|{{\vec m}} \rangle}_{a, r'} ~ _{a, r}\langle {\vec m}|$ transform under the same change of basis vectors as
\begin{equation}\label{transfpbasis}
	P_a \rightarrow U' P_a U^\dagger
\end{equation}
Notice that as the two states that make the projectors $P_a$ in \eqref{projops} generically belong to different Hilbert spaces, the unitary matrices $U'$ and $U^\dagger$, that participate in the `bifundamental' transformation
\eqref{transfpbasis}, are distinct from each other. On the other hand the two states that appear in the projectors ${\hat P}_a$ (see
\eqref{projopsn}) both belong to the same space $(H_{ {\rm in}}$, and so the transformation of these projectors, under a change of basis, is given by the `adjoint' transformation
\begin{equation}\label{transfpbasishat}
	{\hat P}_a \rightarrow U {\hat P}_a U^\dagger
\end{equation}
Note that \eqref{transfpbasishat} leaves the LHS of \eqref{projcomplete} invariant, in agreement with our general expectation that every physical equation is left invariant by a change of basis of CG coefficients.

\section{Crossing and unitarity in matter Chern-Simons theories}\label{ucmcs}

\subsection{The matter path integral as a sum over Wilson lines}

\begin{figure}[h!]
	\centering
	\includegraphics[scale=.25]{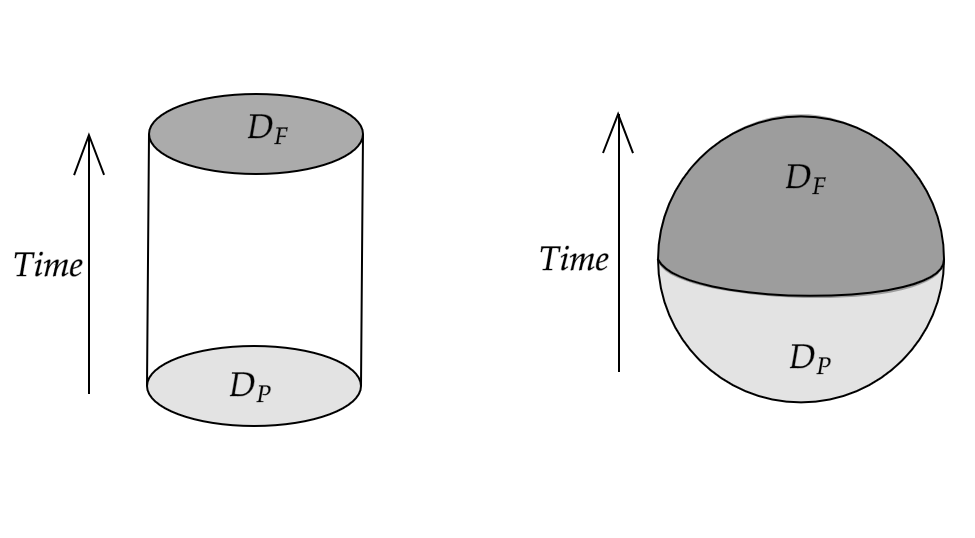}
	\caption{Deformation of pillbox to a sphere}
	\label{spball}
\end{figure}

We work with the regulated version of flat space depicted in Fig. \ref{pillbox}.  The boundary of our spacetime is a Lorentzian cylinder $C$ (see the first of Fig. \ref{spball}) whose curved sides are $S^1 \times I$, and whose `flat' surfaces consist of the future spatial disk $D_F$ and the past spatial disk $D_P$. We assume that the spatial extent of our spacetime $R$ and its temporal extent $T$ are both
very large, ($Rm \gg1$ and $Tm \gg 1$) and also that the spatial extent is much larger than the temporal extent,
$R \gg T$.  This condition ensures that we never have to worry about our scattering particles encountering the curved $S^1 \times I$ boundary of spacetime.

The $S$-matrix is the overlap between a final state $|\psi_f \rangle $ defined at $D_F$ and the past state $|\psi_i \rangle$ defined at $D_P$. The details of these states are encoded in the boundary conditions of the path integral. We work with states in which all matter excitations are well-localized near the centre of $D_P$ and $D_F$.
While the $S^1 \times I$ boundary of the disk hosts boundary chiral WZW excitations, the Hamiltonian for these modes vanishes (recall no particle reaches this boundary). As this Hamiltonian generates translations along $I$, it follows that
nothing happens as we move along $I$, and so $I$ can effectively be shrunk away. After continuing to Euclidean space, the
path integral we need to perform is effectively on a spacetime with the topology of a solid ball, which is schematically depicted in the second of  Fig. \ref{spball}.

To every order in perturbation theory, the matter part of any such path integral can be evaluated in the world line  representation by summing over particle trajectories with possible bulk interactions (see \ref{disc} for some additional discussion). It follows that our path integral of interest is given by the schematic equation
\begin{equation}\label{concepinser}
	\int {\cal D}A_\mu {\cal D} \phi \left(\ldots \right)
	= \int {\cal D}A_\mu  \sum_{{\rm Particle~Trajectories}}
	= \sum_{{\rm Particle~Trajectories}} \int {\cal D}A_\mu ({\rm Wilson~Lines} )
\end{equation}
In the final expression in  \eqref{concepinser} we have interchanged the order of the integral over $A_\mu$ and the sum over particle trajectories. Particle trajectories are weighted by the usual measure ($m$ times the spacetime length of all trajectories plus appropriate factors for each particle- particle interaction). The Wilson lines that appear in the last of
\eqref{concepinser}, in general, includes
configurations in which the Wilson line branches at an interaction point.

Now using the fact that
$$\int {\cal D}A_\mu ({\rm Wilson~Lines} )$$
computes  a Wilson line in pure Chern-Simons theory (i.e., the exactly solvable TFT) it follows that
\begin{equation}\label{smatrewrite}
	S_{fi} \sim  \sum_{{\rm Particle ~Trajectories}} \left \langle {\rm Wilson~Lines} \right \rangle_{{\rm Pure~CS~Theory}}
\end{equation}

The space of particle trajectories that appears in \eqref{smatrewrite} can be decomposed into topologically equivalent topological sectors or chambers. The topological character of pure Chern-Simons theory ensures that the expectation value of
Wilson lines in this theory depends only on the chamber of the trajectory and not its detailed form. It follows that
\begin{equation}\label{smatrewrite}
	S_{fi} \sim  \sum_{{\rm Chambers}} \left \langle {\rm Wilson~Line} \right\rangle_{\rm Chamber}  \sum_{{\rm Ptle ~Trajectories~in~chamber}}
	e^{iS_{L}}
\end{equation}
where $e^{iS_L}$ is the weight associated with any particle trajectory, including possible coupling constant factors associated with interactions.

In summary, the $S$-matrix is given over the weighted sum over Wilson lines that begin and end at the prescribed points on the initial and final surface. Wilson lines are labeled by their interaction structure and winding topologies (and also by their framing structure, see below). The weight in this sum is the volume (in the space of trajectories) of all particle trajectories with the specified topology, together with
coupling constant factors associated with interactions -- which generically cause Wilson lines to bifurcate or trifurcate, etc,
in a gauge invariant manner.
\subsection{Grouping Wilson line topologies into a finite number of blocks}

\eqref{smatrewrite} can be written more compactly as
\begin{equation}\label{sumtop}
	S= \sum_{{\rm Topologies}~t} S_t W_t
\end{equation}
where $W_t$ is the path integral of pure Chern-Simons theory in the presence of Wilson Lines with topology $t$. Roughly speaking, $S_t$ is the $S$-matrix associated with the topology $t$.

While there are an infinite number of topologies $t$, it follows from Witten's classic analysis \cite{Witten:1988hf} that for every $t$
\begin{equation}\label{Tbyt}
	W_t= \sum_i \alpha_t^i G_i
\end{equation}
where the summation over $i$ runs over a basis of the finite-dimensional space of conformal blocks with insertions associated with the end points, on the boundary, of Wilson lines.
Inserting \eqref{Tbyt} into \eqref{sumtop}
and defining
\begin{equation}\label{sti}
	\mathcal S^i = \sum_{{\rm Topologies}~t} \alpha^i_t S_t
\end{equation}
we find
\begin{equation}\label{sumtopn}
	S= \sum_i \mathcal S^i G_i, ~~~~~
\end{equation}

Comparing \eqref{sumtopn} with \eqref{expofsmatrixmt}, we see that conformal blocks play the same role in the current context  (when studying scattering in matter Chern-Simons theories) that invariant index structures played in the previous section (i.e.
in the study of scattering in topologically trivial massive theories with a global symmetry).

 The expression \eqref{sumtopn} has a feature that has no analogue in \eqref{sumtop}. The blocks $G_i$ that appear in
this sum are not single valued (they have cuts). Of course $S$-matrix itself is single valued (as is clear from \eqref{sumtopn}).
This means that $\mathcal S^i$ themselves have cuts to compensate those of $G_i$ Also the precise definition of
$\mathcal S^i$ depends not only on the choice of block $G_i$ but also on the choice of sheet for that block. We elaborate on all these points in Appendix \ref{multisheet}. In this paper, we proceed by following our nose and making natural choices of sheet structure when defining $\mathcal S^i$, leaving a fuller exploration of these `sheet ambiguities' and their implications to future work.
\subsection{Compounding blocks} \label{cSm}

After recasting the $S$-matrix as a sum over invariant tensors in \eqref{expofsmatrixmt}, we were immediately confronted with the issue of defining the multiplication of two invariant tensors in order to make sense of the unitarity equation  $S^\dagger S=I$. In section \ref{csm} we explained how this multiplication was defined.

\eqref{sumtopn} plays the same role in this section that \eqref{expofsmatrixmt} played in the last section, with conformal blocks playing the role of invariant tensors. In order to make sense of the unitarity equation, once again we need a definition of $G_j^\dagger \times G_i$. \footnote{Here $G_i$ and $G_j$ are both blocks with the same insertions in the space $H_{ {\rm in}} \times H_{ {\rm out}}^*$.} \footnote{In order to understand unitarity we also need to understand what  $I$ on the RHS of the unitarity equation means: we postpone this question to the next subsection.} As we have explained in the introduction, the multiplication rule for blocks is determined by physical considerations. We spell out the rule here in some detail.

In order to find the product $G_j^\dagger \times G_i$ one is instructed to proceed as follows.
\begin{itemize}
	\item First find a representation for each of $G_i$ and $G_j$ in terms of a `tangle' of Wilson lines (possibly with interaction vertices: see the next subsection for examples)
	on a solid ball of unit radius. Also make a definite choice for all framing vector fields. All the blocks $G_i$ are taken to have insertions of the same primary operators at the same locations. Consequently, $G_i$ and $G_j$ are both represented by open Wilson line `tangles' with end points at the same locations (and in the same representations). In all these blocks, the framing vector fields are also required to take the same (arbitrarily chosen) values for at the end points of all Wilson lines.
	\item  Next, divide the insertions in both $G_i$ and $G_j$ into initial and final (the product rule depends on this division). We denote the initial representations by $H_{ {\rm in}}$, and the final representations by $H_{ {\rm out}}^*$.
	\item For both $G_i$ and $G_j$ next flatten out the part of the boundary of the ball that hosts the final insertions (see Fig. \ref{prodfig})

	\begin{figure}[h!]
		\centering
		\includegraphics[scale=.25]{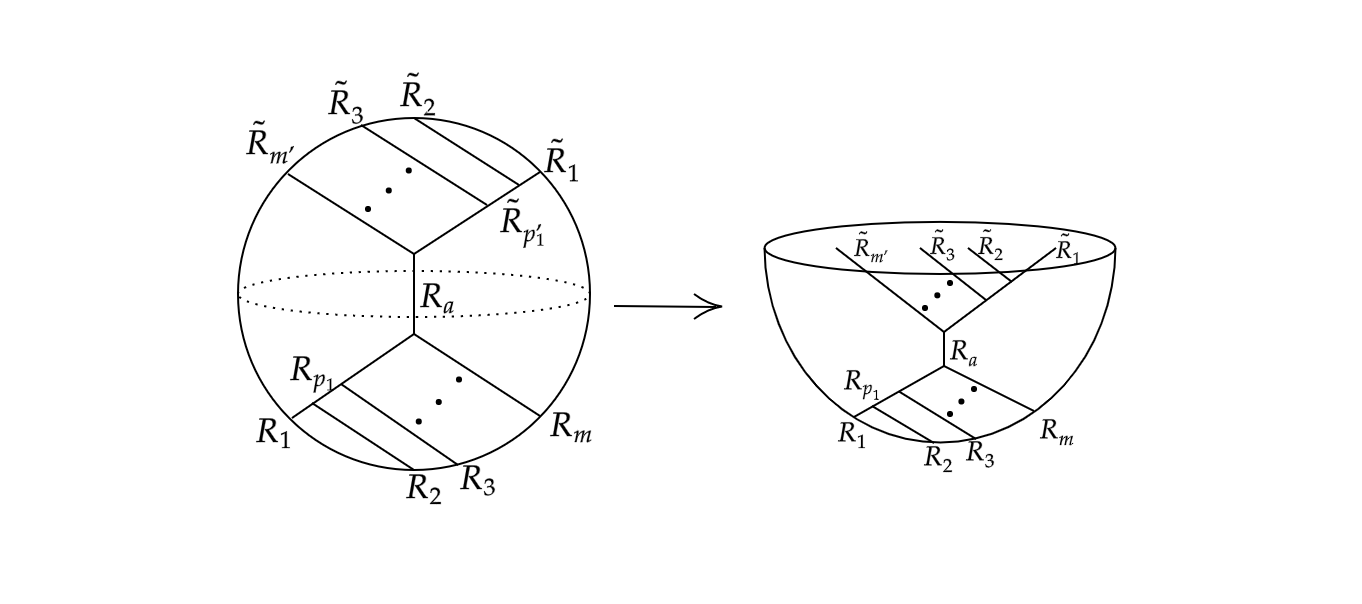}
		\caption{The part of the sphere to be glued is flattened as shown in this figure.}
		\label{prodfig}
	\end{figure}

	\item To produce the path integral for $G_j^\dagger$, reflect the path integral for $G_j$ around its flat surface, and also reflect
	the `arrow' (representing the flow of colour) along each Wilson line. This second reflection turns every boundary insertion
	in representation $R_m$ into a boundary insertion in representation $R_m^*$. The result of such a reflection is depicted in the top left diagram of Fig. \ref{prodfig2}.
	\item
	$G_j^* \times G_i$ is obtained by gluing the two balls along the flat surfaces (see Fig. \ref{prodfig2}). Since $G_i$ and $G_j$ have the same insertions at the same locations,  the resultant  path integral  computes the expectation value of a tangle of Wilson lines that are continuous across the glued flat surfaces. The resultant  Wilson line tangles have ends on the boundary of the new effective solid ball (see Fig. \ref{prodfig2}). The condition that the framing vector field takes the same value at the endpoint of Wilson lines in $G_i$ and $G_j$ ensures that the framing field of this new tangle is also continuous across the flat glued surfaces.

	\begin{figure}[h]
		\centering
		\includegraphics[scale=.25]{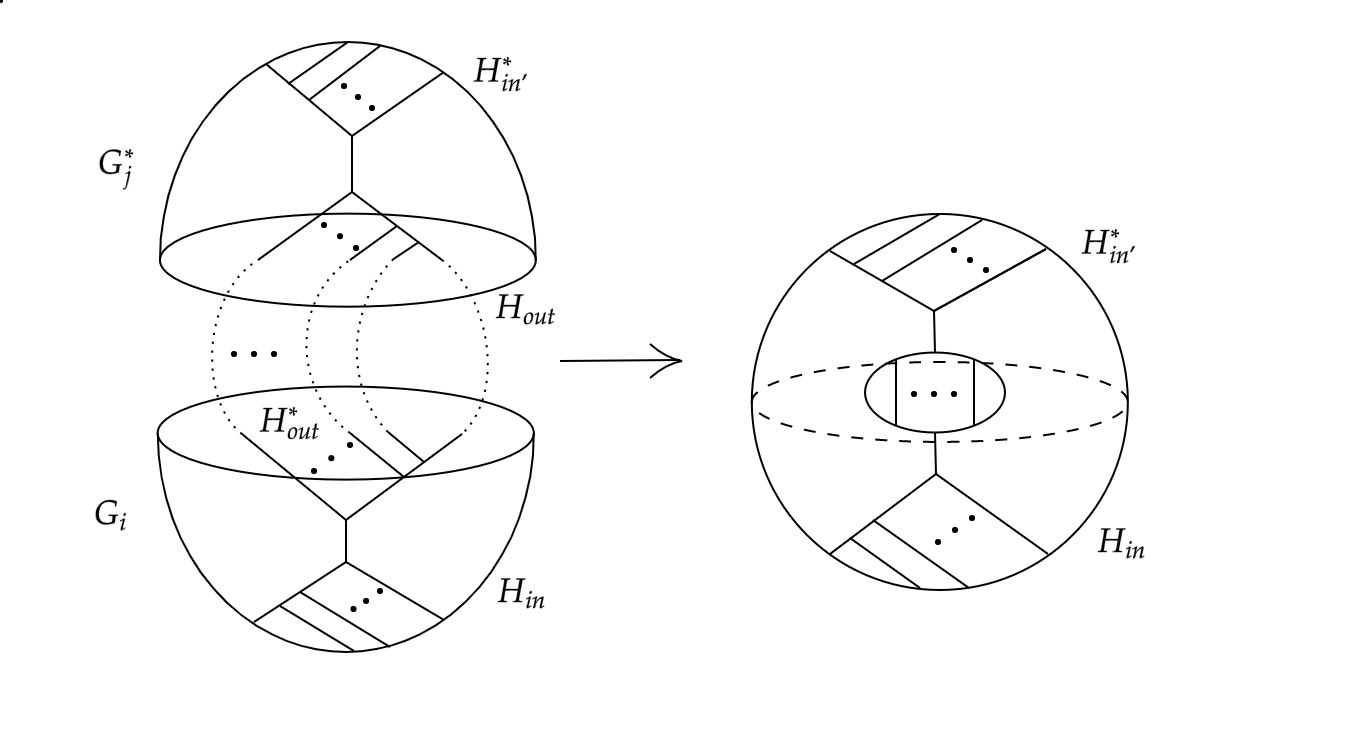}
		\caption{Compounding of blocks}
		\label{prodfig2}
	\end{figure}

	\item  This final result is the Wilson line representation of a block, whose insertions transform in the representations $H_{ {\rm in}} \otimes H_{ {\rm in}}^*$.
\end{itemize}

As we have explained (and as for invariant tensors) the product rule for blocks depends on
the division of block insertions into $H_{ {\rm in}}$ and $H_{ {\rm out}}^*$.  As we now explain, the rule also depends on some additional choices that had no analogue in the previous section.

\begin{figure}[h]
	\centering
	\includegraphics[scale=.25]{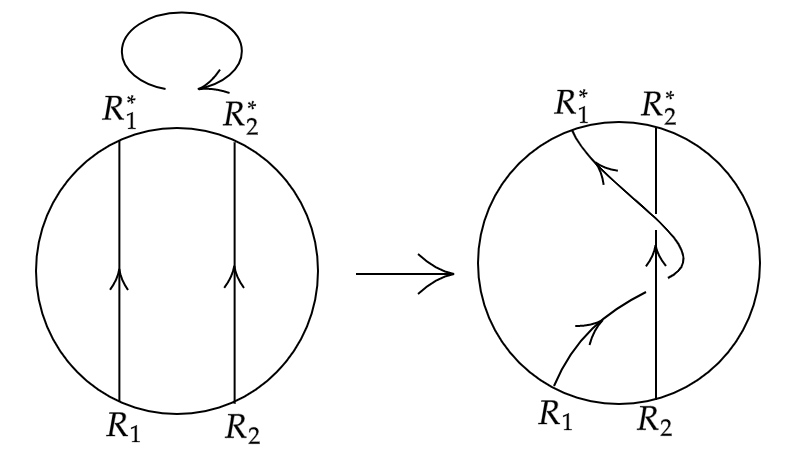}
	\caption{Change in topology due to monodromy}
	\label{entangle}
\end{figure}

Recall that chiral conformal blocks are, in general, multi-sheeted. From the Wilson line representation this comes about because a  monodromy move performed on the end points of Wilson lines generically changes either the Wilson line topology or framing or both. \footnote{An example of the first phenomenon is
	presented in Fig \ref{entangle}.
	An example of the second phenomenon taking
	place without a change in topology is given by the two-point function block of two operators of  dimension $h$. Rotating one of the operators around the other and back to its original position  induces a $4 \pi $ twist in the  framing vector field, compared to its original value. The two-point function thus changes by a factor $e^{-4 \pi i h}$, in agreement with the formula $\frac{1}{z^{2h}}$ for this block. }

A definite Wilson line tangle (with a definite choice of framing field) specifies a conformal block together with a choice of sheet. Our product rule -- which is stated in terms of Wilson line representations -- is thus well defined for blocks together with a choice of sheet.\\
\linebreak

The product we have defined between $G_j^\dagger$ and $G_i$ above may appear to depend on the location of the glued operators. A moment's thought, however, will convince the reader that the  product is actually unchanged by a continuous simultaneous change in the choice of boundary locations (and values of boundary framing fields) of the glued operators. This follows because any such change performed on both glued blocks deforms  the effective product tangle (last of Fig. \ref{prodfig2}) in a continuous -- hence topologically trivial manner, and so leaves the final answer for $G_j^\dagger \times G_i$ (both the final block as well as its branch structure) unchanged. It follows, in particular, that our product rule for blocks  is single valued under simultaneous monodromy moves for both sets of  glued operators,  even though the blocks $G_i$ and $G_j$ both individually change under these monodromy moves. \footnote{In more detail, the product is left invariant by all continuous motions of the locations of glued operators, and so, in particular, by the motions that achieve a monodromy operation (take one operator around the other). Such an operation changes both $G_i$ and
$G_j^*$ by monodromy operators: however these changes apparently cancel out in the product.}.

\subsection{The identity block} \label{ib}

In order to make sense of the unitarity equation we need a definition of
the symbol $I$ on the RHS of $S^\dagger S=I$.
As the LHS of the unitarity equation is a block
with insertions in $H_{ {\rm in}} \otimes H_{ {\rm in}}^*$, the RHS must also be a block with the same set of insertions. Which block represents identity?

\begin{figure}[h]
	\centering
	\includegraphics[scale=.25]{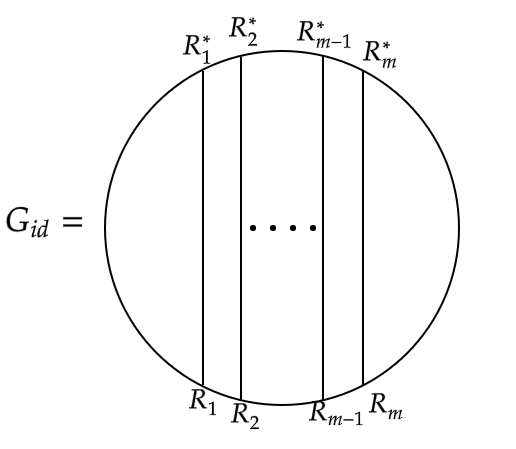}
	\caption{The identity block}
	\label{identityblock}
\end{figure}

\begin{figure}[h]
	\centering
	\includegraphics[scale=.25]{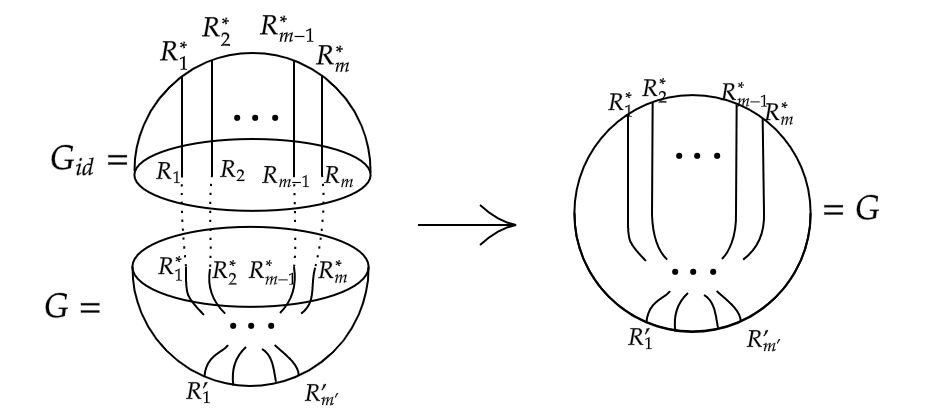}
	\caption{Compounding of identity block with any generic block $G$}
	\label{identity2}
\end{figure}

Recall that the identity invariant tensor $T_{ {\rm id}}$  on $H_{ {\rm in}} \otimes H_{ {\rm in}}^*$ (see around \eqref{projcomplete}) was identity under the tensor multiplication operation. \footnote{ Specifically, we treat $H_{ {\rm in} }$ indices as initial and $H_{ {\rm in}}^*$ indices as final. With this rule, multiplication of the identity invariant tensor with any other tensor $A$ on $H_{ {\rm in}} \otimes H_{{\rm in}}^* $ equals $A$.}

A Wilson line representation of the conformal block $G_{ {\rm id}}$ with the same property depicted in Fig. \ref{identityblock}. As explained in \ref{identity2}, it is obvious that this block is the multiplicative identity. It is also the natural generalization of the collection of contractions that define the identity invariant tensor $T_{ {\rm id}}$. We denote this block by $G_{ {\rm id}}$. We propose that the RHS of the unitarity equation takes the form
\begin{equation}\label{unitarityblockeq}
S^\dagger S= G_{ {\rm id}} \mathcal S_{ {\rm id}}
\end{equation}
where $\mathcal S_{ {\rm id}}$ is the delta function localized at forward scattering.

\subsection{Our conjecture for the structure of $S$-matrices in matter
Chern-Simons theories} \label{coss}

Let us summarize the discussion presented so far in this section. Physical considerations have led us to conjecture that
\begin{itemize}
	\item   $S$-matrices in Matter Chern Simons theories admit an expansion of the form \eqref{sumtopn}, where the index $i$ runs over the space of conformal blocks
	on $S^2$ associated with insertions of primary operators, one for each scattering particle.
	\item The dagger of an $S$-matrix is given by interchanging initial and final momenta and taking complex conjugate in each ${\cal S}_i$, and simultaneously
	implimenting the procedure explained just under Fig. \ref{prodfig}.
	\item The product of $S^\dagger$ and $S$ is obtained by convoluting the component $S$-matrices ${\cal S}_i$, and simultaneously compounding blocks
	in the manner described in subsection \ref{cSm}.
	\item $S$-matrices obey the unitarity equation  \eqref{unitarityblockeq}, where
	the identity block is defined in Fig. \ref{identityblock}.
\end{itemize}

\subsection{An explicit construction for `projector' blocks} \label{pbs}

In this subsection we present an explicit construction of `projector blocks',
the finite $k$ analogues of the explicit basis of projector index structures we constructed in subsection \ref{pis}.

\subsubsection{The conformal block analogues of Clebsch-Gordan coefficients}

Open Wilson line tangles with bulk `interactions', or trivalent vertices, extensively studied by Witten in \cite{Witten:1989wf}, may be used to find a simple and natural conformal block generalization of the previous subsection.

\begin{figure}[h]
	\centering
	\includegraphics[scale=.25]{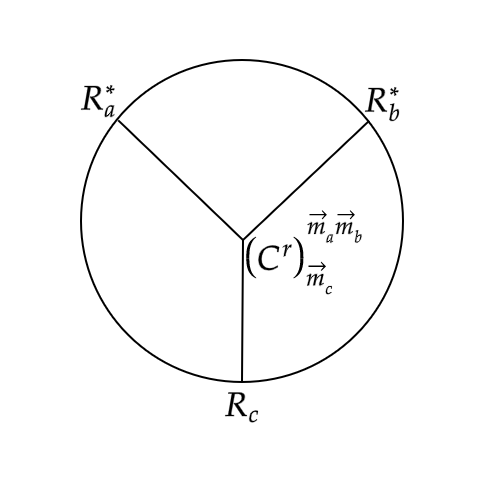}
	\caption{Three point conformal block}
	\label{threepointblock}
\end{figure}

Note that the Chern-Simons analogue of the  Clebsch-Gordan coefficient $(C^r)_{{\vec m}_c}^{{\vec m}_a{\vec m}_b}$ is a three-point block, with `initial' insertion in representation
$R_c$, and final insertions in representations
$R_a^*$ and $R_b^*$. All such blocks admit a Wilson line representation of the form depicted in Fig. \ref{threepointblock}. In Appendix \ref{conventions} we list the conventions (closely following the paper \cite{Witten:1989wf}) that we use all through this paper when giving concrete meaning to diagrams like Fig. \ref{threepointblock}.\footnote{ See also \cite{Moore:1989ni}}. In particular
we adopt the choice of `vertical framing' (see Appendix \ref{conventions} for a definition) of \cite{Witten:1989wf} all throughout.

Note that the three-point blocks depicted  in Fig. \ref{threepointblock} are  parameterized by the choice of Lie algebra Clebsch-Gordan coefficient $(C^r)_{{\vec m}_c}^{{\vec m}_a{\vec m}_b}$ that defines the interaction vertex in Fig. \ref{threepointblock}. \footnote{The precise definition of the block \ref{threepointblock} depends on the cyclic order in which the
three representations are fused at the interaction vertex. In Fig.
\ref{threepointblock}, for instance, we encounter the representations $a, b, c$ in clockwise order, when we view the interactions so that the framing vector fields all point towards us. If we change the cyclical order of the interaction (say by flipping the order of $b$ and $c$ at the interaction vertex) but work with the same operator insertion
positions as Fig. \ref{threepointblock}, then the Wilson line in representation $R_a$ has to pass either over or under the line in representation $R_b$ before reaching its final boundary position. The block defined in this manner is related to the block depicted in
Fig. \ref{threepointblock} by a constant phase (see Fig. 16 of \cite{Witten:1989wf}). }

 As a consequence, at  large enough $k$ (with the representations $a$, $b$ and $c$ held fixed) there is a one to one correspondence between classical Clebsch-Gordan coefficients and WZW three-point function blocks. In particular the number of independent 3 particle blocks, $N^{\rm wzw}_{abc}$, equals $N^{\rm cl}_{abc}$ at large $k$.

For generic fixed representations $a$, $b$ and $c$, on the other hand,  there is always a value of $k$ below which the blocks Fig. \ref{threepointblock} vanish when the CG coefficients in Fig \ref{threepointblock} lie in a particular subspace of the full $N^{\rm cl}_{abc}$  space of CG coefficeints
 \footnote{See the discussion under Eq 3.49 in the classic paper of Gepner and Witten, \cite{Gepner:1986wi}.}. In this case
$N^{\rm wzw}_{abc} < N^{\rm cl}_{abc}$, and
conformal blocks Fig. \ref{threepointblock} are parameterized by equivalence classes of Clebsch-Gordan coefficients: two CG coefficients are equivalent if their difference creates a vanishing block. The precise structure of these equivalence classes depends on $k$. \footnote{See around Pg. 642 of \cite{Witten:1989wf} for a similar discussion.}

In summary, the conformal block analogues of `classical' Lie algebra CG coefficients are the blocks depicted in Fig. \ref{threepointblock}. The
conformal blocks in question are labeled by equivalence classes of classical CG coefficients.
When $k$ is large enough the equivalence class structure trivializes, and blocks are in one to one correspondence with classical CG coefficients, and are labeled by them.

\subsubsection{An inner product on the space of three-point blocks and an orthonormal basis}

\begin{figure}[h]
	\centering
	\includegraphics[scale=.25]{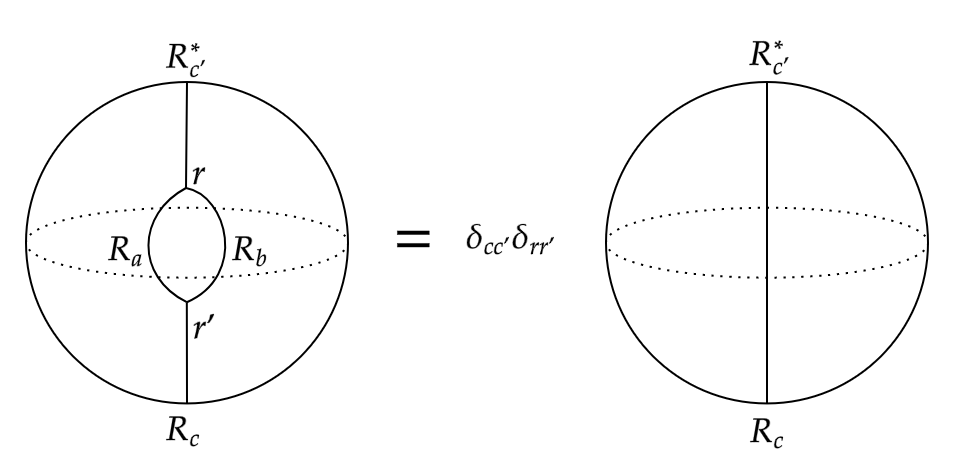}
	\caption{Analogue of \eqref{cgorthog}}
	\label{threeptorthog}
\end{figure}

\begin{figure}[h]
	\centering
	\includegraphics[scale=.25]{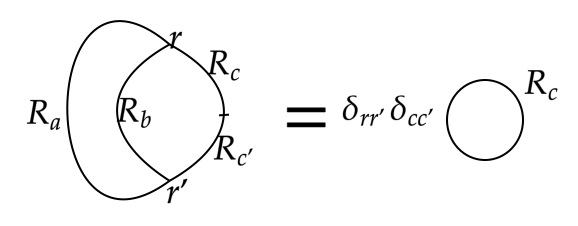}
	\caption{Analogue of \eqref{cgorthognorm}}
	\label{threeptorthogtw}
\end{figure}

In section \ref{pis}  we chose a basis in the space of Clebsch-Gordan coefficients that obey the `orthonormality relations' \eqref{cgorthog}. The conformal block analogue of the contraction of indices in
\eqref{cgorthog} is the compounding of blocks (see subsection \ref{cSm}). Consequently, the block analogue of \eqref{cgorthog} is the equation depicted in
Fig. \ref{threeptorthog}.

The LHS and RHS of  Fig. \ref{threeptorthog} are two-point blocks (with insertions in representations $R_c$ and $R_c^*$). Since this space of blocks is one-dimensional, the inner product of the equation in Fig. \ref{threeptorthog} with any nonzero two-point block carries exactly as much information as the equation Fig. \ref{threeptorthog} itself.
Taking the inner product of this equation with the two-point block that appears on the RHS of fig.
\ref{threeptorthog} we obtain the equation in Fig. \ref{threeptorthogtw}, which is the direct analogue of \eqref{cgorthognorm}.
Notice that the expectation value of the circular Wilson loop in representation $R_c$ (which appears on the RHS of
the equation in Fig. \ref{threeptorthogtw}) equals the quantum dimension $D^k_c$ on the representation $R_c$, and replaces the
classical dimension $D^{\rm cl}_c$ of the same representation in \eqref{cgorthognorm}.

The LHS of Fig.  \eqref{threeptorthogtw} is the inner product
\begin{equation}\label{ip}
\langle r, c' | r', c\rangle
\end{equation}
between the three-point conformal blocks Fig. \ref{threepointblock}.
The inner product is taken in Witten's Hilbert space for blocks, \cite{Witten:1988hf}, and is, in fact, the quantity that Witten called $K(a, b, c, \epsilon_i, {\tilde \epsilon}_j)$ (see Eq. 2.10 of \cite{Witten:1989wf}). It follows \footnote{From the fact that the inner product on Witten's
Hilbert space of conformal blocks is positive definite, see Appendix \ref{conventions} for some more detail.}  that the LHS of Fig. \ref{threeptorthogtw} -- i.e., $K(a, b, c, \epsilon_i, {\tilde \epsilon}_j)$ --  defines a positive definite inner product on the $N^{\rm wzw}_{abc}$ dimensional equivalence classes of CG coefficients. It follows that, just as in the discussion around \eqref{cgorthognorm},  all we need to do in order to solve the equation in Fig.  \ref{threeptorthogtw} is to pick an orthonormal basis (with this definition of the inner product) in the equivalence classes of CG coefficients, and then rescale these basis vectors by a factor of $\sqrt{D_c^{\rm k}}$ (recall $D_c^k$ is a positive number).  Of course this is always possible to do; infact, as in the discussion around \eqref{basischangeh}, there are many different choices -- all related by $N^{\rm wzw}_{abc} \times N^{\rm wzw}_{abc}$ unitary transformations -- that accomplish this.
As in the `classical' discussion of subsection \ref{pis}, we make one choice and proceed, always keeping in mind that we could have made another choice, so  physical results we obtain should be invariant under
\eqref{basischangeh}. \footnote{
	Given any $K(a, b, c, \epsilon_i, {\tilde \epsilon}_j )$ that satisfy Witten's 2.10, the rescaled quantities
	$\frac{ \sqrt{ D^{\rm k}_c}} { \sqrt{ { D^{\rm k}_a}  D^{\rm k}_b}}
	K(a, b, c, \epsilon_i, {\tilde \epsilon}_j )$ obey the equation in
	Fig. \ref{threeptorthogtw}, once we replace Witten's $(\epsilon_i, {\tilde \epsilon}_j)$ with
	our $(r, r')$ .}

\subsubsection{`Orthonormal' basis for more general conformal blocks}

\begin{figure}[h]
	\centering
	\includegraphics[scale=.25]{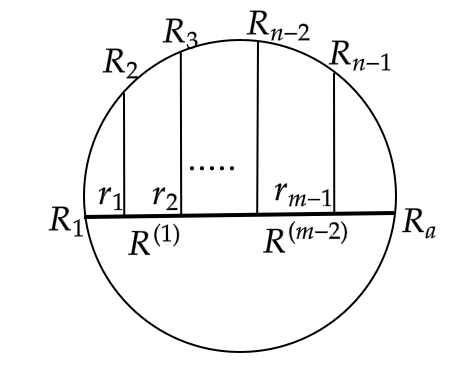}
	\caption{Block obtained by fusion of $R_1 \ldots R_m$ to $R_a$. The three-point Clebsch-Gordan coefficients at interactions are drawn from the orthonormal basis of the previous subsubsection.}
	\label{blockfuse}
\end{figure}

Once we have made a choice of an orthonormal basis in the space of equivalence classes of CG coefficients, the construction
of an orthonormal basis of blocks that describes the fusions of representations $R_1 \ldots R_m$ into $R_a$ proceeds exactly as in section \ref{pis} (see the last two paragraphs of section \ref{pis} and Fig. \ref{classfusion}). The only difference is
that the abstract fusions in Fig. \ref{classfusion} are replaced by actual physical Wilson lines, with ends on the boundary
$S^2$ of our spacetime, as depicted in Fig. \ref{blockfuse}. \footnote{See Fig. 24 of \cite{Witten:1989wf} for an essentially identical construction.}

\begin{figure}[h]
	\centering
	\includegraphics[scale=.25]{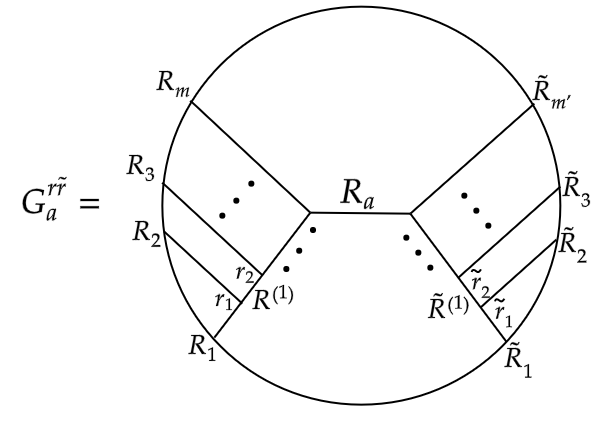}
	\caption{The projector block analogue to the classical counterpart in \eqref{projops}. Once again, the three-point Clebsch-Gordan coefficients at interactions are drawn from the orthonormal basis of the previous subsubsection.}
	\label{projblocko}
\end{figure}

The analogue of the classical projector $P_a^{r r'}$ (see \eqref{projops}) is the block $G_a^{r r'}$ depicted in Fig. \ref{projblocko}. \footnote{The sum over ${\vec m}$ in  \eqref{projops} is replaced by the fusion of  the Wilson line corresponding to the operator $R_a$ in fig. \ref{projblocko}. }

We will now demonstrate that the projector blocks $G_a^{r r'}$  obey the `orthonormal' product rule (analogue of \eqref{projopsnn} )
\begin{equation} \label{projoblock}
	(G_a^{r_1 r_2})^\dagger \times G_{a'}^{r_3 r_4}  = \delta_{a a'} \delta_{r_2, r_4}  {\hat G} _a^{r_3 r_1}
\end{equation}
Here ${\hat G} _a^{r_3 r_1}$ are blocks with insertions in representations in $H_{ {\rm in}}\otimes H_{ {\rm in}}^*$. A similar argument (to the one we are about to present)
demonstrates that the blocks ${\hat G}$ also obey the `orthogonal product rule'
(analogue of \eqref{projopsnn} )
\begin{equation} \label{projblock}
	{\hat G}_a^{r_1 r_2} \times ({\hat G}_{a'}^{r_3 r_4})
	=  \delta_{a a'} \delta_{r_1, r_4}  {{\hat G}}_a^{r_3 r_2}, ~~~~~\left( {\hat G}_a^{r_1 r_2} \right)^\dagger
	= {\hat G}_a^{r_2 r_1}
\end{equation}

\begin{figure}[H]
	\centering
	\includegraphics[scale=.35]{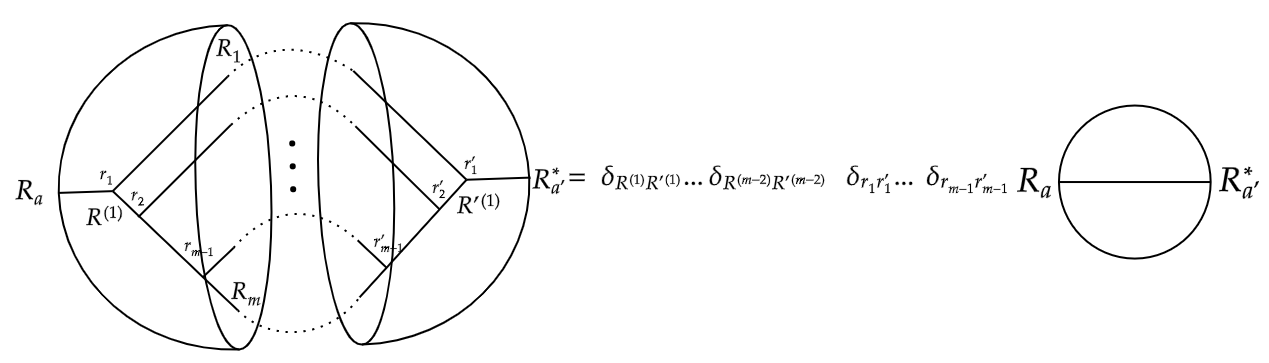}
	\caption{Compounding of two projector type blocks}
	\label{whyorthogtwo}
\end{figure}
In order to demonstrate \eqref{projoblock}, it is necessary and sufficient to demonstrate the equation
asserted in Fig. \ref{whyorthogtwo}.
\footnote{ Note that the product of $\delta$ functions on the RHS of this equation
is the expansion of the symbol $\delta_{r_2, r_4}$, as, in the basis we have adopted, $r$ is a composite index consisting of the
$m-2$ intermediate representations and the $m-1$ choices of CG couplings.}

\begin{figure}[H]
	\centering
	\includegraphics[scale=.35]{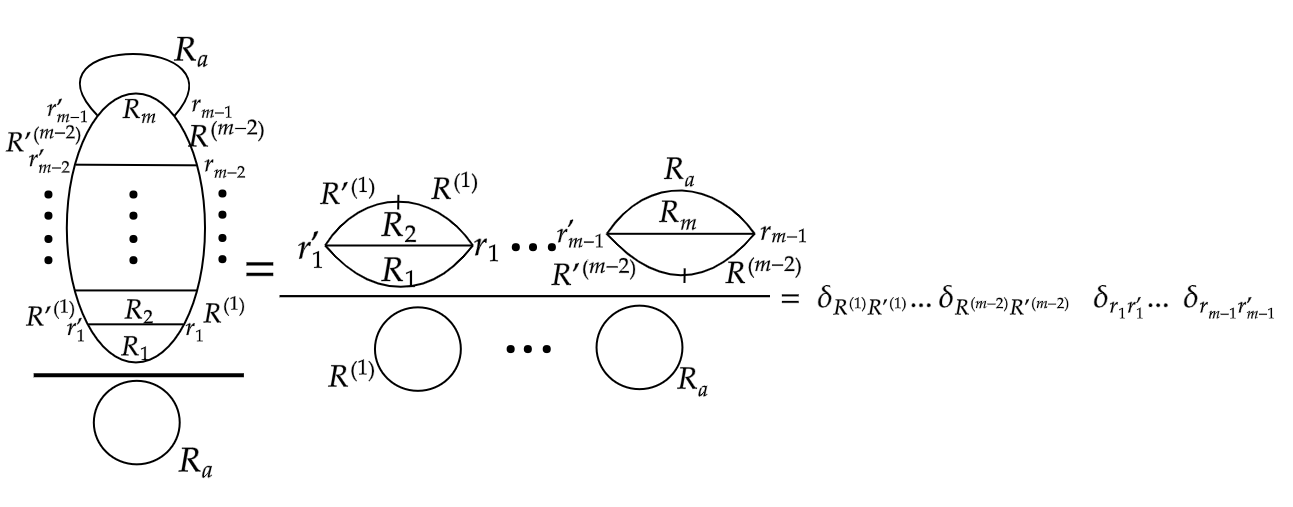}
	\caption{Evaluating the first factor on RHS of fig. \ref{whyorthogtwo}}
	\label{whyorthogthree}
\end{figure}

\begin{figure}[H]
	\centering
	\includegraphics[scale=.25]{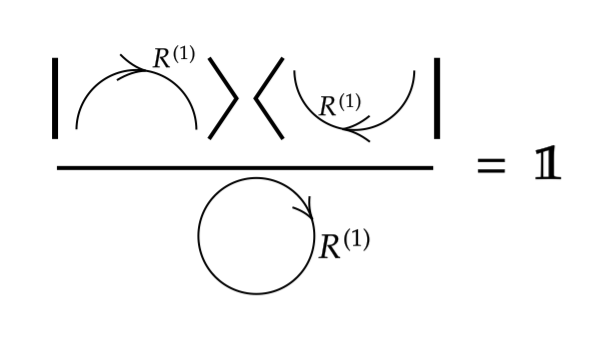}
	\caption{Completeness relation in the one dimensional space of two-point blocks}
	\label{twopointblockcomp}
\end{figure}

\begin{figure}[H]
	\centering
	\includegraphics[scale=.25]{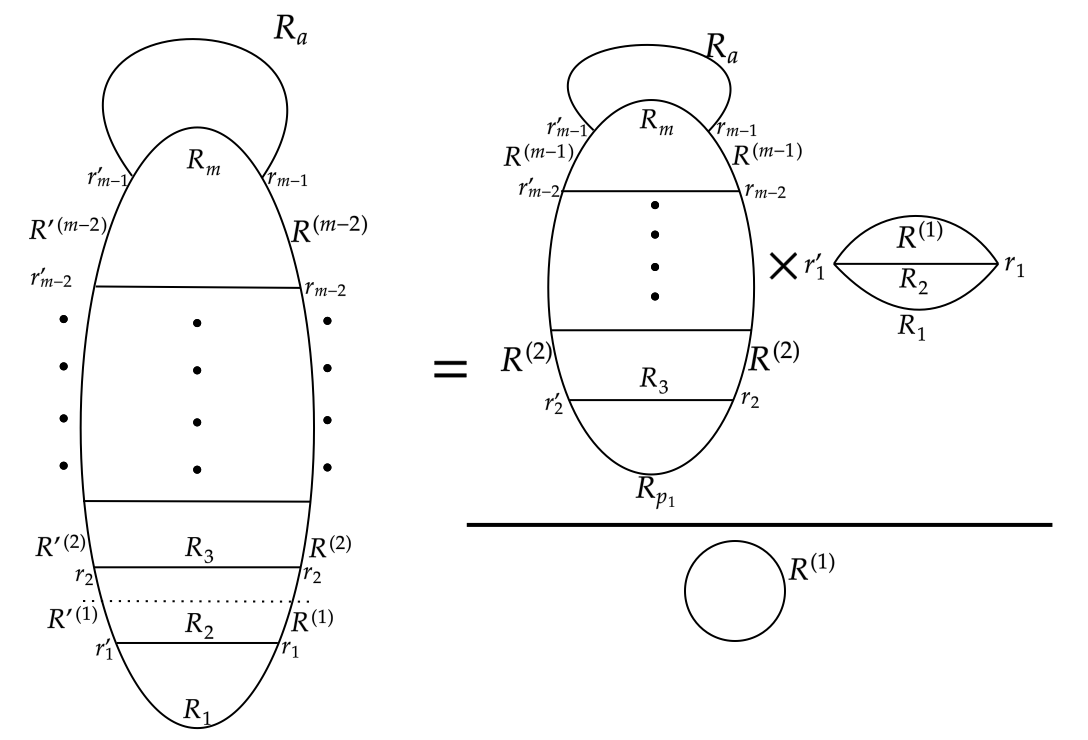}
	\caption{Evaluating LHS of Fig \ref{whyorthogthree} using cutting procedure}
	\label{peeloff}
\end{figure}

As the LHS and RHS of Fig. \ref{whyorthogtwo} are both two-point function blocks (and as the space of two-point function blocks is one dimensional), the equation in Fig. \ref{whyorthogtwo} is exactly equivalent to the inner product of both sides
of Fig. \ref{whyorthogtwo} with the block that appears on its RHS. In other words the Fig. \ref{whyorthogtwo} is exactly equivalent to the assertion that the extreme LHS of Fig \ref{whyorthogthree} equals the product of $\delta$ functions on the  extreme RHS of the same diagram. We will now verify this is the case. Our demonstration proceeds as follows.
We first demonstrate that the product of $\delta$ functions on the RHS of Fig \ref{whyorthogthree} equals the middle diagram in the same Fig. We then demonstrate that the LHS of Fig. \ref{whyorthogthree}
also equals the middle diagram in the same Fig, thereby establishing the desired
result.  \footnote{This argument outlined over the next three paragraphs is essentially the same as that described in Fig. 25 of
\cite{Witten:1989wf} }

The fact that middle diagram in  Fig \ref{whyorthogthree} equals the product
of $\delta$ functions on its RHS is infact obvious. It follows from fact that the special basis of CG coefficients we have been using obey the identities
 Fig. \ref{threeptorthogtw}. Using these identities $m-1$ times, turns the ratio diagrams in the middle term into the product of $\delta$ functions on its RHS.

 The argument that the left most term in Fig \ref{whyorthogthree} equals the middle diagram in the same Fig. is a bit more elaborate. We imitate Witten's analysis
 in \cite{Witten:1989wf} to proceed as follows. We first cut the diagram in the numerator of the LHS along a horizontal line
 just above the line marked $R_2$. This way of cutting the diagram allows us to view it as an inner  product of two 2 point blocks. However the space of two-point function blocks is zero when the insertions are not in conjugate representations, or one dimensional otherwise. As a consequence the diagram either evaluates to zero, or evaluates to  what one gets by inserting the `completeness relation displayed in Fig.
 \ref{twopointblockcomp} into the dotted line. Consequently we obtain the identity displayed in Fig. \ref{peeloff}.

 Now the block in the numerator of the RHS of Fig. \ref{peeloff} is structurally rather similar to the block in the numerator
 of Fig \ref{whyorthogthree}. For this reason we can repeat the process, this time cutting the numerator of the RHS of Fig. \ref{peeloff} just above $R_3$ and using the completeness relation analogous to Fig. \ref{twopointblockcomp}.  Continuing in this manner, the equality of the first and second diagrams of Fig. \ref{whyorthogthree} follows.

The construction of projector blocks, presented in this subsection, is applicable very generally. However, we note that, in the special case that $H_{ {\rm in}}= H_{ {\rm out}}$, (so that the collection of representations that appear in the conformal block in question is self-conjugate), it is always possible to find a second representation of all projector blocks in terms of tangles of Wilson lines without introducing any bulk interaction vertices. In Appendix \ref{quantexamp}, we show how this works in the context of a particular example.

In summary, we have demonstrated that the `projector' blocks $ G_{a'}^{r_3 r_4}$ constructed in this subsection obey precisely the same multiplication  rules as the `classical projectors' $ P_{a'}^{r_3 r_4}$. The reader may wonder why the product algebras of conformal blocks so closely
resemble those of the group projectors. We suspect that the answer to this question lies in the following fact. Recall that
the classical projector $ P_{a'}^{r_3 r_4}$ lived in a tensor product of $H_{ {\rm out}}$ and the dual of $H_{ {\rm in}}$. In a similar manner,
once we slice a conformal block separating the insertions that correspond to $H_{ {\rm in}}$ and $H_{ {\rm out}}^*$, a conformal block may ${\it almost}$ be thought of as living in the tensor product Hilbert spaces associated with the `initial' and `final' disks. The `almost' in this statement stems from the fact that each disk carries a boundary Hilbert space arising from gauge transformations that are nontrivial on the boundary of each disk. In the individual disk Hilbert spaces, these gauge transformations are thought of as `large' and are phase space variables, whose quantization gives rise to the boundary Hamiltonian. In the sphere partition function, on the other hand, the boundary degrees of freedom of the two disks are
treated identified and modded out (the modding out is consequence of the fact that these modes are pure gauge).

Ignoring this subtlety for the moment, the Hilbert space structure of conformal blocks exactly parallels that of \eqref{projops}. Presumably the subtlety plays no essential role in the product algebra, explaining why block multiplication has the same structure as state projector multiplication.  We leave the exploration of this suggestion to future work.

\subsection{Change of basis} \label{cob}

As in subsection \ref{bdcr}, our definition of the special basis of projectors, presented in the previous subsection, involved several arbitrary choices.
We have already emphasized that our choice of orthogonal basis in the space of three
point CG coefficients between operators in representations $A$, $B$ and $M$ was arbitrary upto a $U(N^k_{ABM})$ transformation. In addition, in  constructing the projectors out of three-point fusions, we arbitrarily chose to first fuse $R_1$ with
$R_2$, and then fuse the result with $R_3$, etc. Any other order of fusions would have worked as well, and would have given us a completely different basis.

In general, given one construction of the basis blocks $G_a^{rr'}$, any change of basis
of the form
\begin{equation}\label{formofbach}
U^r_s G_a^{ss'} V_{s'}^{\dagger r'}
\end{equation}
also yields a set of projectors that obey all properties required of the projector basis,
and so could equally well have been used as the starting point of our construction.

When the space of projectors at a given value of $a$ is one dimensional, the ambiguity
\eqref{formofbach} is one of phases. This fact will play an important role in the next section.

The fact that no particular basis of projector blocks is sacrosanct is particularly
clear at finite $k$. This is because a monodromy move on (lets say) the initial insertions in $G_a^{ss'}$ results in a transformation of the form \eqref{formofbach}.
In other words moving our scattering particles around each other - on the `sphere' at infinity - results in an effective basis change on our projector blocks.  We believe that this phenomenon, which has no counterpart at $k=\infty$, is a manifestation
of the fact that the anyonic particles that we scatter at finite $k$ are never really non interacting (there is, really, no such thing as a Fock space of anyons). If we take one anyon around another it picks up a phase, no matter how far the two particles were from each other.

\subsection{Unitarity in the `projector' basis} \label{upb}

The projector blocks on the space with insertions in $H_{ {\rm in}} \otimes H_{ {\rm in}}^*$ obey the  completeness relation (analogue of \eqref{projcomplete})
\begin{equation}\label{projcompleteblock}
	\sum_{a, r} {\hat G}_a^{rr} = G_{ {\rm id}};
\end{equation}
where $G_{ {\rm id}}$ is the identity block defined in section \ref{ib}. In order to see why \eqref{projcompleteblock} holds,
recall that the most general block with insertions in the space  $H_{ {\rm in}} \otimes H_{ {\rm in}}^*$ can be expanded as follows
\begin{equation}\label{blockexp}
	G= \sum_{a, r, r'} \alpha_{a, r, r'}  {\hat G}_a^{r, r'}
\end{equation}
Using the second of \eqref{projblock}, \eqref{blockexp} can be rewritten as
\begin{equation}\label{blockexpn}
	G= \sum_{a, r, r'} \alpha_{a, r, r'}  \left( {\hat G}_a^{r', r} \right)^\dagger
\end{equation}
Using \eqref{projoblock} on the product, term by term it follows that
\begin{equation}\label{prodidd}
	G \times \left( \sum_{a, r} {\hat G}_a^{rr} \right) = \sum_{a, r, r'} \alpha_{a, r, r'}  {\hat G}_a^{r, r'} = G
\end{equation}
By the definition of the identity block, however,
\begin{equation}\label{projiin}
	G \times G_{ {\rm id}} = G
\end{equation}
Since \eqref{projiin} and \eqref{prodidd} both hold for every block $G$, the comparison of these two equations gives us
\eqref{projcompleteblock}.

Because the projector blocks obey exactly the same multiplication rule as the invariant tensors of the previous section, and also admit the same resolution of identity as the invariant tensors of the previous section, the rest of the analysis of this subsection proceeds as in subsection \ref{upbc}.

Briefly, we expand the general $S$-matrix
\begin{equation}\label{smofprese}
	S = \sum_{a, r, r'}\mathcal S_a^{r r'} G_a^{r r'}
\end{equation}
and plug this expansion into the equation  $S^\dagger S= G_{ {\rm id}} \mathcal S_{ {\rm id}}$ we conclude that
\begin{equation}\label{ssdag} \begin{split}
		&S^\dagger  S= G_{ {\rm id}} \mathcal S_{ {\rm id}} ~~~~~\\
		&\implies \sum_{\rm final ~states} \left( \sum_{a, r, r'} \left( \mathcal S\right)_a^{* r r'} \left( G _a^{r r'} \right)^\dagger  \right)
		\left( \sum_{b, t, t'} \left( \mathcal S\right)_b^{ t t'} G_b^{t t'}  \right) = \mathcal S_{ {\rm id}} \left(
		\sum_{a, u} {\hat G}_a^{uu} \right) ~~~~~   \\
		& \implies \sum_{\rm final ~states}~~\sum_{a, r, t, r'}  \left( \mathcal S_a^{* r r'}   \star  \mathcal S_a^{t r'} \right)  \hat G_a^{tr} =
		\sum_{a,t,r} \left( \mathcal{S}_{ {\rm id}} \right) \delta^{tr} {\hat G}_a^{tr} ~~~~~~~~~~~~~~~~~~~~~~~~~~~~~~ \\
		& \implies \sum_{\rm final ~states} \mathcal S_a \star \mathcal S_a^\dagger =\mathcal{S}_{ {\rm id}} \mathcal{I}_a\\
	\end{split}
\end{equation}
where we have defined the matrices $\mathcal S_a$
\begin{equation}\label{sasadag}
	\left( \mathcal S_a \right)_m^{~n}= \mathcal S_a^{mn}, ~~~~
	\left( \mathcal S_a^\dagger\right)_m^{~n}= \mathcal S_a^{*nm}, ~~~~\left( \mathcal I_a \right)_m^{~n} = \delta^n_m
\end{equation}
As anticipated above, \eqref{ssdag} is identical in structure to \eqref{smatrixsuni}.

\subsection{Crossing in the block projector basis}
\label{cpb}

As in subsection \ref{crossing}, the same conformal block can be cut
in various ways, yielding different divisions of the insertions of that block into `initial and final'. These blocks form the basis for the expansion of the $S$-matrix in different scattering channels. The claim of crossing invariance
in matter Chern-Simons theories is simply the assertion that the coefficient $S$-matrices for the same conformal block (but in different scattering channels) are analytic continuations of each other.

The implications of crossing invariance become most concrete when combined with the constraints of unitarity. As we have seen in the previous subsection, projector blocks are defined to ensure that their coefficient $S$-matrices obey the canonical unitarity equations \eqref{ssdag}. It is easy to see how these canonically normalized $S$-matrices are related to each other via crossing.

Consider two different divisions of the insertions of a block into initial and final, and corresponding to these divisions expand the $S$-matrix as
\begin{equation}\label{smofprese}
	S = \sum_{a, r, r'} \left( \mathcal S\right)_a^{r r'} G_a^{r r'}
\end{equation}
and
\begin{equation}\label{smofprese2}
	{ S} = \sum_{a, r, r'}\tilde{\mathcal S}_a^{r r'} { \tilde G}_a^{r r'}
\end{equation}
where $G_a^{r r'}$ and ${\tilde G}_a^{r r'}$ are the `projector' type blocks appropriate to the divisions of block insertions into initial and final.

As $G_a^{r r'}$ and  ${\tilde G}_a^{r r'}$ each constitute a basis for the same vector space of blocks, it follows that
\begin{equation}\label{basisinbol}
	G_a^{r r'} = M_{a r r'}^{b s s'} { \tilde G}_b^{s s'}
\end{equation}
for some constant matrices $M_{a r r'}^{b s s'}$\footnote{As explained above the matrices $M_{a r r'}^{b s s'}$ are well defined only once we have also chosen a particular branch for each block.}

Crossing symmetry is the assertion that
\begin{equation}\label{claimcross}
	\sum_{a r r'}  M_{a r r'}^{b s s'}\left( \tau\right)_a^{r r'} =\tilde{\tau}_b^{s s'}
\end{equation}
where the equality in \eqref{claimcross} holds after the appropriate analytic continuation in cross ratios.

The coefficients $M_{a r r'}^{b s s'}$ may be computed as follows. Let us view ${ \tilde G}_a^{r r'}$ as elements in the vector space of conformal blocks, which we denote by $| {\tilde G}_a^{r r'} \rangle $. Now let us consider the inner product
\begin{equation}\label{inprodu}
	\langle {\tilde G}_b^{s s'}|{\tilde G}_a^{r r'} \rangle.
\end{equation}
In the Wilson line representation, this inner product is computed by multiplying ${\tilde G}_b^{\dagger s s'}$ with ${\tilde G}_a^{r r'}$ and then fusing the corresponding Wilson lines. We can perform this fusion operation in two steps. In the first step we fuse those representations corresponding to the `final insertions'. This is precisely the compounding operation we have defined above,
and equals $\delta_{ab} \delta_{r' s'}$ times the Wilson line representation of the block
${\tilde G}^{rs}_a$, whose insertions lie in the space $H_{ {\rm in}} \otimes H_{ {\rm in}}^*$. In the second step, we fuse the end points of the remaining Wilson lines and their complex conjugates. This fusion process gives $\delta_{rs} D^k_a$ where
$D^k_a$ is the quantum dimension (expectation value of a circular Wilson loop) of the representation $R_a$. Putting it all together,
we conclude that
\begin{equation}\label{inproduans}
	\langle {\tilde G}_b^{s s'}|{\tilde G}_a^{r r'} \rangle
	= \delta_{sr} \delta_{s'r'} \delta_{ab} D^k_a
\end{equation}

Now \eqref{basisinbol} may be rewritten, in vector language, as
\begin{equation}\label{basisinbolvec}
	|G_a^{r r'} \rangle = M_{a r r'}^{b s s'} |{\tilde  G}_b^{s s'} \rangle
\end{equation}

Taking the inner product of \eqref{basisinbolvec} with $\langle {\tilde G}_b^{s s'}|$ and using \eqref{inproduans}, we conclude that
\begin{equation}\label{formm}
	M_{a r r'}^{b s s'} =  \frac{ \langle {\tilde G}_b^{s s'}|G_a^{r r'} \rangle}{D^k_a}
\end{equation}
The quantity $\langle {\tilde G}_b^{s s'}|G_a^{r r'} \rangle$ that appears on the RHS of \eqref{formm} is the expectation value of
the {\it closed} Wilson line tangle obtained by fusing the Wilson line representations of the
two states in the inner product,
and may be evaluated explicitly using the
techniques of \cite{Witten:1989wf}. We will see how this works concretely in the context of a particularly simple example in the next section.

Let us summarize. The $S$-matrices in matter Chern-Simons theories obey the crossing relations \eqref{claimcross}, with the matrix coefficients $M_{a r r'}^{b s s'}$ given by
the expectation value of the Wilson line tangle denoted in \eqref{formm}.

The discussion presented of this section is modified somewhat in the case that two or more inserted operators are identical. In this Appendix \ref{ipfk}
we outline the nature of these  modifications focussing on the case of $2 \rightarrow 2$
scattering. While these modifications are interesting in their own right, as far as we can tell, they play no role in the discussion of crossing symmetry, the topic of principal interest to us in this paper.

\subsection{$S_a^{rr'}$ in terms of `Wilson lines at infinity' }

In \eqref{smofprese} we have presented a convenient expansion of the $S$-matrix. We may regard $S$ in \eqref{smofprese} as a vector in the space of conformal blocks. Taking the inner product of \eqref{smofprese} with the bra $\langle G_a^{rr'}|$ and using
the relation \eqref{inproduans}
we find that
\begin{equation} \label{invdef}
{\mathcal S}_a^{r r'} = 	\frac{ \langle G_a^{rr'}| S \rangle}{D_a^k}
\end{equation}
Now recall that, in Wilson line language, the inner product of the $S$-matrix with the block $G_a^{rr'}$ is obtained by completing the world lines that make up the $S$-matrix
$S$ with the Wilson line representation of $G_a^{rr' \dagger}$. It follows that
the `invariant' definition of ${\mathcal S}_a^{r r'}$ is given by attaching the inserted operators to Wilson lines that extend to infinity in a very particular way - the way that makes up the Wilson line representation of the block  $G_a^{rr'\dagger}$.

\section{Scattering involving fundamental-fundamental, antifundamental-antifundamental insertions} \label{fundnew}

In this section we study the matter Chern-Simons analogue of the example of subsection
\ref{Classexamp}; i.e., crossing symmetry
for the $S$-matrices captured by a correlation function of two fundamentals and two antifundamentals of $SU(N)_k$ theory (also of the $U(N)_k$ Type I and Type II theories).

In the context of the example of this section, we will present an explicit and detailed work out of the material of subsections \ref{pbs} and \ref{cpb}, and find completely explicit crossing symmetry rules for the relevant $S$-matrices.

\subsection{Fundamental-fundamental scattering} \label{ffs}

Let us first consider the scattering of two fundamentals to two fundamentals
(see \eqref{scpro}).

\begin{figure}[H]
	\centering
	\includegraphics[scale=.25]{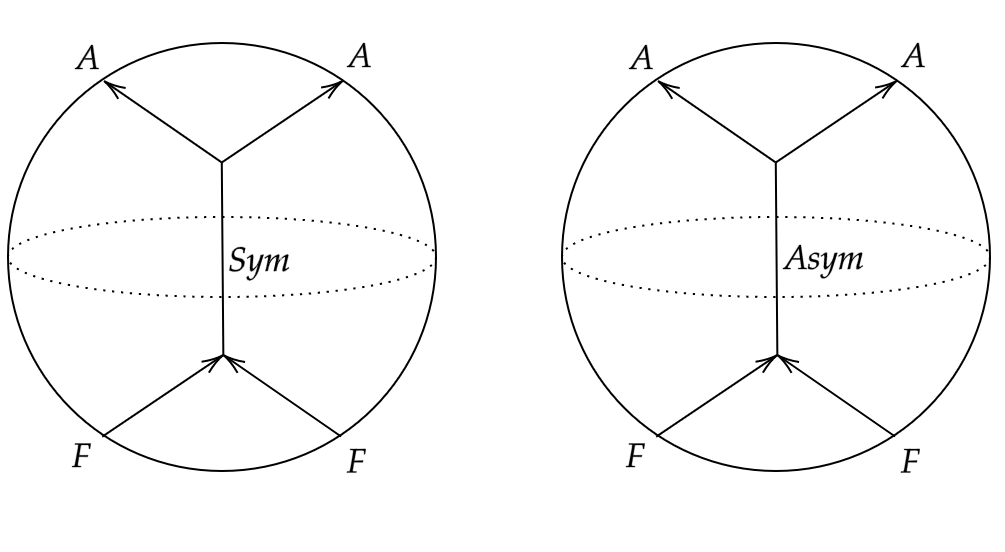}
	\caption{Symmetric and antisymmetric blocks}
	\label{symasym}
\end{figure}

The two orthogonal basis
blocks for this process are depicted in Fig.
\ref{symasym}. The vertex factors in the diagrams of Fig. \ref{symasym} are proportional to the unique
CG coefficients for the coupling of two fundamentals to the symmetric and antisymmetric, also to the unique CG coefficient for the coupling of two antifundamentals to the symmetric or antisymmetric of antifundamentals.
Following the discussion in Appendix \ref{conventions} we choose these two CG coefficients to be
complex conjugates of each other. This convention ensures that the blocks $G_{s}^\dagger$ and $G_{a}^\dagger$ take the
same form as $G_{s}$ and $G_{a}$ respectively, (but with the locations of the fundamental and antifundamental
insertions interchanged).
 At this stage (and unlike in the analysis of subsection \ref{pbs}) we have not chosen a particular normalization for these couplings. In our exposition below we will make amends for this by suitably normalizing our blocks; the net result of this will be
to move to using blocks normalized as instructed by the analysis of \ref{pbs}. In practical terms,
it will be a consistency check on our computation, that the unknown normalizations of CG coefficients in Fig.
\ref{symasym}  cancel out in all final formulae.

\begin{figure}[h]
	\centering
	\includegraphics[scale=.15]{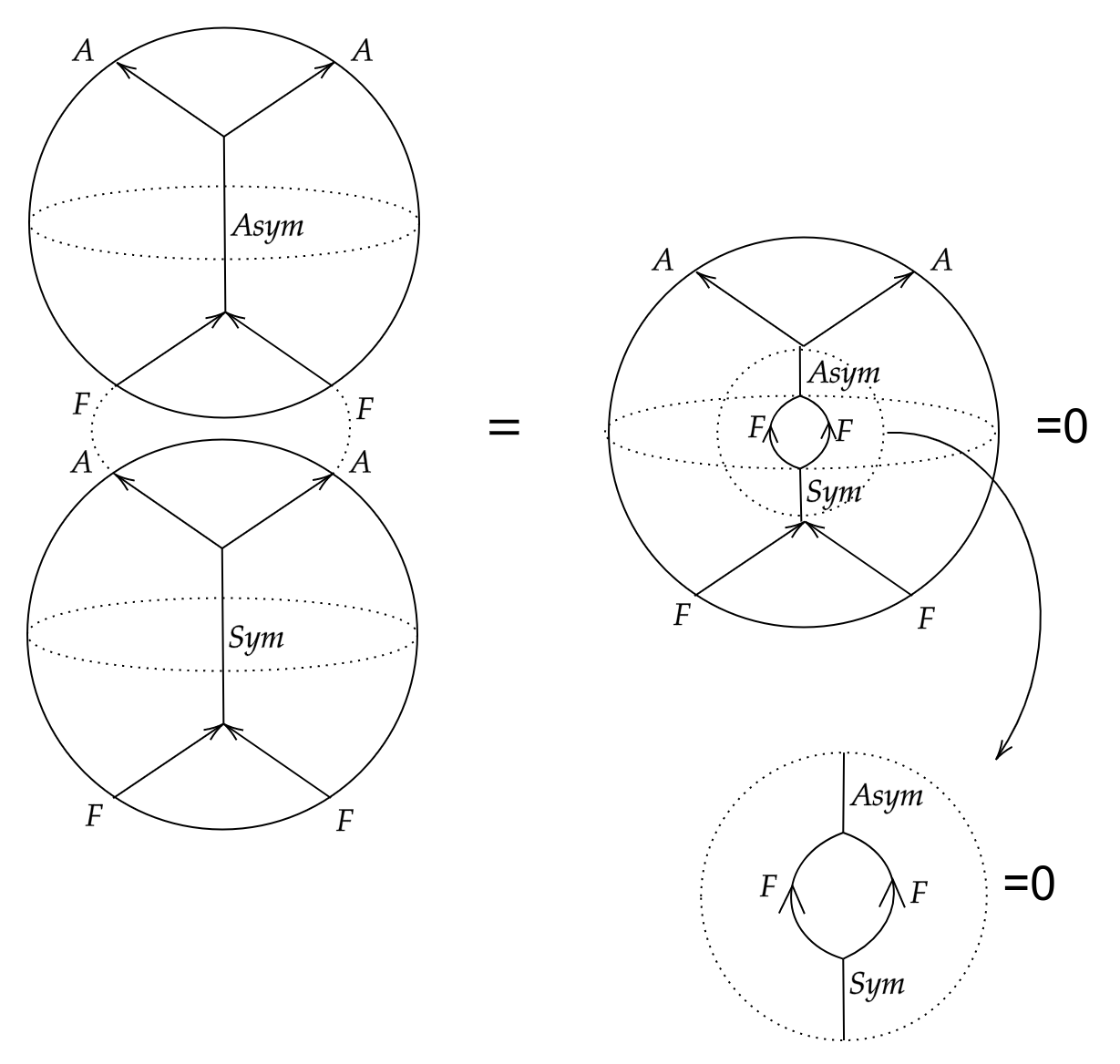}
	\caption{Compounding of symmetric with antisymmetric}
	\label{sasorth}
\end{figure}

The fact that the two blocks in Fig. \ref{symasym} are orthogonal under compounding  (more precisely that $G^\dagger_{s} \times G_{a}=G^\dagger_{a} \times G_{s} = 0$) is demonstrated in Fig. \ref{sasorth}. The argument is that the part of the path integral enclosed by the dotted sphere vanishes, as it evaluates to a two-point function block with insertions of non conjugate representations, and no such blocks exist.

\begin{figure}[h]
	\centering
	\includegraphics[scale=.15]{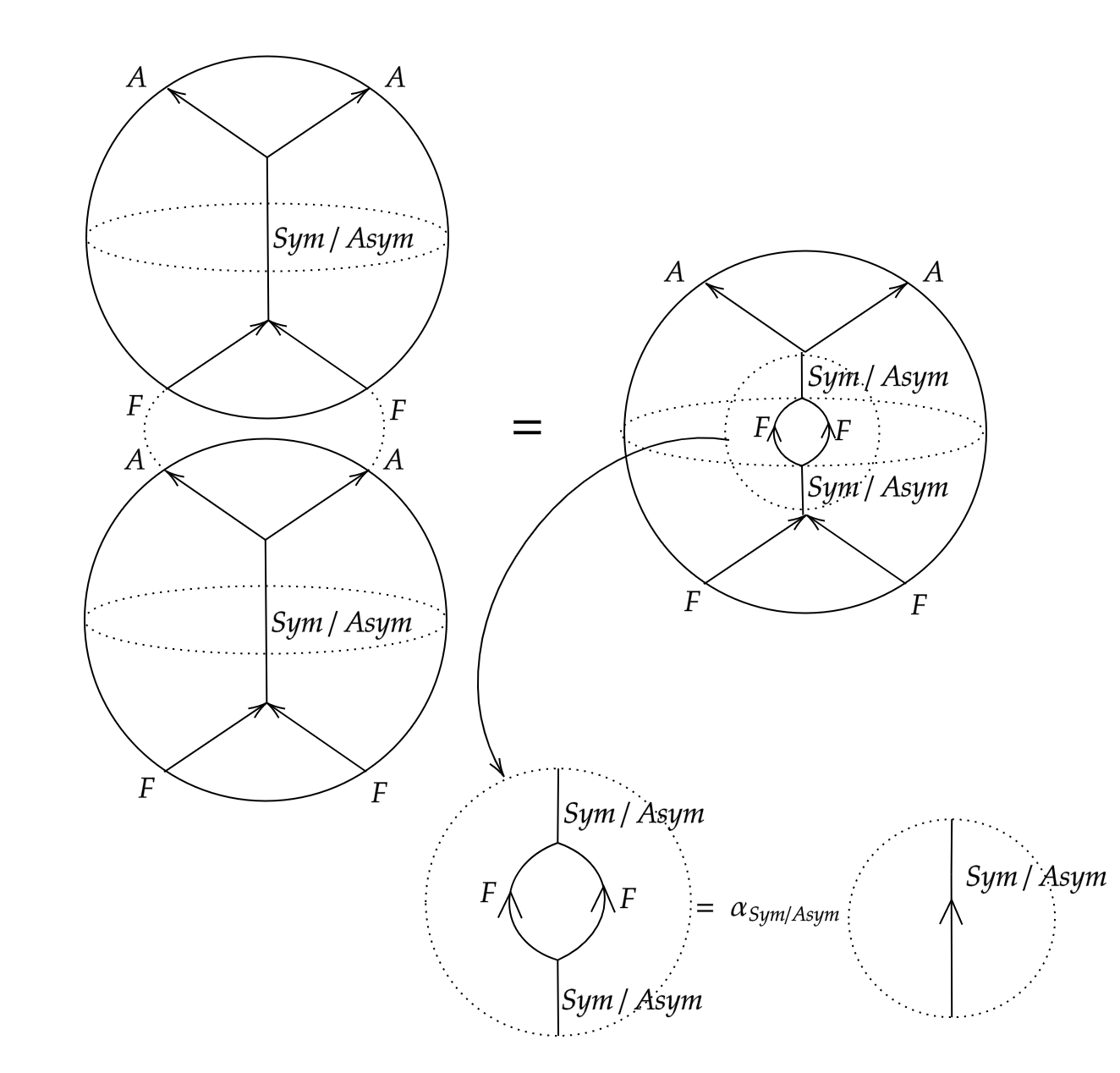}
	\caption{Compounding of (anti)symmetric with (anti)symmetric}
	\label{ssasas}
\end{figure}

Fig \ref{ssasas} explains that
the compounding of $G^\dagger_{s/a}$ with $ G_{s/a}$ equals
${\hat G}_{s/a}$ \footnote{As in the previous subsection, we denote blocks that live in $H_{ {\rm in}} \times H_{ {\rm in}}^*$ with a hat, as in ${\hat G}$.} times a number, $\alpha_{Sym/Asym}$. The value of
$\alpha_{Sym/Asym}$ is computed in Fig. \ref{cnorm} below.
\begin{figure}[h]
	\centering
	\includegraphics[scale=0.25]{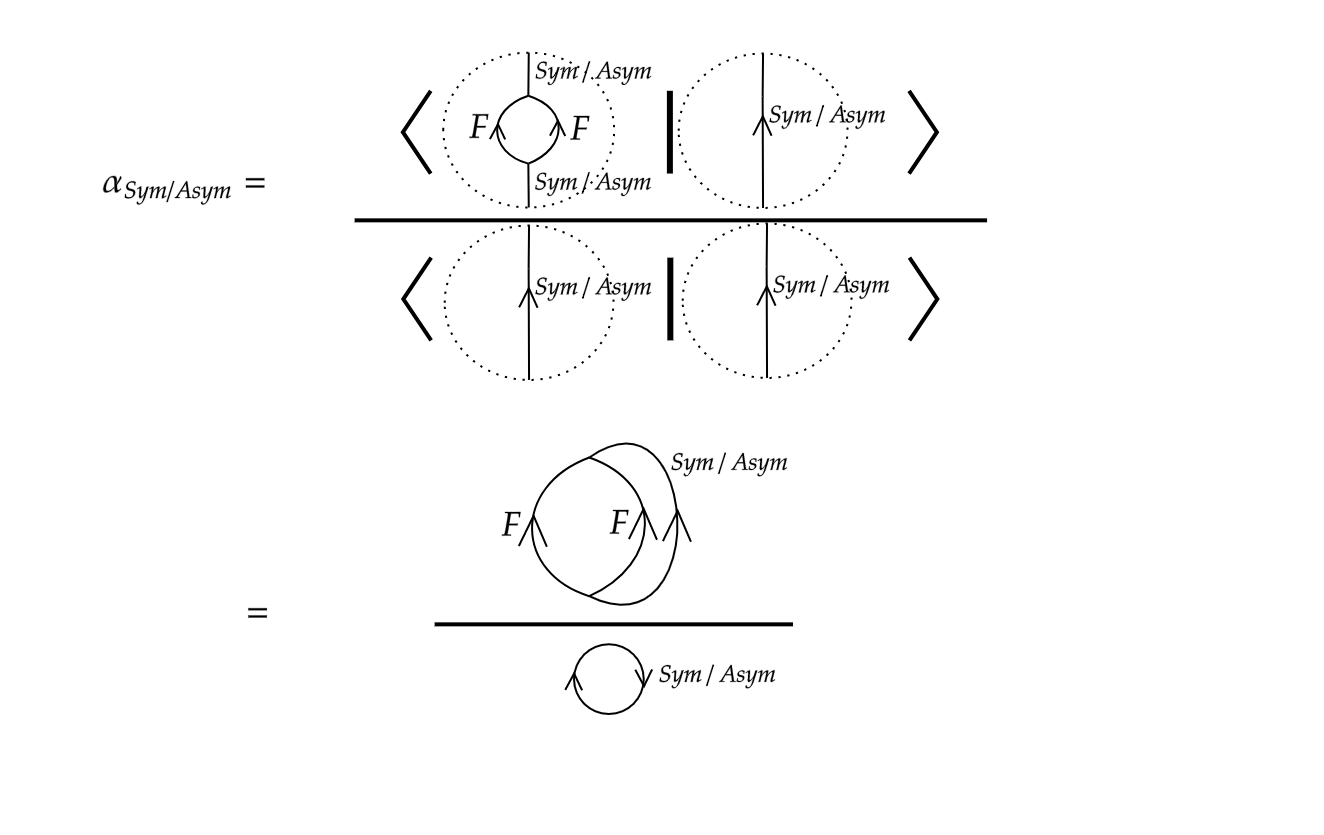}
	\caption{Normalization of the symmetric/antisymmetric blocks under compounding}
	\label{cnorm}
\end{figure}
The final answer for $\alpha_{Sym/Asym}$ is presented in the last line of Fig. \ref{cnorm}.
It follows that the normalized Sym/Asym block (normalized so that the block squares to itself)
is given by Fig. \ref{nblocks1}.
\begin{figure}[h]
	\centering
	\includegraphics[scale=.25]{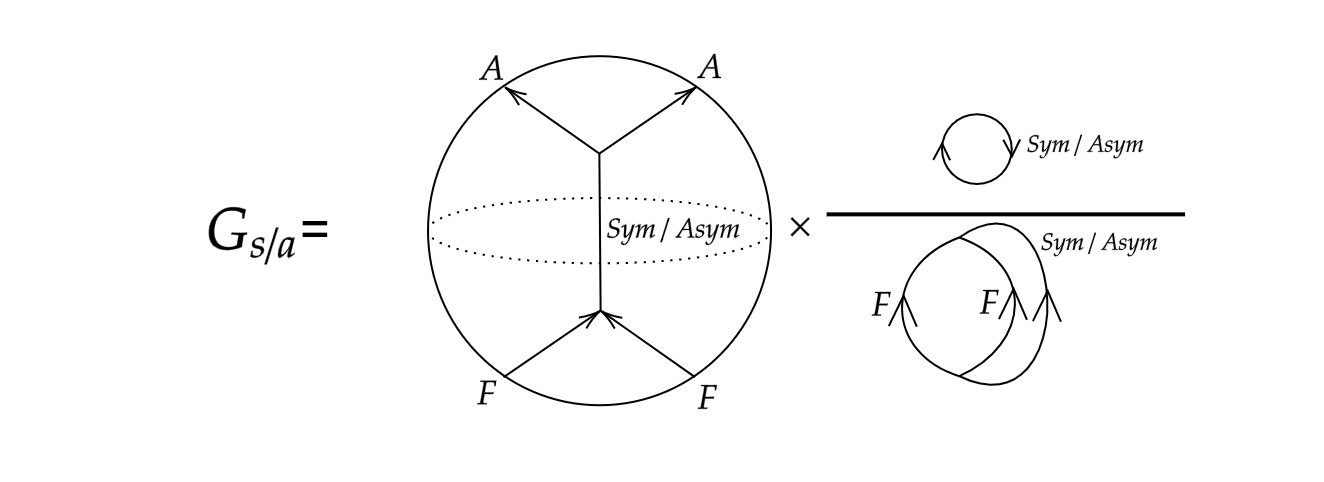}
	\caption{Normalized symmetric/antisymmetric blocks}
	\label{nblocks1}
\end{figure}

The blocks $G_{s/a}$ depicted in Fig \ref{nblocks1} are the conformal block analogues of the invariant tensors $I_{s/a}$
(see \eqref{ppscatff})  and reduce to the same in the limit $k \to \infty$.

The block depicted in Fig. \ref{nblocks1} could also have been obtained directly following the method outlined in subsection \ref{pbs}. In that subsection we worked with CG coefficients that were normalized in the specific manner depicted in Fig. \ref{threeptorthogtw}.  It is not hard to convince oneself that the suitably normalized
CG coefficient for each interaction vertex can be obtained, starting with an arbitrarily normalized
CG coefficient, and then dividing  by the square root of the factor that appears in Fig. \ref{nblocks1}. \footnote{Multiplying the original coefficient by a constant $\beta$ leaves the normalized coefficient invariant, as the normalization factor is an expression proportional to one over the square root of an expression quadratic in the CG coefficient.}. Since the diagram in
Fig. \ref{nblocks1} has two vertices, it needs to be multiplied by this normalization factor squared, explaining the factor in Fig. \ref{nblocks1}.

In the specially simple example under study,
$H_{ {\rm in}}=H_{ {\rm out}}$. As a consequence
$H_{ {\rm in}} \otimes H_{ {\rm out}}^*= H_{ {\rm in}} \otimes H_{ {\rm in}}^*$, and so the $G$ (defined to have insertions in $H_{ {\rm in}} \otimes H_{ {\rm out}}^*$) and the blocks ${\hat G}$
(defined to have insertions in $H_{ {\rm in}} \otimes H_{ {\rm in}}^*$) belong to the same vector space. Consequently, \eqref{projcompleteblock}
asserts that
\begin{equation}\label{gagsident}
	G_s+G_a= G_{ {\rm id}}.
\end{equation}
\footnote{Actually \eqref{projcompleteblock} really assert that ${\hat G}_s+{\hat G}_a= {\hat G}_{ {\rm id}}$.
As $H_{ {\rm in}}=H_{ {\rm out}}$ in the example of this section, however, hatted and unhatted blocks are
structurally identical, and so \eqref{projcompleteblock} also implies \eqref{gagsident}.}

where $G_{ {\rm id}}$ is the block defined by the Wilson line configuration in which the two initial fundamentals directly connect to the two final fundamentals. Infact, $G_{ {\rm id}}$ is the block we name  $\theta$ in Fig. \ref{thph} in the Appendix, and so we continue to use this notation here.

In this simple example it is easy to directly check that \eqref{gagsident} holds. In order to do this we view the blocks that appear in \eqref{gagsident} as vectors in the Hilbert space of conformal blocks, and take the inner products of \eqref{gagsident}  with $G_{s/a}$.
The inner products $\langle G_{s/a}| G_{s/a} \rangle$ and $\langle G_{s/a}| \theta \rangle$ are computed in Fig \ref{ss}:
\begin{figure}
	\includegraphics[scale=.25]{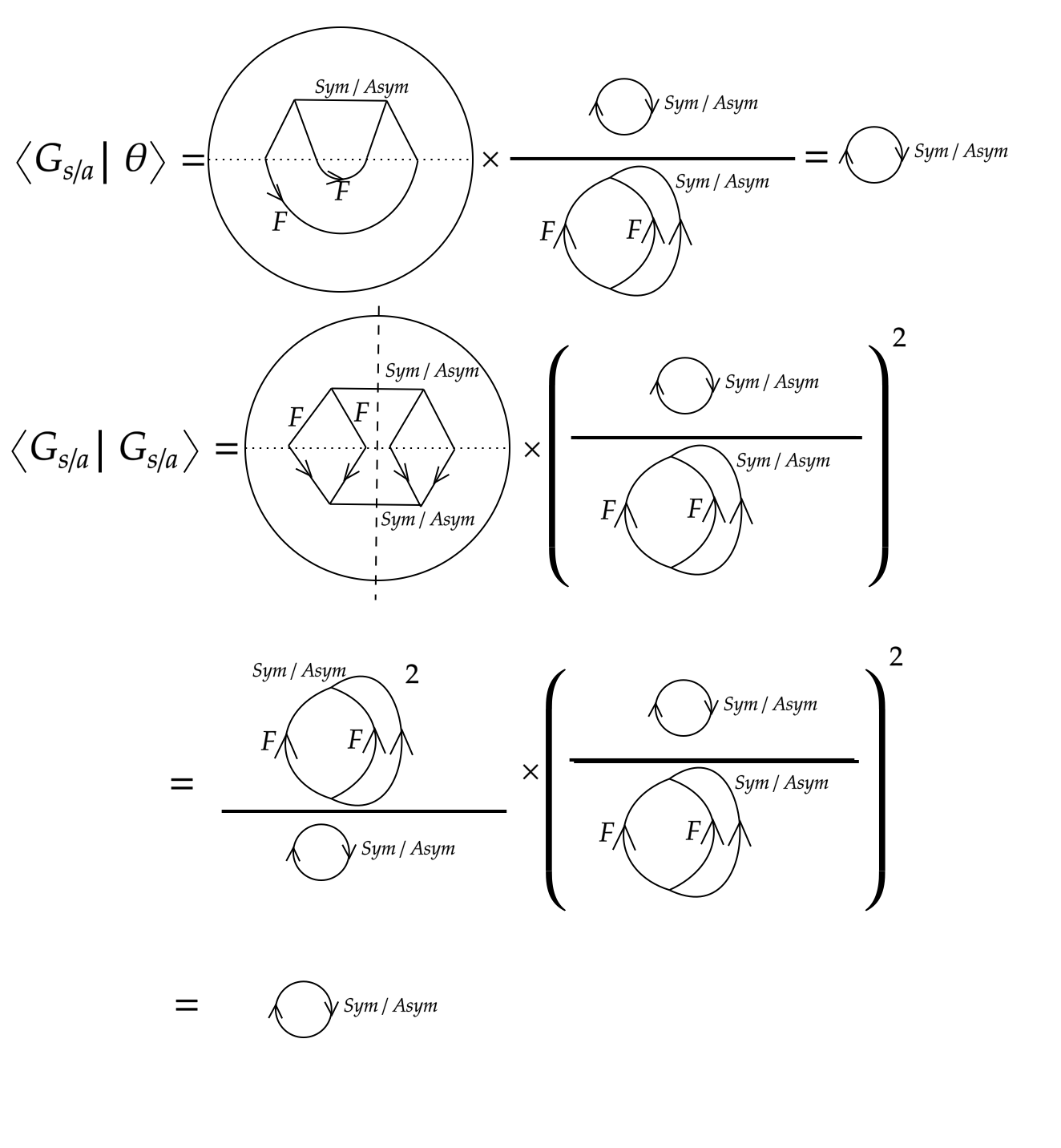}
	\caption{Computation of $\langle G_{s/a}| G_{s/a} \rangle$ and $\langle \theta| G_{s/a} \rangle$}
	\label{ss}
\end{figure}
From fig \ref{ss}, we see that
$$\langle G_{s/a}| \theta \rangle = \langle G_{s/a}| G_{s/a} \rangle $$
It follows that
\begin{equation}\label{gagsidentvec}
	|G_s \rangle +|G_a \rangle= |\theta \rangle
\end{equation}
(recall that $|G_s\rangle$ and $|G_a \rangle$ are orthogonal to each other) and so \eqref{gagsident} holds.

Notice that in the course of our demonstration,
we have also checked that
\begin{equation}\label{dsa}
	\langle G_{s/a}| G_{s/a} \rangle =D^k_{s/a}
\end{equation}
(recall that the circular Wilson loop in any representation equals its quantum dimension). It follows that we have also verified \eqref{inproduans} for this particular case.

As we have noted above, in the example under study in this section, $H_{ {\rm in}}= H_{ {\rm out}}$ and so
the collection of representations that appears in all blocks is self conjugate. It follows that
there exists a representation for all blocks in terms of Wilson line configurations with no explicit bulk interactions. In Appendix \ref{quantexamp} we demonstrate that the normalized symmetric and antisymmetric blocks depicted in Fig \ref{nblocks1}
can equally well be represented as in this alternate language by  \eqref{symasb}. In Appendix  \ref{quantexamp} we also repeat all the computations presented in the rest of this subsection using the representation \eqref{symasb}, and
verify we obtain the same result for all final formulae.

\subsection{Fundamental-antifundamental scattering}

Let us now consider fundamental antifundamental scattering, \eqref{papscat}.

\begin{figure}[h!]
	\centering
	\includegraphics[scale=.15]{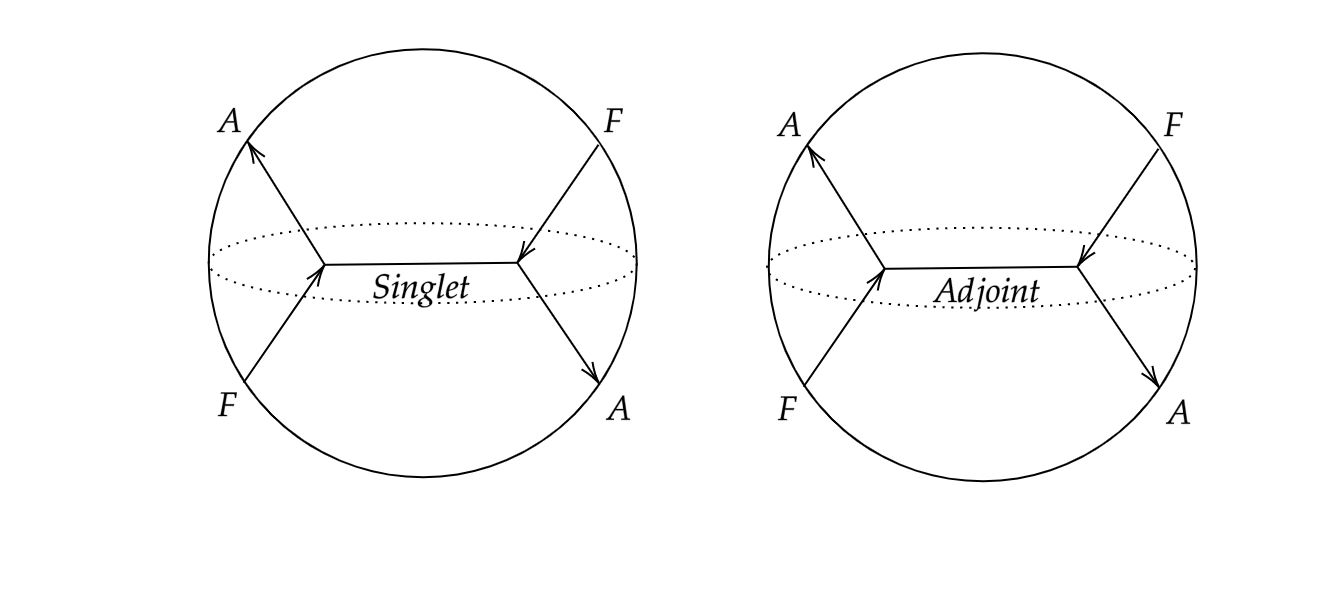}
	\caption{Singlet and adjoint blocks}
	\label{singadj}
\end{figure}

An orthogonal (though not yet orthonormal) basis of blocks for this scattering process is
depicted in Fig \ref{singadj}.

As in the previous subsection, at this stage the interaction vertices that appear in Fig.
\ref{singadj} are arbitrarily normalized.

\begin{figure}[H]
	\centering
	\includegraphics[scale=.25]{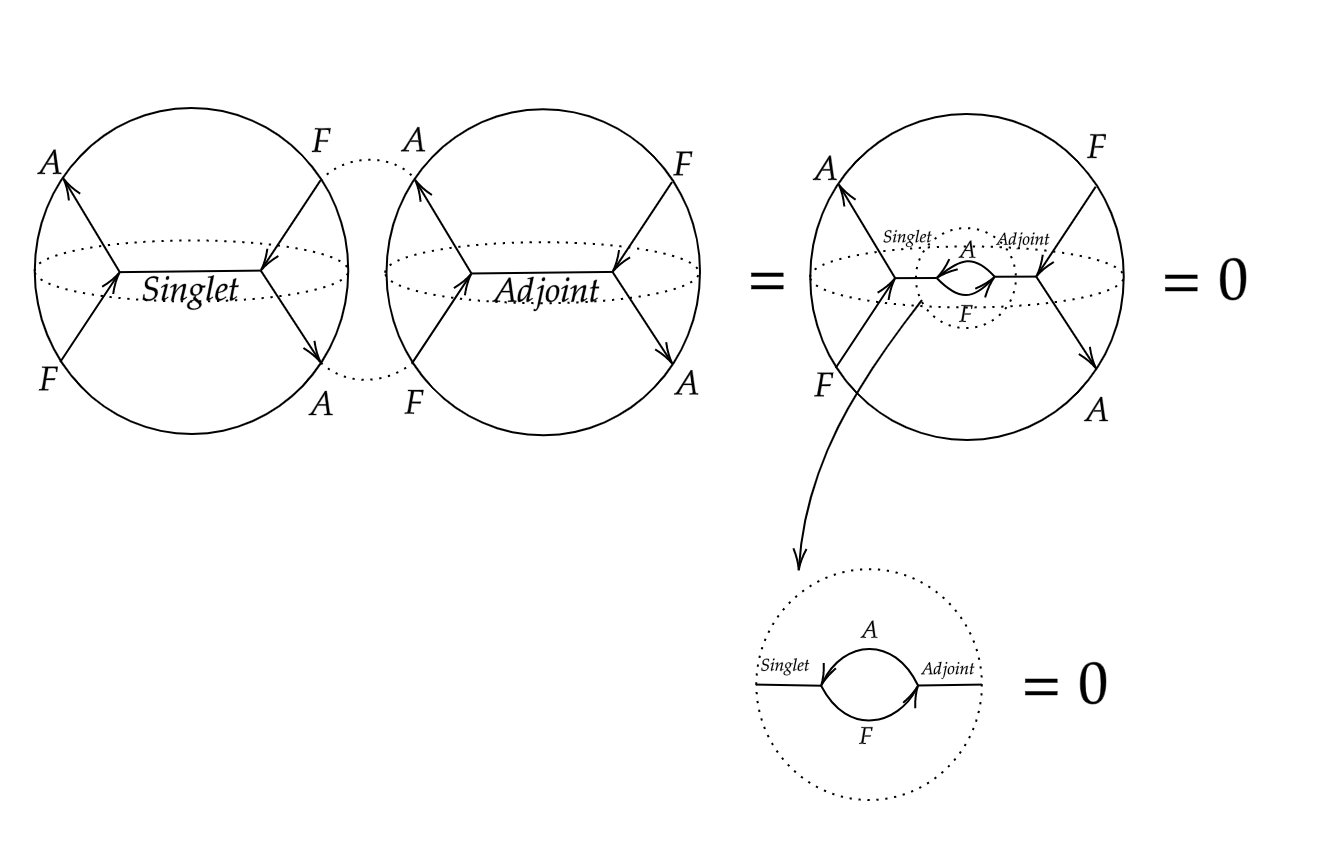}
	\caption{Compounding of Singlet with Adjoint}
	\label{ortho2}
\end{figure}

The fact that the blocks depicted in Fig. \ref{singadj} are orthogonal (i.e., that $G_{I/{\rm Adj}}^\dagger G_{{\rm Adj}/I}=0)$  is verified in Fig. \ref{ortho2}
(once again the argument is that the two-point block enclosed by the dotted sphere vanish).

\begin{figure}[H]
	\centering
	\includegraphics[scale=.25]{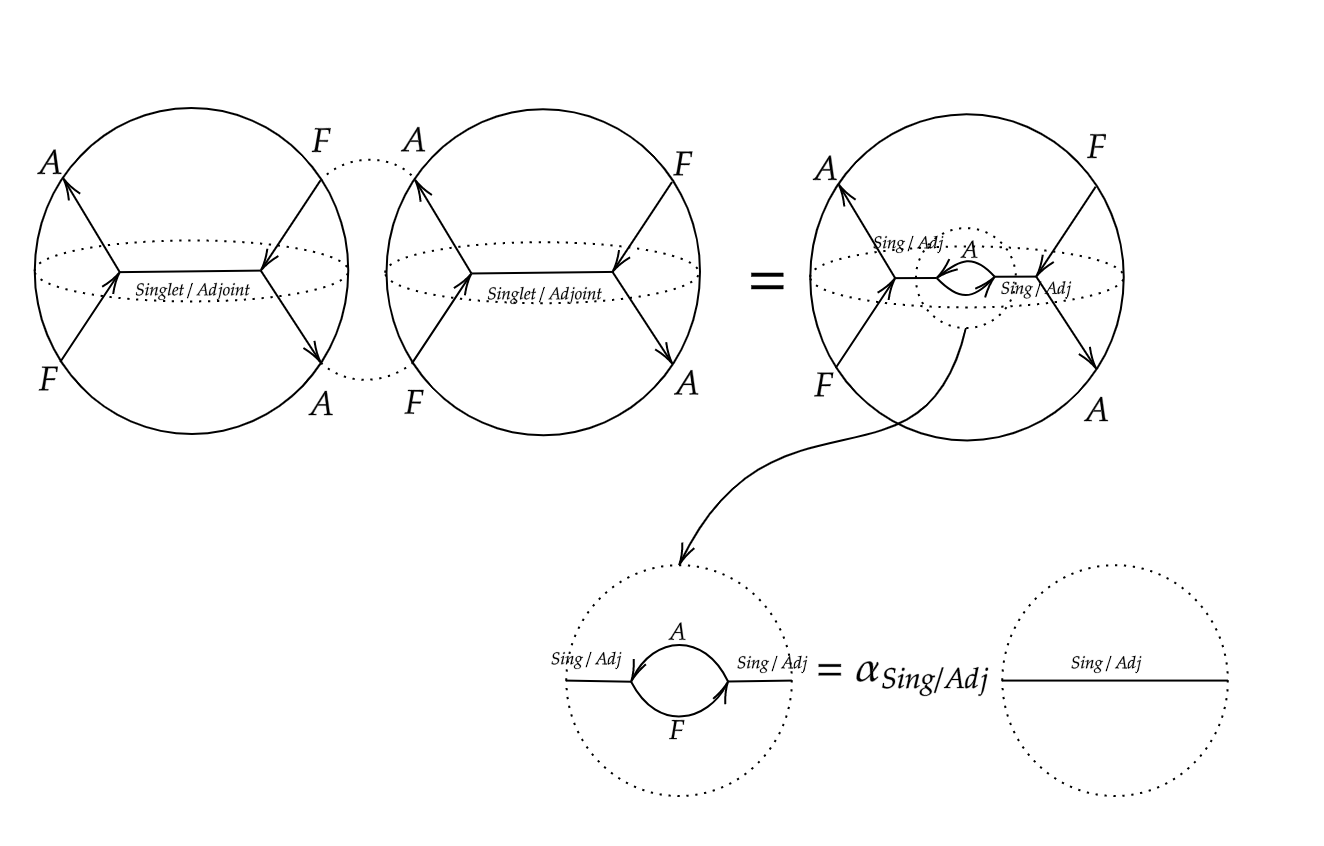}
	\caption{Compounding of Singlet(Adjoint) with Singlet(Adjoint)}
	\label{ortho3}
\end{figure}

\begin{figure}[h]
	\centering
	\includegraphics[scale=.25]{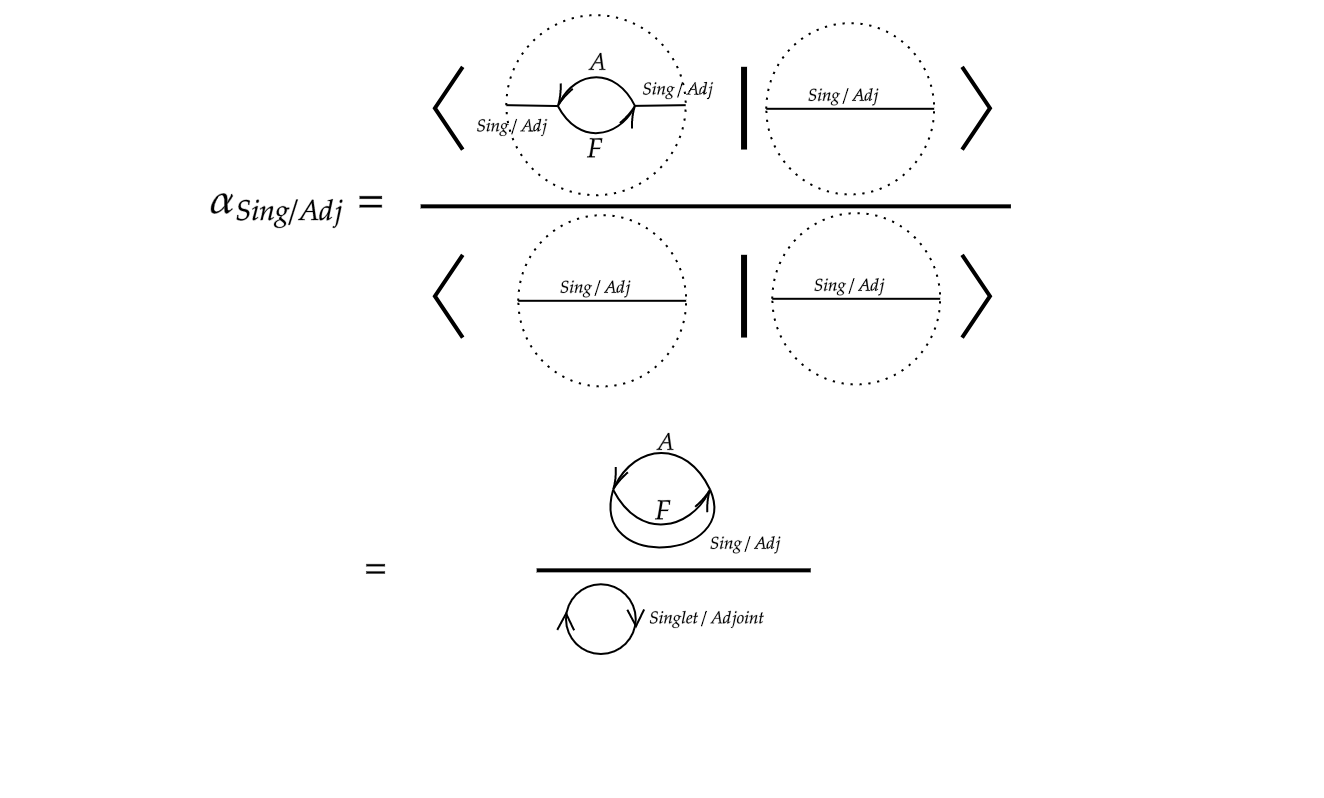}
	\caption{Normalization under compounding}
	\label{normsa}
\end{figure}
\begin{figure}[h]
	\centering
	\includegraphics[scale=.25]{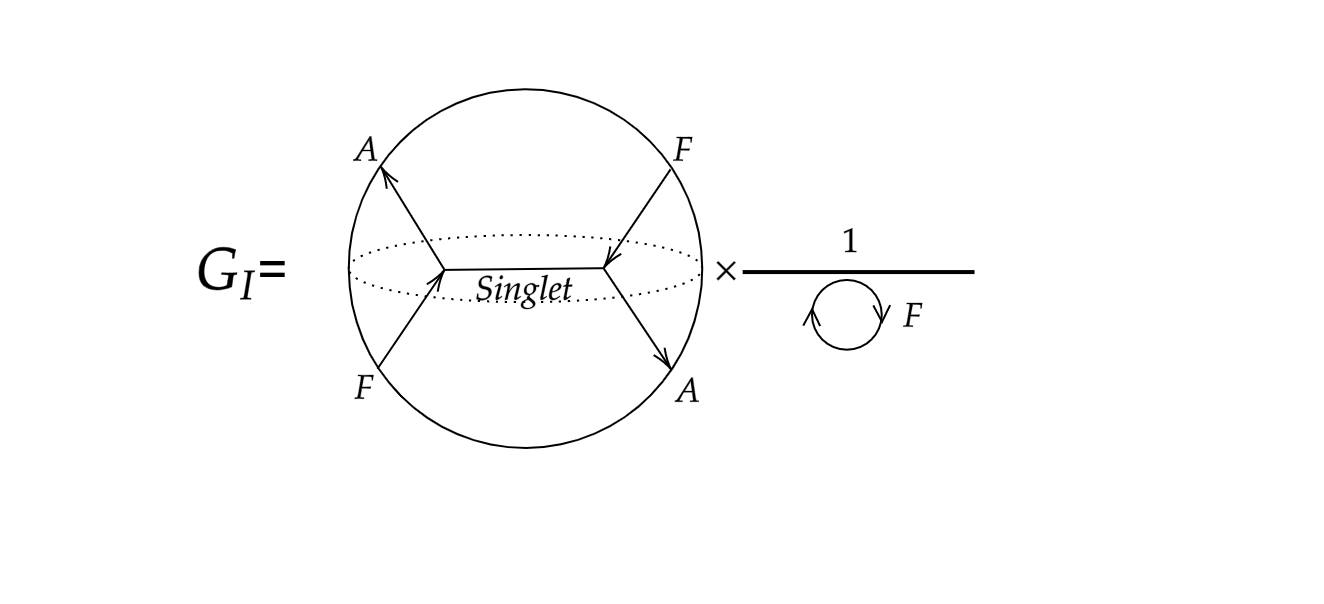}
	\includegraphics[scale=.25]{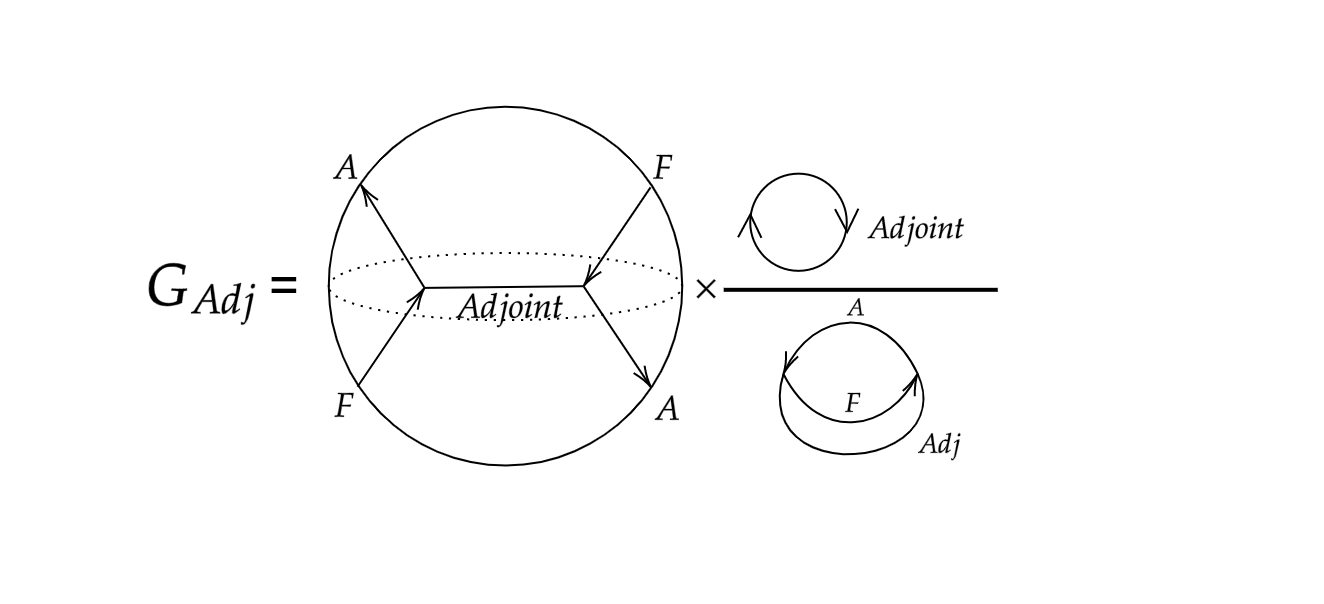}
	\caption{Normalized singlet and adjoint blocks}
	\label{nblocks2}
\end{figure}

As in the previous subsection, the blocks
depicted in Fig \ref{singadj} square to themselves \footnote{More precisely that
	$G_{{\rm Adj}/I}^\dagger \times {G}_{{\rm Adj}/I}= {\hat G}_{{\rm Adj}/I}$.} (under compounding) only after an
appropriate normalization. This normalization
is computed in Fig. \ref{ortho3} and Fig. \ref{normsa}. The final normalized `projector block' -- the matter Chern-Simons analogue of the projector invariant tensors $T_I$ and $T_{ {\rm Adj}}$ --  are depicted in Fig. \ref{nblocks2}.

As in the previous subsection, the factors that appear in Fig. \ref{nblocks2} can be thought of as a product of two square roots. Associating one of each of these square roots with the interaction vertices in Fig. \ref{nblocks2}, turns these vertices into the canonically normalized vertices of subsection \ref{pbs}.

As in the previous subsection, it is easy to show that
\begin{equation} \label{idap}
	\tilde G_{ {\rm id}}=X=G_I + G_{ {\rm Adj}}
\end{equation}
Here $X=\tilde{G}_{ {\rm id}}$ is the identity block, whose Wilson line representation is a straight Wilson line from the initial $F$ insertion to the final $A$ insertion, and a similar straight
Wilson line from the initial $A$ insertion to the final $F$ insertion. \footnote{Note that the block
${\tilde G}_{ {\rm id}}$ is different from the block $G_{ {\rm id}}$ that appears in \eqref{gagsident}, as the
straight Wilson lines that define these two distinct identity blocks connect different insertions.
${\tilde G}_{ {\rm id}}$ is the identity for compounding for $FA \rightarrow FA$ scattering, while
${G}_{ {\rm id}}$ was identity for scattering in the $FF \rightarrow FF$ channel.}

The block $X$ is depicted in Fig \ref{comp3} in the Appendix.

Finally, as in the previous subsection it is
easy to explicitly verify that
\begin{equation}\label{dIA}
	\langle G_{I}| G_{I} \rangle =1, ~~~\langle G_{ {\rm Adj}}| G_{ {\rm Adj}} \rangle =D^k_{ {\rm Adj}}
\end{equation}
in agreement with  \eqref{inproduans}.

As in the case of FF scattering, in Appendix \ref{quantexamp} we find an alternate representation
of the normalized blocks of Fig \ref{nblocks2} in terms of Wilson lines with no bulk interactions (see \eqref{adjsingprojnint}).
In the Appendix we use this alternate representation to repeat all the computations presented in this section with identical final results.

\subsection{Crossing}\label{cross}

As in subsection \ref{cpb}, we determine the crossing relations by expressing the symmetric/ antisymmetric projector blocks in terms of the
singlet/adjoint projector blocks. \begin{equation}
	\begin{split}
		&G_s=a_s^I G_I + a_s^{ {\rm Adj}} G_{ {\rm Adj}}\\
		&G_a=a_a^I G_I + a_a^{ {\rm Adj}} G_{ {\rm Adj}}
	\end{split}
\end{equation}
As in subsection \ref{cpb}, the coefficients in this expansion may be determined by viewing the blocks as vectors in the Hilbert space of blocks. Using \eqref{dIA} and \eqref{dsa} it follows that
\begin{equation}\label{asa}
	a_{s/a}^{ {\rm Adj}}=\frac{\langle G_{ {\rm Adj}}|G_{s/a}\rangle}{D_{ {\rm Adj}}},
	~~~~a_{s/a}^I=\langle G_{ {\rm Adj}}|G_{s/a}\rangle
\end{equation}

\begin{figure}[h]
	\centering
	\includegraphics[scale=.2]{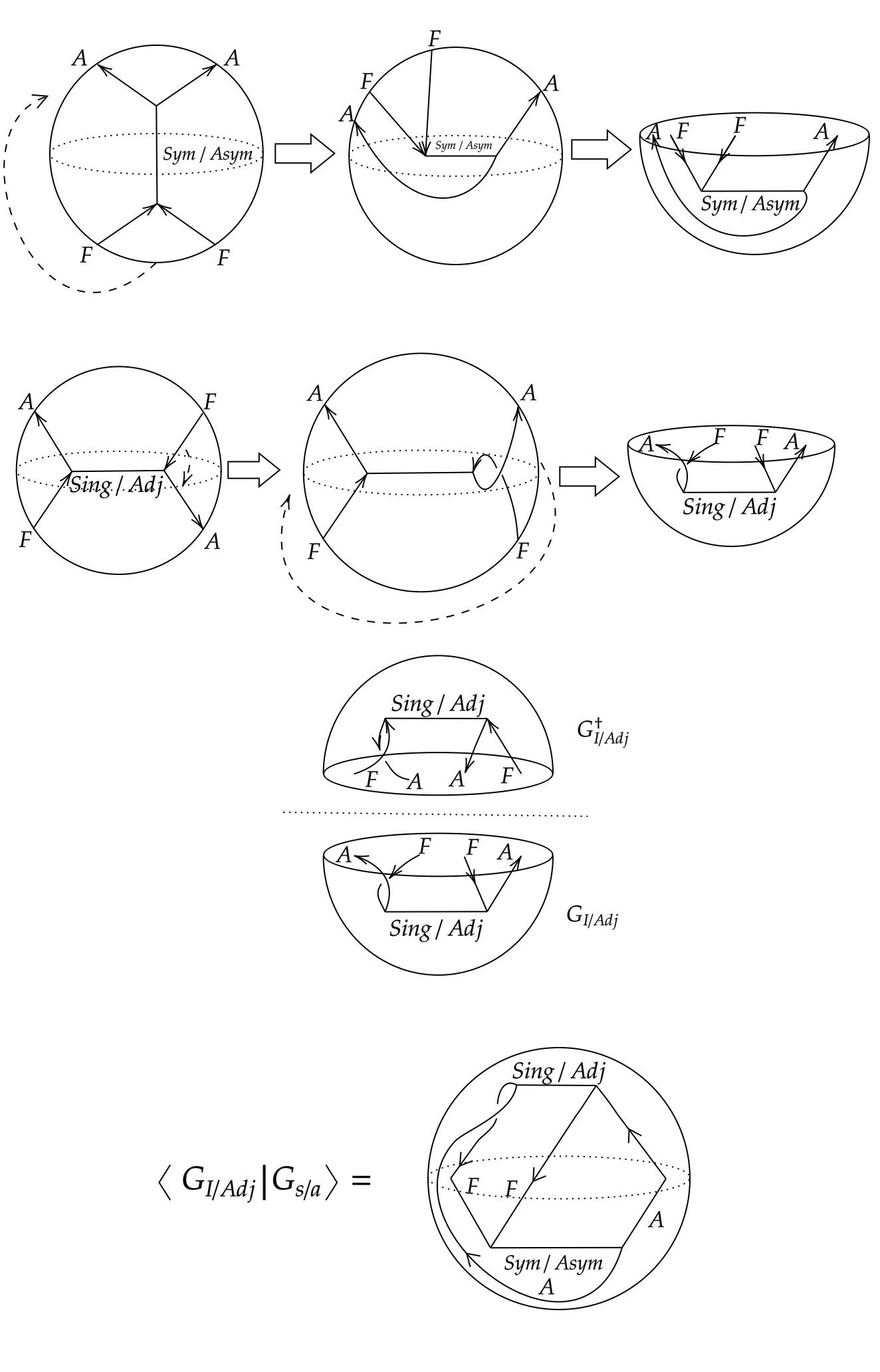}
	\caption{Manipulations on the blocks to take the inner product}
	\label{defadjsing}
\end{figure}

In order to take the inner product
in \eqref{asa}, we need to glue
the ket corresponding to $G_{s/a}$
with the `bra' corresponding to the reflection of $G_{{\rm Adj}/I}$. The ket corresponding $G_{s/a}$ is obtained starting from Fig \ref{nblocks1}. We then deform this diagram so that both fundamental insertions and the rightmost antifundamental insertion are moved to the left top of the figure. We finally flatten out the top boundary (where all insertions are located), as depicted in Fig
\ref{defadjsing}.

The ket corresponding to $G_{ {\rm Adj}}$ or $G_I$ is obtained starting from the diagrams in Fig. \ref{nblocks2}.
The first step here is to  `interchange' the location of
the fundamental and antifundamental insertions on the left of this diagram (see Fig. \ref{defadjsing}). This interchange  is necessary to ensure that the bra we finally obtain has insertions at locations that allow for a smooth gluing with the ket we have already
obtained above. The interchange above can be done in two ways
\footnote{More precisely it can be done in an infinite number of ways, by twisting by arbitrary additional multiples of $2 \pi$. In the main text below we only refer to the
two simplest ways of achieving this interchange.}
 by taking the fundamental either above or below the antifundamental. These two choices differ only by a phase (see below for details). As this phase is presumably physically unimportant (see below for a discussion) we make one choice aribitrarily: we choose to take the fundamental line above the
antifundamental, as depicted in Fig.
\ref{defadjsing}.

Once we have performed this twisting operation, we now have a block with
two fundamental insertions at the bottom and two antifundamental insertions at the top. This is the same configuration we started with for the blocks $G_{s/a}$ and the remaining manipulations we perform on this block are the same as for that case. As depicted in Fig. \ref{defadjsing} we  move the top right and two bottom insertions to the top left of the diagram, and flatten out the top (the part with all insertions). Finally, in order to obtain the bra corresponding to this ket, we reflect the ket about the flat surface (again see Fig \ref{defadjsing}).

In order to compute the inner product \eqref{asa} we now simply glue the ket in the last of Fig. \ref{defadjsing} with the bra in the last of Fig. \ref{defadjsing}. The closed Wilson tangle that we obtain from this process is depicted in the first line of Fig. \ref{sasadj}.

We then process the Wilson line tangle in the first line of Fig. \ref{sasadj} as follows. In going from the first to the second line of Fig. \ref{sasadj} we have used the identity in Fig. 16 of \cite{Witten:1989wf} (this identity allows us to flip the cyclical order of interactions that appear in an interaction vertex, at the cost of a phase). The closed Wilson line tangle obtained at the end of this process is precisely the `symmetric
tetrahedron' that appears in
Fig. 28 of \cite{Witten:1989wf}.
In going from the second to the third line of Fig \ref{sasadj} we follow the steps outlined in Fig. 29 of \cite{Witten:1989wf}. The $\pm$ that appears in this diagram is plus when we are working with $G_s$ and minus if we are working with $G_a$ (the reason for this is mentioned at the end of the fourth last paragraph in Appendix \ref{conventions}). These steps convert the symmetric tetrahedron to a tetrahedron that is less symmetric, but easier to explicitly evaluate. The manipulations that take us from the third to the fourth line of Fig.
\ref{sasadj} are given in Fig.
31 of \cite{Witten:1989wf}. The constant ${\cal N}_{s/a~I/{\rm Adj}}$ that appears in the fourth line of
Fig. \ref{sasadj}, was evaluated in
\cite{Witten:1989wf} for the case of the $SU(N)_k$ theory. We have generalized Witten's computation to obtain a result that applies equally well to the $SU(N)_k$, $U(N)_k$ Type 1 and Type II theories. We find
\begin{equation}\label{Nsa}
\mathcal{N}_{s/a~ I/{\rm Adj}}=\frac{e^{-\frac{\pi i}{2} (4h_F-h_s-h_a)}e^{\pi i (h_{s/a}+h_{I/{\rm Adj}}-2h_F)}(e^{\pi i \frac{h_s-h_a}{2}}-e^{-\pi i \frac{h_s-h_a}{2}} )}{1-e^{-\pi i (4h_F-h_s-h_a)}e^{2\pi i (h_{s/a}+h_{I/{\rm Adj}}-2h_F)}}
\end{equation}

\begin{figure}[h]
	\centering
	\includegraphics[scale=.2]{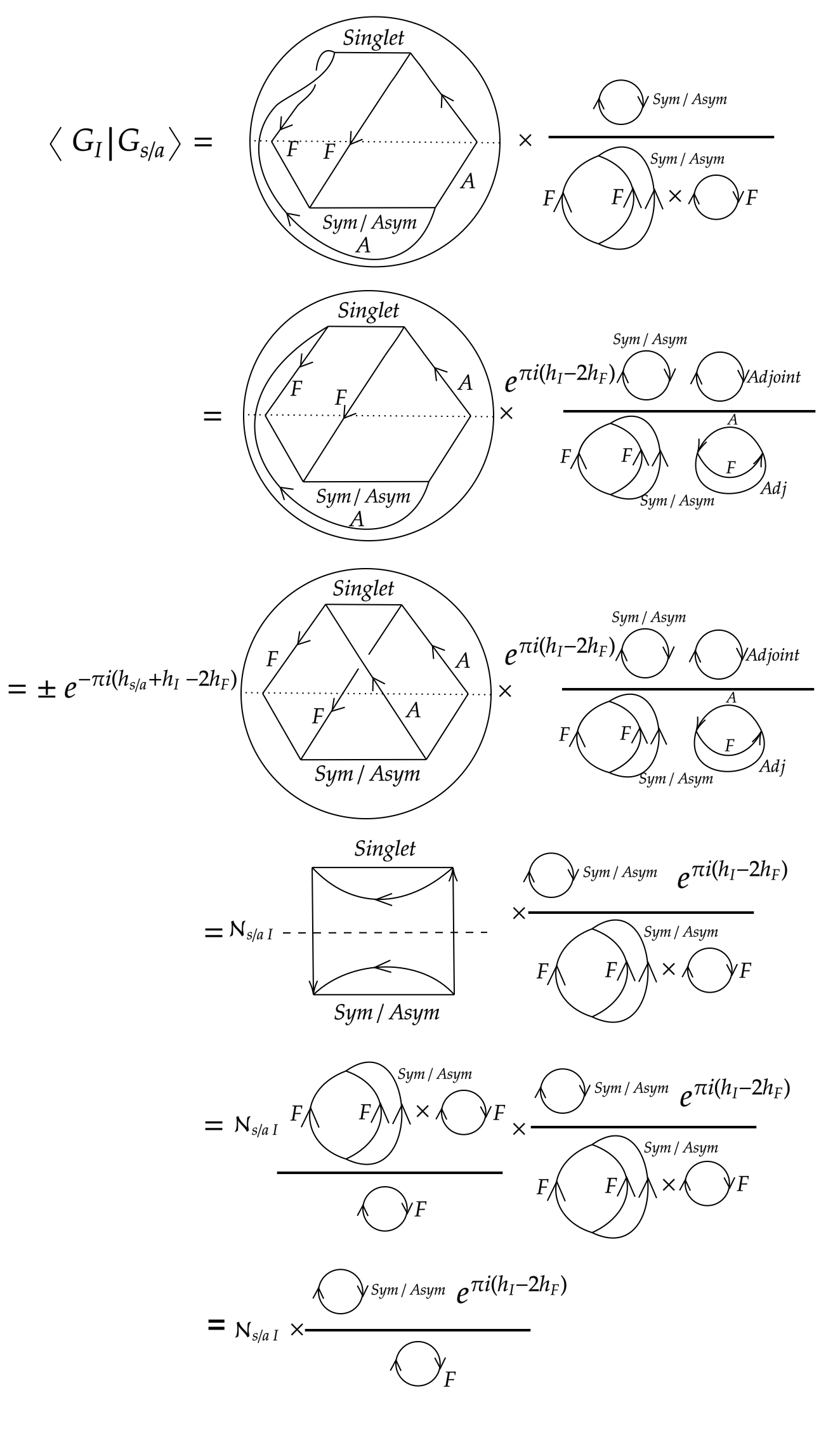}
	\caption{$\langle G_{s/a}|G_I\rangle$}
	\label{sassing}
\end{figure}

\begin{figure}[h]
	\centering
	\includegraphics[scale=.2]{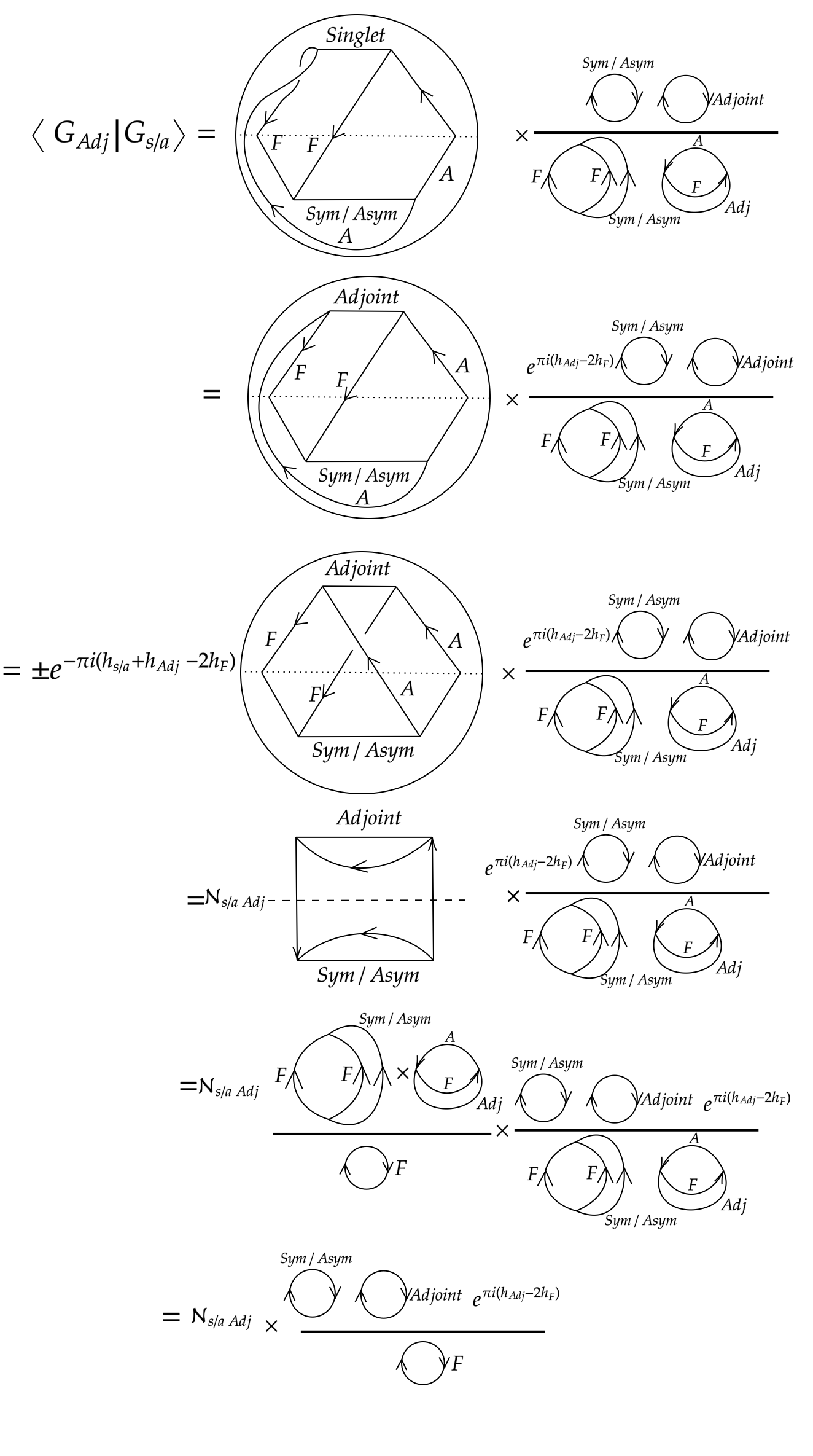}
	\caption{$\langle G_{s/a}|G_{ {\rm Adj}}\rangle$ }
	\label{sasadj}
\end{figure}

In going from the fourth to the fifth line of Fig. \ref{sasadj}, we insert the completeness relation Fig \ref{twopointblockcomp} along the dotted line in the fourth line of Fig. \ref{sasadj}. In going from the fifth to the sixth line of Fig. \ref{sasadj} we have cancelled equal diagrams in the numerator and denominator. Note that all diagrams involving three-point vertices cancel at this stage, demonstrating that our final answer does not depend on normalization we used for the CG coefficients in Figs
\ref{symasym} and Fig. \ref{singadj}. The final answer for the inner products -- presented on the sixth line of Fig. \ref{sasadj}. We see that
\begin{equation}\label{intresalpha}
a_{s/a}^{{\rm Adj}/I }=\frac{ \mathcal{N}_{s/a~ I/{\rm Adj}} ~D^k_{s/a}}{D^k_{F}}
\end{equation}
Plugging in for $\mathcal{N}_{s/a~ I/{\rm Adj}}$, and performing a bit of algebra, we find
\begin{equation}\label{blockstrnfmain}
\begin{split}
&G_s=\frac{G_I e^{-2\pi i h_F}\lfloor N+1 \rfloor_q  + G_{ {\rm Adj}}e^{\pi i(h_{ {\rm Adj}}-2h_F)}}{\lfloor 2 \rfloor_{q} } \\
&G_a=\frac{G_Ie^{-2\pi i h_F}\lfloor N-1 \rfloor_{q} - G_{ {\rm Adj}} e^{\pi i(h_{ {\rm Adj}}-2h_F)}}{\lfloor 2\rfloor_{q}}
\end{split}
\end{equation}
where the $q$ number $[m]_q$ was defined in \eqref{smatrixformintro}.

The result \eqref{blockstrnfmain} applies equally well to the $SU(N)_k$, Type I and Type II $U(N)_k$ theories.

Satisfyingly, \eqref{blockstrnfmain} is in perfect agreement with the expectations of level rank duality. Let us recall that under either Type I, Type II or  $SU(N)$ level rank duality we have
\begin{equation}\label{dualranklev}
N'=|k|, ~~~~k'= -{\rm sgn}(k)N, ~~~~\kappa'= -\kappa,
~~~q'=q^{-1}
\end{equation}
where we have used a prime to denote dual quantities.

It follows immediately from definitions that for any  number $x$
$$\lfloor x \rfloor _q=\lfloor x \rfloor_{q^{-1}}.$$
Moreover it is easy to verify that
\begin{equation}\label{qnumbtrans}
\lfloor |k|+1 \rfloor_{q^{-1}}= \lfloor |k|+1 \rfloor_{q}= \lfloor N-1 \rfloor_{q}
\end{equation}
Finally, it follows from \eqref{dimendiff} that
\begin{equation}\label{linkbetld}
e^{ - 2 \pi i h_F'}= -e^{ - 2 \pi i h_F},
~~~e^{ \pi i \left( h_{ {\rm Adj}}' - 2  h_F' \right)}= e^{ \pi i \left( h_{ {\rm Adj}} - 2  h_F \right) }
\end{equation}
It follows that the level rank dual version of
\eqref{blockstrnfmain}- when rewritten in terms of the variables
of the original frame ($N$, $k$, etc) takes the form
\begin{equation}\label{blockstrnfmainld}
\begin{split}
&G_s'=\frac{-G_I' e^{\pi i \left( h_I -2h_F \right) }\lfloor N-1 \rfloor_q  + G'_{ {\rm Adj}}e^{\pi i(h_{ {\rm Adj}}-2h_F)}}{\lfloor 2 \rfloor_{q} } \\
&G_a'=\frac{-G_I' e^{\pi i \left( h_I -2h_F \right)}\lfloor N+1 \rfloor_{q} - G'_{ {\rm Adj}} e^{\pi i(h_{ {\rm Adj}}-2h_F)}}{\lfloor 2\rfloor_{q}}
\end{split}
\end{equation}
(here $h_I=0$ is the dimension of the identity operator).
Comparing \eqref{blockstrnfmain} and \eqref{blockstrnfmainld}, we see that the  transformation rules that relate $(-G_a', -G_s', G_I', G_{{\rm Adj}}')$ are identical to those
that relate $(G_s, G_a, G_I, G_{{\rm Adj}})$. In other words the `crossing relations' between blocks are the same as those  between level rank dual blocks, accompanied by the additional interchange
$$G_s \leftrightarrow - G_a$$
Of course this interchange is expected on physical grounds as level rank duality interchanges the symmetric and antisymmetric representations.

As we have mentioned before, in Appendix \ref{quantexamp} we have presented an alternate
construction of the blocks $G_{s/a}$ and $G_{{\rm Adj}/I}$
in terms of Wilson line configurations that have no bulk interactions. In Appendix \ref{quantexamp} we use this alternate representation to rederive \eqref{blockstrnfmainld} (without making use of the technology introduced in \cite{Witten:1989wf}).

\subsubsection{Transformation of $S$-matrices under crossing}

Using \eqref{blockstrnfmain}, \eqref{smofprese}, and \eqref{smofprese2}, we can find the transformation of S-matrices in FA channel in terms $S$-matrices in FF channel as in \eqref{claimcross}. We find that
\begin{equation}\label{Smattrq}
	\begin{split}
		&\tau^{(0)}_I=e^{\pi i (h_I-2h_F)}\left(\tau_s \left( \frac{ {\lfloor N+1 \rfloor}_{q}}{{\lfloor 2\rfloor}_{q}} \right) + \tau_a  \left( \frac{ {\lfloor N-1 \rfloor}_{q}}{{\lfloor 2\rfloor}_{q}} \right)\right)  \\
		&\tau^{(0)}_{ {\rm Adj}}=e^{\pi i (h_{ {\rm Adj}}-2h_F)} \left(\frac{\tau_s - \tau_a}{{ \lfloor 2 \rfloor}_{q}}\right)
	\end{split}
\end{equation}
(Note that $h_I=0$; we have presented the phase in the first line of \eqref{Smattrq} in this slightly redundant form to emphasize its structural form.)

We have put the superscript $(0)$ on the $\tau$ matrices in the singlet and adjoint channel obtained via our crossing relations. The reason for this will become clear in the next subsection.

As in the previous subsubsection, the crossing relations are consistent with level rank duality,
accompanied by the additional interchange $S_a \leftrightarrow S_s$, an interchange that was expected anyway on physical grounds.

Following the discussion of subsection \ref{moi}, we define
the quantities
\begin{equation}\label{fluxnu}
\nu_I= 2 h_F-h_I= 2h_F, ~~~~~\nu_{ {\rm Adj}}= 2 h_F-h_{ {\rm Adj}}
\end{equation}
in terms of which the crossing relations \eqref{Smattrq} can be rewritten as
\begin{equation}\label{Smattrqo}
\begin{split}
&\tau_I^{(0)}=e^{-\pi i \nu_I}\left(\tau_s \left( \frac{ {\lfloor N+1 \rfloor}_{q}}{{\lfloor 2\rfloor}_{q}} \right) + \tau_a  \left( \frac{ {\lfloor N-1 \rfloor}_{q}}{{\lfloor 2\rfloor}_{q}} \right)\right)  \\
&\tau_{ {\rm Adj}}^{(0)}=e^{-\pi i \nu_{ {\rm Adj}}} \left(\frac{\tau_s - \tau_a}{{ \lfloor 2 \rfloor}_{q}}\right)
\end{split}
\end{equation}

\subsection{Phases in the crossing relations} \label{pcr}

The crossing relations \eqref{Smattrq} involve both real numbers (like $\lfloor N-1 \rfloor_q$) as well as phases like ($e^{  \pi i \left( h_{I/{\rm Adj}}-2 h_F \right)})$.  Ignoring phases for a moment, our final results for crossing can be stated very simply:
all the explicit real numbers that appear in the classical crossing relations \eqref{crossclass}
are simply replaced by their $q$ number analogues in
\eqref{Smattrq}. This is, of course, an extremely simple and satisfying result.

We will now discuss the meaning of the phases that appear in \eqref{Smattrq}. To begin this discussion, note that there is a sense in which these phases are ambiguous. Recall that we defined the overlap in \eqref{asa} by the diagram
depicted in the first of Fig. \ref{sasadj}, in which
the antifundamental line goes over the fundamental line. As we have noted in the text describing the manipulations depicted in Fig. \ref{sasadj}, we could as well have chosen to have the antifundamental line pass under the fundamental line. It is easy to verify that this choice would have led to all phases that appear in \eqref{blockstrnfmain} and \eqref{Smattrq} to be replaced by their complex conjugates. More generally, a joining involving arbitrary twists of the fundamental and antifundamental line would have led to all phases in \eqref{blockstrnfmain} and \eqref{Smattrq} being raised to the power $(2n+1)$ for an arbitrary integer $n$
(the case discussed just above corresponds to $n=-1$).

The ambiguity in phases noted above has its origin in the fact that conformal blocks are multivalued. The inner product between two blocks defined by two different tangles of Wilson lines is completely well defined if the end points of Wilson lines in the two tangles coincide. If the end points of lines in the two blocks do not coincide, however, one has to
transport the end points of one of the blocks
to the locations of the end points of the other block. The multivaluedness of conformal blocks (equivalently the fact that the Knizhnik-Zamolodchikov connection is flat only away from points where two insertions collide) introduces an ambiguity in this motion.

The discussion in the paragraph above suggests that the phases in the crossing relations \eqref{Smattrq}
have no physical significance, and this is physically reasonable. Let us first recall that, in quantum mechanics,  overall phases of states (as opposed to relative phases within a superposition) have no physical meaning, as they change all amplitudes by an overall phase, and so leave all probabilities unchanged. As an $S$-matrix is an overlap between an in and an out state, it follows also that the overall phase of an $S$-matrix has no physical importance.
\footnote{For instance, a gauge transformation on the scattering wave functions change the phase of the $S$-matrix: this is particularly easy to see in the Schrodinger formalism of the non-relativistic theory.}
This general fact explains why the phases in \eqref{Smattrq} are physically irrelevant,  upto one important subtlety. In the special case of a scattering process in which the
initial and final particles are the same, the full $S$-matrix is a linear sum of $i\tau$ and a term proportional to identity (see \eqref{smatrixformintro}). A rephasing of $\tau$ is physically inconsequential only if accompanied by a
simultaneous rephasing of the term proportional to identity. \footnote{ The physical nature of the relative phase between identity and $\tau$ is  illustrated by the optical theorem which determines the total cross section in terms of the imaginary part of this phase at forward scattering, in a convention in which the term proportional to $I$ term is real.} We conjecture that the ambiguous phases that appear in our crossing relations are unphysical precisely because they multiply the full $S$-matrix (the identity term as well as the $\tau$ term). \footnote{The intuition behind our conjecture goes as follows. In order to evaluate the
overlap between two blocks, we had to move the positions of the insertions in one of these blocks
to a different location: it is precisely this maneuver that was responsible for the ambiguous phases in crossing. At least naively, it would appear that this motion applies equally to both the $I$ and the analytic part of scattering, and so would appear to phase up both terms in the $S$-matrix.} \footnote{This conjecture is supported by the analysis of Appendix \ref{upw}, in which it is demonstrated, in particular, that the 
phase of the delta function is tightly related to the phase of that part of $\tau$ that is singular at forward scattering.} 

In other words,  we conjecture that the $\tau$ matrices obtained from \eqref{Smattrqo} are completed to full $S$-matrices via \footnote{Recall that, according to the discussion of subsection \ref{moi} and Appendix \ref{mstn}, the $S$-matrix in the singlet and adjoint channels include a piece proportional to $\cos\left( \pi \nu_{I/{\rm Adj}} \right)$ times identity.}
\begin{equation}\label{smatrixformo}
\begin{split}
&{\mathcal S}^{(0)}_{I} = e^{-i \pi \nu_I} \cos \left(\pi \nu_{I} \right) \mathcal S_{ {\rm id}} + i \tau_{I}^{(0)} \\
&{\mathcal S}_{ {\rm Adj}}^{(0)} = e^{-i \pi \nu_{ {\rm Adj}}} \cos \left( \pi \nu_{ {\rm Adj}} \right) \mathcal S_{ {\rm id}} + i \tau_{ {\rm Adj}}^{(0)}, \\
\end{split}
\end{equation}

More generally, we conjecture the $\tau$ matrices obtained via the crossing relations \footnote{ Recall that our derivation of the crossing relations \eqref{Smattrqo} made an arbitrary choice (for the branch of $G_{I/{\rm Adj}}$ that we land up on when we continue the locations of insertions of these blocks to match those of $G_{s/a}$). As we have discussed above, these various choices are parameterized by an integer $n$, and different choices yield the crossing relations \eqref{Smattrqt} below.}
\begin{equation}\label{Smattrqt}
\begin{split}
&\tau^{(n)}_I=e^{(2n-1)\pi i \nu_I}\left(\tau_s \left( \frac{ {\lfloor N+1 \rfloor}_{q}}{{\lfloor 2\rfloor}_{q}} \right) + \tau_a  \left( \frac{ {\lfloor N-1 \rfloor}_{q}}{{\lfloor 2\rfloor}_{q}} \right)\right)  \\
&\tau^{(n)}_{ {\rm Adj}}=e^{(2n-1)\pi i \nu_{ {\rm Adj}}} \left(\frac{\tau_s - \tau_a}{{ \lfloor 2 \rfloor}_{q}}\right)
\end{split}
\end{equation}
are completed to full $S$-matrices together with a term proportional to identity that has a similar phase, i.e.,
\begin{equation}\label{smatrixformn}
\begin{split}
&{\mathcal S}^{(n)}_{I} = e^{i (2n-1)\pi \nu_I} \cos \left(\pi \nu_{I} \right) \mathcal{S}_{ {\rm id}} + i \tau_{I}^{(n)} \\
&{\mathcal S}^{(n)}_{ {\rm Adj}} = e^{i (2n-1) \pi \nu_{ {\rm Adj}}} \cos \left( \pi \nu_{ {\rm Adj}} \right)\mathcal S_{ {\rm id}} + i \tau_{ {\rm Adj}}^{(n)}, \\
\end{split}
\end{equation}
Note that \eqref{Smattrqt} and \eqref{smatrixformn} reduce to \eqref{Smattrqo} and \eqref{smatrixformo} at $n=0$.

Clearly, the matrices $\tau_{I/{\rm Adj}}^{(n)}$ and $S_{I/{\rm Adj}}^{(n)}$ are proportional to each other (the proportionality constant is a phase). In other words
\begin{equation}\label{staun} \begin{split}
&\tau_I^{(n)}= e^{i (2n-1)\pi \nu_I} \tau_I, \\
&{\mathcal S}_I^{(n)}= e^{i (2n-1)\pi \nu_I} {\mathcal S}_I, \\
&\tau_{ {\rm Adj}}^{(n)}= e^{i (2n-1)\pi \nu_{ {\rm Adj}}} \tau_{ {\rm Adj}}, \\
&{\mathcal S}_{ {\rm Adj}}^{(n)}= e^{i (2n-1)\pi \nu_{ {\rm Adj}}} {\mathcal S}_{ {\rm Adj}}, \\
\end{split}
\end{equation}
where the quantities $\tau_I$, ${\mathcal S}_I$,
$\tau_{ {\rm Adj}}$ and ${\mathcal S}_{ {\rm Adj}}$ are defined by
\eqref{staun}.

The quantities $\tau_{I}$ and $\tau_{ {\rm Adj}}$ are obtained from $\tau_s$ and $\tau_a$ via the phase free crossing relations
\begin{equation}\label{Smattrqth}
\begin{split}
&\tau_I=\left(\tau_s \left( \frac{ {\lfloor N+1 \rfloor}_{q}}{{\lfloor 2\rfloor}_{q}} \right) + \tau_a  \left( \frac{ {\lfloor N-1 \rfloor}_{q}}{{\lfloor 2\rfloor}_{q}} \right)\right)  \\
&\tau_{ {\rm Adj}}= \left(\frac{\tau_s - \tau_a}{{ \lfloor 2 \rfloor}_{q}}\right)
\end{split}
\end{equation}
and are completed to the full $S$-matrices
${\mathcal S}_I$ and ${\mathcal S}_{ {\rm Adj}}$ via
\begin{equation}\label{smatrixformth}
\begin{split}
&{\mathcal S}_{I} = \cos \left(\pi \nu_{I} \right)I  + i \tau_{I} \\
&{\mathcal S}_{ {\rm Adj}} = \cos \left( \pi \nu_{ {\rm Adj}} \right)I  + i \tau_{ {\rm Adj}}, \\
\end{split}
\end{equation}

The $S$-matrices \eqref{smatrixformo}, \eqref{smatrixformn} and \eqref{smatrixformth} are phase proportional to each other, and so are physically identical. However the crossing relations take the simplest form when expressed in terms of
$\tau_I$ and $\tau_{ {\rm Adj}}$ (which make up the $S$-matrices ${\mathcal S}_{I}$ and ${\mathcal S}_{ {\rm Adj}})$
In the discussion that follows below, we will thus work with these $S$-matrices.

The conclusions of this subsection hinge crucially on the validity of the (in our opinion plausible) conjectures \eqref{smatrixformo} (and the related conjecture \eqref{smatrixformn}).
The simplicity of the final answer -- the economy and beauty of the final crossing relations \eqref{Smattrqth} appear to us to provide a posteriori evidence for the conjectures \eqref{smatrixformo} and \eqref{smatrixformn}. However it is certainly important to better understand these (hopefully valid) conjectures from first principles. We leave this to future work.

\subsubsection{Large $N$ limit of the crossing relations}

In the limit where $N$ and $\kappa$ are large, while $\lambda$ is held fixed, \eqref{Smattrqth} simplifies to
\begin{equation}
	\begin{split}
		&\tau_I=N  \left( \frac{  \sin\pi\lambda} {\pi \lambda} \right)
		\left( \frac{\tau_s+\tau_a}{2} \right) \\
		&\tau_A= \left( \frac{\tau_s-\tau_a}{2} \right)
	\end{split}
\end{equation}
Contrast this with the transformation of classical $S$-matrices as given in \eqref{StSs}. We see that we get an extra factor $\frac{\sin \pi\lambda}{\pi\lambda}$ in the singlet channel in the quantum $S$-matrix as compared with classical one, in precise agreement with the conjectures of \cite{Jain:2014nza}.

\subsubsection{Large $k$ limit of the crossing relations}

In the large $k$ limit, \eqref{Smattrq} reduces to \eqref{StSs} as expected on general grounds.

\section{Discussion}\label{disc}

In this paper we have presented a conjecture for the crossing symmetry rules in Chern-Simons gauge theories coupled to massive matter. While our central conjectures,
listed in subsection \ref{coss},  are motivated by (in our opinion) reasonably compelling physical arguments, it would certainly be useful to clarify and  tighten our reasoning and to make our arguments more rigorous. \footnote{It would also be  useful to find a clearer justification (or refutation) of the conjecture presented in section \ref{pcr}.}

One way to check the structure of $S$-matrices outlined in subsection \ref{coss} would be to
retreat to the non-relativistic limit. Consider, for example,
the case of non-relativistic $2 \rightarrow 2$ scattering. We can, as usual, analyse this problem in centre of mass and relative coordinates. The motion of the centre of mass is
trivial. The relative coordinate obeys a Schrodinger equation for the motion of a particle in the background of a non-abelian point  magnetic field localized at the origin. As the relevant background connection is flat everywhere away from the origin, we can work in a gauge in which this background connection vanishes everywhere
except on a cut (say the positive $x$ axis). In this `irregular' choice of gauge
the wave function has a discontinuity across the cut. The nature of this discontinuity
becomes clearer if we work in a basis that diagonalizes ${\vec J}_A \cdot {\vec J}_B$ (here $A$ and $B$ are the initial particles, and ${\vec J}_A$ and ${\vec J}_B$ denote the
group generators in the representations in which the particles $A$ and $B$ transform).
Moving to this basis abelianizes the scattering problem, and we find that our $S$-matrices
have the discontinuity $e^{2 \pi i ( h_a +h_B- h_M)}$ across the cut \footnote{Here
	$M$ is the representation to which $A$ and $B$ have coupled in the process of diagonalizing ${\vec J}_A$ and ${\vec J}_B$.}.
This is exactly the same discontinuity we encounter in the coefficient functions ${\cal S}_i$ in \eqref{sumtopn}, when ${\mathcal S}_i$ multiplies $G_i$, the block
in which $AB$ fuse to $M$. In other words, wave function at large $r$ (working with the Schrodinger equation in the relevant irregular gauge) have exactly the same
structural properties as the coefficients of the associated conformal blocks with the same cuts.
It would be interesting to analyse this connection further, and completely recast the
scattering problem of Schrodinger quantum mechanics in the language of conformal blocks. We leave this to future work.

The statement of crossing symmetry involves an analytic continuation in momenta. Now the $\tau$ function generically has singularities (at minimum the singularities required by unitarity: e.g. pole type singularities corresponding to the exchange of a stable particle). In order to make the statement of crossing invariance completely precise, we thus also need to specify a path in complex momentum space along which this analytic continuation is performed. A similar question arises in the study of crossing for trivially gapped theories; in that case (atleast for $2\rightarrow 2$ scattering) the answer to this question is well understood \cite{Bros:1965kbd} (see also the more recent papers \cite{DeLacroix:2018arq, Mizera:2021fap}). It would be interesting to generlize this analysis to the case of scattering of Chern Simons theories coupled to massive matter.\footnote{ We thank D. Jain and A. Sen for related discussions. As a wild thought, perhaps, in this case, there are many possible analytic continuations - each of which lead to the distinct but phase related $\tau$ matrices \eqref{staun}.}

In this paper, we have expressed the full $S$-matrix as a sum over component $S$-matrices multiplying
conformal blocks with specified sheet structures. The fact that the full $S$-matrix is single-valued --  while the conformal blocks $G_i$ have cuts --  tells us that the component $S$-matrices ${\mathcal S}_i$ also have cuts: it would be useful to understand the physical origins and consequences of this observation. We suspect that the effective multivaluedness of ${\mathcal S}_i$ is a consequence of the fact that anyonic particles are never really non interacting (taking one anyon around another always produces a monodromy, no matter how far from each other the anyons are). It would be useful to understand this point better. It would also be interesting to understand how particular Feynman diagram computations of $S$-matrices in a given channel decide which sheet to `live' on. It is possible that the answer to this question depends on the gauge employed in the Feynman diagram computation in question.

The starting point the analysis of section \ref{ucmcs} was a formula for the (past to future) transition amplitude of matter Chern-Simons theories in terms of a sum-over-worldlines, which now turn into Wilson lines in pure Chern-Simons theories. It would be useful carefully arrive at this formulation starting from the more standard representation of the matter Chern-Simons path integral as a sum over field configurations. In particular, Wilson lines in pure Chern-Simons theory are labelled by (the topology and initial and final values of) a framing vector field in addition to their trajectories. This framing vector field will have to show up in the sum over Wilson lines; it would be interesting to see how this works in detail. It is possible that an investigation triggered by these considerations could lead the way to a first principles derivation of Bose-Fermi duality in these models.

The analysis of this paper has been entirely structural (as opposed to computational). It would be very interesting to understand how our arguments relate to explicit Feynman diagrammatic calculations of  $S$-matrices. We have already explained that our results are in perfect agreement with Feynman diagram computations in the large $N$ limit
of Chern-Simons theories coupled to fundamental matter \cite{Jain:2014nza, Inbasekar:2015tsa, Gabai:2022snc}; it would be nice to perform a similar comparison at finite $N$ and $k$.
\footnote{Recall that  scattering
at large $N$ lacks some of the complexities of finite $N$ scattering. For instance, scattering is
effectively anyonic in only one of the four channels at large $N$, but is anyonic in every one of these channels at finite $N$.} In this regard, mass-deformed ABJM theory is a clear target of opportunity. Recall that ABJM theory is highly supersymmetric. For this reason the computation of ABJM $S$-matrices -- at least order by order in the coupling constant -- should be a particularly tractable proposition. Indeed there have already been several attempts in this direction (see \cite{Agarwal:2008pu}, \cite{Bargheer:2012cp}, \cite{Bianchi:2011fc}, \cite{Chen:2011vv}, \cite{Bianchi:2011dg}), but the results of these papers seem confusing. In particular, these studies appear to establish both that the one-loop scattering amplitude vanishes in ABJM theory and that the two-loop scattering amplitude has a nontrivial unitarity cut. Put together these results appear to violate unitarity. As mentioned in the discussion section of \cite{Jain:2014nza}, it seems likely to us that the reasons for these confusions lies in the fact that $S$-matrices in matter Chern-Simons theories are structurally different from
those of trivially-gapped theories. It is possible that the confusing aspects of \cite{Agarwal:2008pu}, \cite{Bargheer:2012cp}, \cite{Bianchi:2011fc}, \cite{Chen:2011vv}, \cite{Bianchi:2011dg} (atleast in the mass deformed case) are all a
consequence of these structural novelties. The generalization of the explicit crossing results presented in \eqref{Smattrqintro} to the scattering of bifundamentals in
$U(N)_k \times U(N)_{-k}$ theories is a straightforward exercise. It would be very interesting to compare the results of this exercise to explicit finite $N$ perturbative results in mass deformed ABJM  theory.

Of course massless ABJM theory has a well studied holographic dual. It would be very interesting to find a similar dual description of the mass deformed ABJM theory. If such a description proves possible to find, then it presumably provides dual realization the crossing results predicted by this paper; this would be fascinating to investigate.\footnote{Since the mass deformed theory reduces to pure Chern Simons theory at low energies, part of the bulk dual to this theory must include the bulk dual to pure Chern Simons - perhaps a closed topological string theory - which would be responsible for the structural modifications to crossing that we have observed in this paper. It would be facinating to understand in detail how this works. We thank O. Aharony for this suggestion.}. 

Even though we do not really have a very good reason to expect this, it is possible that the massless ABJM theory happens to enjoy the same crossing properties as its massive counterpart. It may then prove possible to use the existing bulk dual description of massless ABJM theory to verify the crossing relations (and other structural properties) discussed in this paper,  perhaps by generalizing \cite{Alday:2007hr} Alday Maldacena to the ABJM context. 

As mentioned above, it would be very interesting to understand the modified crossing properties
of $S$-matrices in massive matter Chern-Simons theories directly in terms of an analysis of Feynman diagrams. From this point of view, it is possible that the modifications in crossing arise as
a consequence of IR effects -- that replace the IR divergences that we would encounter when studying
more traditional gauge theories that have a massless gluon as  perturbative excitations. Along similar lines it is also possible that an explanation of our modified crossing symmetry rules may also be found within the framework of Celestial Holography. In the current paper we have carefully chosen the width of the spacetime pillbox of Fig. \ref{pillbox} to be much larger than its height so as to absolve ourselves of the responsibility of studying the dynamics of the massless WZW modes on the curved boundary of the cylinder of Fig. \ref{pillbox}. These massless modes may be thought of as being produced by the action of large gauge transformations, and are the direct analogues of the soft modes that play so prominent a role in the Celestial holography programme. Perhaps there is an interesting way of re-obtaining the results of this paper on a spacetime cylinder that is taller than it is wide, but accounting for the dynamics of boundary modes as Wilson lines (representing dynamical particles) enter and leave the boundary. Such an analysis could make direct contact with Celestial holography.

In this paper, we have studied only Chern-Simons theories coupled to massive matter fields. As mentioned in the introduction, however, it seem plausible that a slight generalization of the discussion of this paper applies to all massive $2+1$ dimensional theories interacting with
topological field theories. The Wilson lines of pure Chern-Simons theories are replaced by topological defects (anyons) in a general TFT (see, e.g.\ \cite{Witten:1989wf, PhysRevB.100.115147} and references therein). It seems plausible that the rules of crossing symmetry in topological gapped theories can be formulated in terms of TFT data, in particular the braiding statistics and fusion rules of anyons, and it would be interesting to formulate this more concretely.

Along these lines, it would be interesting to obtain a more structural understanding of
the crossing symmetry matrices presented in this paper, perhaps in terms of 6j symbols -- or Racah W factors --  that appear in the study of the quantum group (see e.g. \cite{RamaDevi:1992np}, Ch 11 of \cite{Chari:1994pz}).

Either using the possible connection with 6j symbols of the quantum group, or otherwise, it would be
interesting to use the formalism developed in this paper to find completely explicit results for the crossing matrices for more general gauge groups and representations than the one example studied
explicitly in this paper. Recall that our final answer for crossing with two fundamental and two antifundamental insertions, \eqref{Smattrqintro},  took a remarkably simple form. The finite $k$ crossing matrices were simply $q$ deformations of their large $k$ counterparts. It is possible that this structural connection holds more generally (i.e., in the case of other gauge groups and other representations).
We think that it would be interesting to explore this exciting suggestion.

As we have mentioned at the beginning of this paper, S matrices are among the best studied observables in 
quantum field theory (and quantum gravity). In particular, there has recently been a revival of the old programme to discover useful constraints on the S matrix using only general universal features: analyticity, unitarity and crossing. It would be interesting to insert the modified crossing rules of
this paper into the general programme described above and to analyse the consequences. Perhaps it will turn out that large $N$ matter Chern Simons theories saturate 
 the relevant bounds (see \cite{Chowdhury:2017vel} for a hint that this might be the case).

It is possible that the modified structural properties of $S$-matrices, discussed in this paper,  might even turn out to have measurable consequences of relevance to
condensed matter physics. Recall that transport phenomena in relatively weakly coupled theories are often well described by the Boltzmann transport equation.
This equation uses the $S$-matrix as a basic input. It seems likely to us that the new structural
properties of the $S$-matrices studied in this paper (upon generalization to $S$-matrices in the presence of a finite temperature and chemical potential bath) will have qualitatively important consequences for the dynamics of the Boltzmann transport equation. At the intuitive level it feels that this should be the case. Recall that the Boltzmann transport equation for bosons has different structural properties from the analogous equation for fermions. As the particles in matter Chern-Simons theories are effectively anyonic (with statistics that, in some sense, interpolate between bosonic and fermionic), it seems natural
to expect the Boltzmann transport equations in these theories also to interpolate between the bosonic and fermionic transport equations. Perhaps, in particular, it is the universal structural properties of these equations that ensure that the equilibrium solutions of these transport equations take the
form described in \cite{Minwalla:2022sef}. It would be interesting to investigate this further.

\section*{Acknowledgments}

We would like to thank A. Gadde, D. Gaiotto, I. Haldar, D. Jain,  S. Jain, P. Mitra, O. Parrikar, N. Prabhakar, P. Ramadevi, N. Seiberg, A. Sen,  S. Wadia, and E. Witten for very useful discussions. We would also like to thank O. Aharony, S. Giombi, S. Jain, Z. Komargodski, S. Mizera, N. Prabhakar, O. Parrikar, P. Ramadevi, T. Sharma, D. Tong and S. Wadia for comments on the manuscript.  The work of all authors was supported by the Infosys Endowment for the study of the Quantum Structure of Spacetime. The work of S.M. and C.P. is supported by the J C Bose Fellowship JCB/2019/000052. The work of S.P. was partially supported by a DST grant MTR/2018/0010077. The work of U.M. is supported in part by the U.S. DOE grant No. DE-FG02-13ER41958 and a Simons Investigator grant (PI: Dam Thanh Son). We would all also like to acknowledge our debt to the people of India for their steady support to the study of the basic sciences.

\appendix

\section{Modification of the structure  $\mathcal S_{ {\rm id}}+i\tau$ in topologically-nontrivial theories} \label{mstn}

\subsection{$I$ in trivially-gapped theories}

As we have reviewed above, $S$-matrices in trivially-gapped theories (with mass gap $m$) take the form

\begin{equation}\label{smatform}
\mathcal S_{ {\rm id}}+i\tau,
\end{equation}

where $\mathcal S_{ {\rm id}}$ is the `identity' $S$-matrix and $\tau$ is an analytic function of scattering momenta. In this section, we first pause to remind the reader about the origin of the term $\mathcal S_{ {\rm id}}$ in \eqref{smatform}. We focus on the case of  $2 \times 2$ scattering for simplicity.

In the centre of mass frame, the initial scattering state is a plane wave at some fixed relative momentum. The probability flux in this plane wave flows orthogonal to a $D-2$ dimensional spatial plane parameterized by the impact parameter ${\vec b}$. As the theory is trivially-gapped, the part of the incident wave with $|{\vec b}|\gg \frac{1}{m}$ does not scatter and  continues undeviated. This part of the plane wave carries all but a finite part of the infinite probability flux of the incoming wave and gives rise to the $\mathcal S_{ {\rm id}}$ in \eqref{smatform}. The term $\tau$ in \eqref{smatform} captures the evolution  of that finite part of the plane wave with $|{\vec b} m |$ of order unity.
\footnote{As this part of the wave carries only a finite amount of probability flux, it does not lead to a diminishing of the forward flux, and so does not backreact on $\mathcal S_{ {\rm id}}$.}

The fact that $\tau$ is an analytic function of scattering momenta may be understood as follows.  Recall that the $S$-matrix may be written in a partial wave expansion
\begin{equation}\label{schemformsm}
S= \sum_{l} a_l P_l({\hat p}. {\hat  p}')
\end{equation}
where $P_l$ is the unique spherical harmonic of $SO(D-1)$ that preserves the
$SO(D-2)$ invariance (rotating ${\hat p}'$ around the fixed vector ${\hat p}$), ${\vec p}$ is the initial relative momentum and ${\vec p}'$ the final relative momentum in centre of mass frame, and $a_l$ are coefficients.
Let $a_l^0$ be the coefficients corresponding to no scattering (i.e.,  to the expansion of the plane wave in the direction ${\hat p}$ in spherical coordinates) so that
\begin{equation}\label{Iipw}
\mathcal S_{ {\rm id}}= \sum_{l} a_l^0  P_l({\hat p}\cdot {\hat  p}').\\
\end{equation}
Let the coefficients that characterize the actual $2\rightarrow 2$ scattering process be given by
\begin{equation} \label{formofcoeff}
a_l=a_l^o e^{i \delta_l}.
~~~{\rm so~ that}~~~
S= \sum_{l} a_l^0 e^{i \delta_l}  P_l({\hat p}\cdot {\hat  p}')
\end{equation}
It follows that
\begin{equation}\label{sfno}
\begin{split}
S&= \mathcal S_{ {\rm id}}+i\tau\\
\tau&= i \sum_{l} a_l^0 \left(1- e^{i \delta_l}\right)  P_l({\hat p}. {\hat  p}')\\
\end{split}
\end{equation}
Because all interactions are short range, $\delta_l$ decays rapidly to zero as $l \to \infty$.
\footnote{ At momentum scale $|p|$ -- which is determined by the energy scale of the scattering process --  scattering in the $l^{th}$ partial wave occurs dominantly at impact parameters $|{\vec b}| \sim \frac{l}{|p|}$. Consequently, scattering `switches off' for $l \gg \frac{p}{m}$, and so $\delta_l \to 0$ for $l \gg \frac{p}{m}$.} It follows that the summation parameter in the third line of \eqref{sfno} effectively receives contributions only from a finite number of $l$
\footnote{More precisely, the contribution from large enough $l$ decays sufficiently fast to ensure the analyticity of $\tau$}. As each $P_l$ is a polynomial,  $\tau$ is an analytic function of scattering momenta.

It follows from unitarity that the total incoming and  outgoing probability fluxes must equal each other in any scattering experiment. When the $S$-matrix is described by \eqref{smatform} this works as follows. The infinite probability flux of the incoming plane wave is perfectly balanced by the infinite flux described by $\mathcal S_{ {\rm id}}$ on the RHS of \eqref{smatform}. Consequently, in this situation $\tau$, carries finite flux \footnote{Moreover the finite flux from the interference of $\mathcal S_{ {\rm id}}$ and $i\tau$ (this piece is $\propto {\rm Im}(\tau)$) must balance the finite flux purely from $\tau$ (this term is $\propto |\tau|^2$). This requirement is the optical theorem.}.

\subsection{Modification in topologically-gapped theories} \label{ngt}

The discussion presented in the previous subsubsection is modified in 2+1 dimensional topologically-gapped QFTs \footnote{The discussion of this subsection is an elaboration of the material in section 2.5 of \cite{Jain:2014nza},  and has been included in this paper mainly for completeness. }. In $2+1$ dimensions the `plane' orthogonal to the incoming probability flux (parameterized by ${\vec b}$) is actually a line. The part of this line with $|{\vec b}| m \gg 1$
consists of two disconnected asymptotic regions separated by the scattering region centered around ${\vec b}=0$. In each of these asymptotic regions the incoming plane wave propagates forward (unaffected by interactions) except for one detail; upon crossing the scattering region, the wave function in these two disconnected asymptotic regions pick up a relative phase $e^{2 \pi i \nu}$. This is simply the Aharonov-Bohm effect: $e^{2 \pi i \nu}$  is the Aharonov-Bohm phase that the incident particles pick up upon going around one another \footnote{Note that the absolute phase -- unlike the relative phase -- can be
	changed by a choice of gauge and so is not physical.}. This Aharonov-Bohm phase has the following rather dramatic consequence. In a choice of gauge that distributes this phase symmetrically between the two unscattered parts of the plane wave, the contribution of the unscattered part of the plane wave to the $S$-matrix changes from $\mathcal S_{ {\rm id}}$ to  $ \frac{ e^{\pi i \nu} + e^{ -\pi i \nu}}{2}~\mathcal S_{ {\rm id}}= \cos (\pi \nu) ~\mathcal S_{ {\rm id}}$. It follows that the $S$-matrix now takes the form
\begin{equation}\label{smatrixform}
S= \cos \left( \pi \nu\right) \mathcal S_{ {\rm id}} + i\tau,
\end{equation}
instead of \eqref{smatform}, where, once again, $\tau$ is an analytic function of scattering momenta.

This phase $e^{2 \pi i \nu}$ is an `operator' rather than a number: it depends on the
`scattering channel'. We explained how this works in section \ref{ucmcs}. Briefly, recall that the low energy TFT sees all  scattering particles as Wilson lines, and every Wilson line has a partner chiral operator in the holomorphic CFT associated with the TFT.
Let the operators associated with the incident scattering particles $A$ and $B$ be $O_A$ and $O_B$. Let us suppose that under fusion
\begin{equation}\label{oaob}
O_A O_B = \sum_M  N_{ABM} O_M
\end{equation}
Then the scattering process $AB \rightarrow AB$ proceeds in channels labeled by the distinct exchange operators $O_M$
\footnote{ The number of channels associated with the exchange of operator $O_M$ operator is $N_{ABM}$: see  section \ref{ucmcs} for details. Recall the operator $M$ occurs
	$N_{ABM}$ times in the fusion rule.}.  If the chiral dimensions of the operators $O_A$, $O_B$ and $O_M$ are given by $h_A$, $h_B$ and $h_M$ respectively,  the Aharonov-Bohm phase difference in the scattering channel associated with $O_M$, $e^{i 2 \pi \nu_M}$, is given by the monodromy made by $O_A$ as it circles $O_B$ in any conformal block in which $AB$ fuses to $O_M$. It follows that
\begin{equation}\label{monophase}
\nu_M=h_A+h_B -h_M
\end{equation}

\subsection{Unitarity and Partial Waves} \label{upw}

Let us now investigate how incoming and outgoing probability fluxes balance each other in \eqref{smatrixform}. The infinite outgoing flux carried by the term $\cos\pi \nu~\mathcal S_{ {\rm id}}$  in \eqref{smatrixform} is smaller than the infinite incoming flux by a factor of $\cos^2\pi \nu$. The difference between these must be made up by the flux carried by $i\tau$, which consequently must blow up somewhere. Infact our intuitive understanding of the scattering process allows us to understand how this must work in a quantitative manner.

Let us work with the S matrix in the gauge and phase convention in which it takes the form \eqref{smatrixform} \footnote{Working in a different gauge or phase convention can result in the S matrix \eqref{smatrixform} picking up an overall phase.}. As we have explained above \eqref{smatrixform}, the factor of $\cos \pi \nu$ is a consequence of the fact that the part of the plane wave that passes  far `above' the collision picks up a phase 
$e^{ i \pi \nu}$, while the part of the plane wave that passes 
far below the collision picks up the phase $e^{- i \pi \nu}$.
Let us now see what this physical expectation implies for the partial wave expansion. At values of $n \gg \frac{k}{m}$, partial waves $e^{i n \theta}$ dominantly pass `far above' the collision (here $k$ is the modulus of the momentum, and $m$ is the mass). Consequently we expect that the only effect of the collision on such partial waves is that they pick up the phase 
$e^{i \pi \nu}$. Similarly, partial waves with  $n \ll  -\frac{k}{m}$ pass far below the collision, and so should only pick up the phase $e^{- i \pi \nu}$. It follows that the 
incident plane wave 
\begin{equation}\label{incipw}
e^{i {\vec k}.{\vec r}}= \sum_{n=-\infty}^\infty i^n J_n(kr) e^{i n \theta}
\end{equation} 
transits to the final scattered wave function 
\begin{equation}\label{incipwln} \begin{split} 
& \psi(r, \theta)= \sum_{n=-\infty}^\infty i^n J_n(kr) e^{i n \theta} e^{i \phi(n)} \\
&\phi(n)=  \pi \nu ~~~~~~~~{n \gg \frac{k}{m}}\\
&\phi(n)= - \pi \nu ~~~~~~~~{n \ll -\frac{k}{m}}
\end{split} 
\end{equation} 
Using the asymptotic expansion of Bessel functions at large values of their argument, 
\begin{equation}
	J_n(kr)\approx \frac{1}{\sqrt{2\pi k r}}\left(e^{ikr-i\pi/4-in\pi/2}+e^{-ikr+i\pi/4+in\pi/2}\right),
\end{equation}
we find that at large values of $r$
\begin{equation} \label{wavefnlarger} 
\psi(r,\theta)\approx \sum_n \frac{1}{\sqrt{2\pi k r}}\left(e^{ikr-i\pi/4+in\theta+ i\phi(n)}+e^{-ikr+i\pi/4+in\pi+in\theta+i\phi(n)}\right)=\psi_+(r,\theta)+\psi_-(r,\theta)
\end{equation} 
In the extreme RHS of \eqref{wavefnlarger}, we have decomposed the wave function 
$\psi(r, \theta)$ into a `radially outgoing' part $\psi_+(r,\theta)$ (proportional to $e^{ikr}$) and a radially ingoing part $\psi_-(r,\theta)$ (proportional to 
$e^{-i k r}$). Since wavepackets are expanded in the $\psi(r,\theta)$  basis only  at late times, it is $\psi_+(r,\theta)$ that determines the S matrix.
If we work with the $S$ matrix $h(\theta)$,  normalized so that identity equals 
$2 \pi \delta(\theta)$ and the unitarity equation takes the form 
\begin{equation}\label{smatrnorm} 
\int d\alpha  h^*(\theta- \alpha ) h(\alpha)  = 2 \pi \delta(\theta)	
\end{equation} 
then \footnote{The invariant $T$ matrix is obtained by multiplying the non identity part of $h(\theta)$ by $4 \pi \sqrt{s}$. See section 2.7 and Appendix C of \cite{Jain:2014nza}.}     the relationship between $\psi_+(r,\theta)$ and $h(\theta)$ is 
(see e.g. C.13 of \cite{Jain:2014nza})
\begin{equation}\label{phrel} 
\psi_+(r,\theta) \approx \frac{e^{ikr-i\pi/4} h(\theta) }{\sqrt{2\pi k r}}
\end{equation} 

Usng \eqref{phrel} and \eqref{wavefnlarger}, it follows that in the current situation 
\begin{equation}\label{psip}
	\begin{split}
h(\theta) &= \sum_{n>0}e^{i n\theta+ i\pi \nu}+ \sum_{n<0}e^{i n\theta-i\pi\nu} + \cos \pi\nu \\
&+\sum_{n>0} e^{in\theta}\left(e^{i\phi(n)}-e^{i\pi\nu}\right) +\sum_{n<0} e^{in\theta}\left(e^{i\phi(n)}-e^{-i\pi\nu}\right)-\cos(\pi\nu)    
 \end{split}
\end{equation}
In \eqref{psip} we have, for convenience, added and subtracted $\cos \pi \nu$ in the first and second 
lines. \footnote{The term $\cos \pi \nu$ has no $\theta$ dependence, and so only contributes to the partial wave at $n=0$. }

The summations in the first line of \eqref{psip} are easily evaluated, using the identities \footnote{In the vicinity of its singularity at $\theta=0$, the function $\cot \frac{\theta}{2}$ that appears in these identities should be taken to be 
defined by the principal value prescription. See e.g. section 2.7 of \cite{Jain:2014nza}.} 	  
\begin{equation}
	\begin{split}
		\delta(\theta)=\sum_{n=-\infty}^\infty \frac{1}{2\pi}e^{in\theta}\\
		\cot \frac{\theta}{2}=\frac{1}{i}\left(\sum_{n>0} e^{i n \theta} - \sum_{n<0} e^{in\theta}\right),
	\end{split}
\end{equation}
We find 
\begin{equation}\label{psipn}
	\begin{split}
		h(\theta) &= 2\pi \delta(\theta)\cos(\pi\nu)-\sin(\pi\nu)\cot(\theta/2)  \\
		&+\sum_{n>0} e^{in\theta}\left(e^{i\phi(n)}-e^{i\pi\nu}\right) +\sum_{n<0} e^{in\theta}\left(e^{i\phi(n)}-e^{-i\pi\nu}\right)-\cos(\pi\nu)    
	\end{split}
\end{equation}

Using \eqref{incipwln}, we see that second line of \eqref{psip} effectively does not receive  contributions from partial waves at large $|n|$. It follows that the second line of \eqref{psip} is an analytic functions of $\theta$. Let us call this function 
$f(\theta)$, it follows that 
\begin{equation}\label{psipnn}
	\begin{split}
		h(\theta) &=  2\pi \delta(\theta)\cos(\pi\nu)-\sin(\pi\nu)\cot(\theta/2) 
		+ f(\theta)
		 \\   
	\end{split}
\end{equation}

Let us summarize. The physical picture of scattering that forced the modification of the 
coefficient of the forward scattering $\delta$ function also guarantees that the S matrix takes the form \eqref{psipnn}. The form \eqref{psipnn} guarantees that the 
outgoing scattering wave carries the same infinite flux as the incoming scattering 
wave (this follows from the fact that the partial wave representation of \eqref{psipnn} is simply a phase at large $n$, see also the analysis of section 2.7.3 of \cite{Jain:2014nza}). An important conclusion of this subsection is that the modified identity piece $2\pi \delta(\theta)\cos(\pi\nu)$ is necessarily accompanied by the 
term $-\sin(\pi\nu)\cot(\theta/2)$ which has a singularity 
near $\theta=0$. Like the coefficient of the identity piece, the coefficient of the singular piece is also exactly determined
by the effective anyonic phase $\nu$.

The S matrix of the non relativisitic Aharonov-Bohm problem precisely takes the form \eqref{psipnn} (see e.g. subsection 2.7.3 of \cite{Jain:2014nza});  this fact can be regarded 
as a consistency check of the work out of this subsection. 

As foreseen earlier in this Appendix, it is the fact that the $\tau$ matrix includes a term blows up as $\theta \to 0$ that allows the total outgoing flux to equal the total incoming flux despite a reduction in the coefficient of identity.  \footnote{It thus follows that the S matrices \eqref{smatrixform} and \eqref{smatform} are not as different from each other as they first appear to be. The flux in an angular cone of opening angle $\epsilon$ around forward scattering  - upto finite corrections - is the same in each of these S matrices. While this flux is precisely localized around $\theta=0$ in \eqref{smatform}, it is a little more smeared around $\theta=0$ in \eqref{smatrixform}. However a detector placed in the forward direction would detect the same flux (upto finite corrections) in both these situations, provided the detector has a finite angular resolution, no matter how small.}

As the difference between the infinite probability flux of the incoming plane wave and the outgoing probability flux from $\cos(\pi \nu) \mathcal S_{ {\rm id}}$ is balanced by a divergence in the flux near the forward limit,  there is a sense in which the difference  between $\mathcal S_{ {\rm id}}$ in \eqref{smatform} and $\cos \pi \nu$ in \eqref{smatrixform} has simply been smeared out around $\theta=0$. The situation may seem vaguely reminiscent of scattering in  QED in the forward limit, but the similarities are very limited. In particular, despite its unusual infrared properties, the $S$-matrix \eqref{smatrixform} is point wise well defined (independent of any IR cut off), so one obtains finite results without retreating to inclusive cross sections.

\section{Identical particles} \label{ipfullap}

\subsection{Identical particles at infinite $k$} \label{ipg}

In the analysis presented so far in this section, we have assumed that all our scattering particles
are distinct, i.e., that all operator insertions are of distinct fields. Even if two particles
happen to transform in the same group representation, they are distinguished by some other
quantum number (e.g., a flavour quantum number).

In situations of interest, however, it is often the case that two or more of field insertions
(that characterize the scattering process in question) are identical. In such a situation
some (or possibly even all) of the scattering processes described by the correlator in question
involve identical particles. The identical particles involved may be either bosons or fermions, and the requirement that all scattering processes respect Bose or Fermi statistics forces a slight modification of the structures outlined thus far in this section. In this subsection we  outline
the nature of this modification for the specially simple case of $2 \rightarrow 2$ scattering. The generalization to multiparticle scattering is technically more complicated,  but the method to be followed will, hopefully, be clear.

Let us consider a scattering in which the initial particles both transform in the representation $R$,
while the final particles transform in the representation $R'$.

Because the two initial and two final representations are identical, the Clebsch-Gordan decomposition
\eqref{classdecomp} -- for both initial and final particles --  can be graded by symmetry under interchange. Let us suppose that
\begin{equation}\label{classdecompident} \begin{split}
		&{\rm Sym} \left( R \otimes R \right) = \sum_{a} Q^e_a \tilde R_a \\
		&\Lambda\left( R \otimes R \right) = \sum_{a} Q^o_a \tilde R_a \\
		&{\rm Sym} \left( R' \otimes R' \right) = \sum_{a} Q^{'e}_a \tilde R_a \\
		&\Lambda\left( R' \otimes R' \right) = \sum_{a} Q^{'o}_a \tilde R_a \\
	\end{split}
\end{equation}
where the symbols ${\rm Sym}$ and $\Lambda$ signify the symmetric and antisymmetric products, respectively, of the two representations. The superscripts $e$ and $o$ stand for even and odd respectively.

Correspondingly, we can also grade the projectors
$P_a^{rr'}$ according to the symmetry under interchange of initial and final particles. We thus now have four distinct types of projectors, namely
\begin{equation}\label{symsplitproj} \begin{split}
		&P_{ee, a}^{rr'}~~~r=1 \ldots Q^e_a, ~~~r'=1\ldots Q^{'e}_a \\
		&P_{eo, a}^{rr'}~~~r=1 \ldots Q^e_a, ~~~r'=1\ldots Q^{'o}_a  \\
		&P_{oe, a}^{rr'}~~~r=1 \ldots Q^o_a, ~~~r'=1\ldots Q^{'e}_a  \\
		&P_{oo, a}^{rr'}~~~r=1 \ldots Q^o_a, ~~~r'=1\ldots Q^{'o}_a \\
	\end{split}
\end{equation}
The expansion \eqref{expofsmatrixmt} is now replaced by the more complicated looking equation
\begin{equation}\label{expofsmatrixmtid}
	S = \sum_{a, r_1, r_2} \left( \mathcal{S}^{r_1, r_2}_{ee, a} P_{ee, a}^{r_1 r_2} +  \mathcal{S}^{r_1, r_2}_{eo, a} P_{eo, a}^{r_1 r_2} + \mathcal{S}^{r_1, r_2}_{oe, a} P_{oe, a}^{r_1 r_2} + \mathcal{S}^{r_1, r_2}_{oo, a} P_{oo, a}^{r_1 r_2} \right)
\end{equation}

Upto this point we have only  assumed that the two initial representations and two final representations (and not necessarily the initial, and/or the final particles themselves ) are identical.
Let us now specialize to the case that the initial particles are actually identical bosons/fermions. The only effect of this on the expansion of the $S$-matrix is that the functions $\mathcal{S}^{r_1, r_2}_{ee, a}$ and
$\mathcal{S}^{r_1, r_2}_{eo, a}$ are now required, by Bose/Fermi statistics, to be symmetric/antisymmetric under the interchange of spacetime quantum numbers (momenta, polarizations, etc) of the two initial particles. Similarly the functions  $\mathcal{S}^{r_1, r_2}_{oe, a}$ and $\mathcal{S}^{r_1, r_2}_{oo, a}$ are required to be antisymmetric/symmetric under interchange of the spacetime quantum numbers of  initial particles. Completely analogous conditions apply if the two final particles are identical.

When the initial particles are identical bosons/fermions, the identity operator that appears on the RHS of the unitarity equation changes to account for statistics. If we denote the momenta and global symmetry indices of the initial particles by
$(p_1, i)$, $(p_2, j)$, then, schematically
\begin{equation}\label{iinp}
	\langle (p_1', i'), (p'_2, j')| I | (p_1, i), (p_2, j)
	\rangle =
	\delta_{i i'} \delta_{j j'} \delta(p_1-p_1') \delta(p_2-p_2') \pm \delta_{i j'} \delta_{j i'} \delta(p_1-p_2') \delta(p_2-p_1')
\end{equation}
(the sign $\pm$ depends on whether we are dealing with bosons or fermions). In other words the RHS of the unitarity equation is replaced by the RHS of \eqref{iinp} when the initial particles are identical
bosons/fermions.

Let us define
\begin{equation}\label{idob} \begin{split}
		&M^e_{ {\rm id}}=  \frac{ \delta_{i i'} \delta_{j j'}+\delta_{i j'} \delta_{j i'} }{2}\\
		&M^o_{ {\rm id}}= \frac{\delta_{i i'} \delta_{j j'}-\delta_{i j'} \delta_{j i'} }{2} \\
		&S^e_{ {\rm id}}= \delta(p_1-p_1') \delta(p_2- p_2')+\delta(p_1-p_2') \delta(p_2-p_1') \\
		&S^o_{ {\rm id}}= \delta(p_1-p_1') \delta(p_2- p_2')-\delta(p_1-p_2') \delta(p_2-p_1') \\
	\end{split}
\end{equation}
(the last two equations are schematic). In terms of these quantities the RHS of the unitarity equation equals
\begin{equation}\label{rhsunitbos}
	{M^e_{ {\rm id}}S^e_{ {\rm id}}+M^o_{ {\rm id}}S^o_{ {\rm id}}}
\end{equation}
for initial identical bosons,
but equals
\begin{equation}\label{rhsunitbos}
	{M^e_{ {\rm id}}S^o_{ {\rm id}}+M^o_{ {\rm id}}S^e_{ {\rm id}}}
\end{equation}
for initial identical fermions.

The equation \eqref{projcomplete} can be decomposed into its odd and even pieces
\begin{equation}\label{projcompleteid} \begin{split}
		&\sum_{a, r} {\hat P}_{ee, a}^{rr} = {M^e_{ {\rm id}}}\\
		&\sum_{a, r} {\hat P}_{oo, a}^{rr} = {M^o_{ {\rm id}}} \\
	\end{split}
\end{equation}
(adding the two equations in \eqref{projcompleteid} gives
\eqref{projcomplete}).
Consequently, the unitarity equation \eqref{smatrixsuni}
becomes
\begin{equation}\label{smatrixsuniidbos}\begin{split}
		&\sum_{\rm final ~states} \sum_{r'} ( \mathcal S_{ee,a}^{r_1r'})^* \star  \mathcal S_{ee,a}^{r_2r'} + (\mathcal S_{eo,a}^{r_1r'})^* \star  \mathcal S_{eo,a}^{r_2r'}
		= \mathcal S^e_{ {\rm id}} \delta_{r_1, r_2} \\
		&\sum_{\rm final ~states} \sum_{r'} ( \mathcal S_{oe,a}^{r_1r'})^* \star  \mathcal S_{oe,a}^{r_2r'} +{(\mathcal S}_{oo,a}^{r_1r'})^* \star  \mathcal S_{oo,a}^{r_2r'}
		= {\mathcal S}^o_{ {\rm id}} \delta_{r_1, r_2} \\
		&\sum_{\rm final ~states} \sum_{r'} ( \mathcal S_{ee,a}^{r_1r'})^* \star  \mathcal S_{oe,a}^{r_2r'} + (\mathcal S_{eo,a}^{r_1r'})^* \star  \mathcal S_{oo,a}^{r_2r'}
		= 0 \\
		&\sum_{\rm final ~states} \sum_{r'} ( \mathcal S_{oe,a}^{r_1r'})^* \star  \mathcal S_{ee,a}^{r_2r'} +{(\mathcal S}_{oo,a}^{r_1r'})^* \star  \mathcal S_{eo,a}^{r_2r'}
		= 0 \\
	\end{split}
\end{equation}
when the initial particles are identical bosons, while it takes the form
\begin{equation}\label{smatrixsuniidbos}\begin{split}
		&\sum_{\rm final ~states} \sum_{r'} ( \mathcal S_{ee,a}^{r_1r'})^* \star  \mathcal S_{ee,a}^{r_2r'} + ({\mathcal S}_{eo,a}^{r_1r'})^* \star  \mathcal S_{eo,a}^{r_2r'}
		= \mathcal S^o_{ {\rm id}} \delta_{r_1, r_2} \\
		&\sum_{\rm final ~states} \sum_{r'} ( \mathcal S_{oe,a}^{r_1r'})^* \star  \mathcal S_{oe,a}^{r_2r'} + ({\mathcal S}_{oo,a}^{r_1r'})^* \star  \mathcal S_{oo,a}^{r_2r'}
		= \mathcal S^e_{ {\rm id}} \delta_{r_1, r_2} \\
		&\sum_{\rm final ~states} \sum_{r'} ( \mathcal S_{ee,a}^{r_1r'})^* \star  \mathcal S_{oe,a}^{r_2r'} + (\mathcal S_{eo,a}^{r_1r'})^* \star  \mathcal S_{oo,a}^{r_2r'}
		= 0 \\
		&\sum_{\rm final ~states} \sum_{r'} ( \mathcal S_{oe,a}^{r_1r'})^* \star  \mathcal S_{ee,a}^{r_2r'} +{(\mathcal S}_{oo,a}^{r_1r'})^* \star  \mathcal S_{eo,a}^{r_2r'}
		= 0 \\
	\end{split}
\end{equation}
when they are identical fermions.

Finally, when the two initial particles are also identical to the two final particles (so that, in particular, $R=R'$), the precise form of the $S$-matrix (including the identity piece, see Appendix \ref{mstn}) is given by the following replacement of
\eqref{compsmat}
\begin{equation}\label{compsmatid}
	\begin{split}
		&{\mathcal S}^{rr'}_{ee,a}= \delta_{rr'} \mathcal S^e_{ {\rm id}}
		+ i \tau^{rr'}_{ee, a}\\
		&{\mathcal S}^{rr'}_{eo,a}= i \tau^{rr'}_{eo, a}\\
		&{\mathcal S}^{rr'}_{oe,a}=
		 i \tau^{rr'}_{oe, a}\\
		&{\mathcal S}^{rr'}_{oo,a}= \delta_{rr'} \mathcal S^o_{ {\rm id}}
		+ i \tau^{rr'}_{oo, a}\\
	\end{split}
\end{equation}

In spite of all the modifications described in this subsection, the crossing relations for $\tau$ matrices -- which are determined completely by group theoretic considerations -- are unaffected by the identical nature of the scattering particles. They take exactly the same form in the case that the particles are identical, as in the case when the (say) two initial particles transform in the same
representation $R$, but are distinguishable for some other reason.

\subsection{Identical particles at finite $k$} \label{ipfk}

Once again we focus on the case of $2 \rightarrow 2$ scattering, with the initial particles chosen to transform in representation $R$ and the final particles chosen to transform in representation $R'$. Our discussion closely follows that of subsection \ref{ipg}.

The WZW$_k$ fusion rules for $R \times R$ (here $R$ represents the primary operator in representation $R$)
are captured by the diagram Fig. \ref{threepointblock},
with the representations $R_a$ and $R_b$ replaced by $R$
and the representation $R_c$ replaced by ${\tilde R}_a$.
The space of fusions is described by the equivalence class
of CG coefficients (see the discussion under Fig. \ref{threepointblock}). On the LHS of Fig. \ref{threeptorthogtw} we defined an inner product in the space of equivalence classes of CG coefficients, and chose our basis of blocks to be orthonormal w.r.t. this inner product. Now it is easy to convince oneself that the inner product defined in Fig. \ref{threepointblock} gives $\langle C'| C\rangle=0$ if $C$ is a CG coefficient that is even/odd
under the interchange of the two $R$ factors, while
$C'$ is a CG coefficient that is odd/even under this interchange. It follows that it is possible to separately choose a basis of orthogonal even three-point functions $C_{RR {\tilde R}_a}$ and a basis of odd orthogonal three-point functions. Having made such a choice, the $S$-matrix admits an expansion analogous to \eqref{expofsmatrixmtid}
\begin{equation}\label{expofsmatrixmtidq}
	S = \sum_{a, r_1, r_2} \left( \mathcal{S}^{r_1, r_2}_{ee, a} G_{ee, a}^{r_1 r_2} +  \mathcal{S}^{r_1, r_2}_{eo, a} G_{eo, a}^{r_1 r_2} + \mathcal{S}^{r_1, r_2}_{oe, a} G_{oe, a}^{r_1 r_2} + \mathcal{S}^{r_1, r_2}_{oo, a} G_{oo, a}^{r_1 r_2} \right)
\end{equation}
Here, for instance, $G_{eo}$ denotes a block of the form depicted in
Fig. \ref{symasym}, with $F$ replaced by $R$, $A$ replaced by
$R'$, $Sym$ replaced by ${\tilde R}_a$ and the CG coefficient between $RR {\tilde R}_a$ being even, while the CG coefficeint between ${\tilde R}_a AA$ is odd.

\eqref{expofsmatrixmtidq} applies provided the two initial representations are identical (similar remarks apply to final representations). We now specialize to the case that
the two initial operators are, themselves, identical bosons/fermions. In this case the
form of the $S$-matrix is further restricted by the requirements of Bose/Fermi symmetry.
Recall that the blocks that appear in \eqref{expofsmatrixmtidq} are equipped with
a cut structure. We could, for instance, choose the cut conventions of Seiberg
and Moore. With this (or any other convenient) choice of cut structure, for any
given locations of the insertions of the two $R$ representations, we can interchange these locations by making either a clockwise or an anticlockwise rotation by $\pi$ about
a point on the line joining these two insertions. We can, however,  make only one of these two rotations without crossing a cut: which one depends on the locations of our insertions. We now elaborate on this point.

Let $L_1$ denote the class of configurations for which insertions can be interchanged by a clockwise rotation of $\pi$ (around any point on the line joining the two insertions) without crossing a cut in the conformal blocks. In the complimentary class of insertions (which we name $L_2$), it is not possible to perform a clockwise rotation by $\pi$ without crossing a cut. It is, however, possible to perform a rotation by $\pi$ anticlockwise, to accomplish the interchange of the two identical insertions, without crossing a cut. For locations  that lie in $L_2$, \eqref{braidingong} and \eqref{braidingongsm} apply, but with all
phases replaced by their inverses (i.e., complex conjugates).

Let us suppose that we start with insertion positions that lie in the class $L_1$.
The interchange of insertions leads to a new configuration. It is clear from definitions that this new configuration lies in the class $L_2$  (because an anticlockwise rotation performed on the new configuration simply undoes the original rotation that orginally led us to this configuration, and that rotation, by definition, did not
involve crossing a cut). It is also clear that this new insertion configuration does not lie in $L_1$ (because if it did, a further $\pi$ rotation clockwise on the new configuration would take us back to the original configuration, but upto a phase contradicting the single valuedness of $G$ on the cut plane). We conclude that $L_2$ and
$L_1$ are disjoint sets. Atleast with reasonable choices of cuts (like the one adopted by Seiberg and Moore) we believe that $L_1$ is the compliment of $L_2$ (and vica verca)

It is useful to define a quantity $\epsilon(p_1,p_2)$ (see \cite{Furlan:2014pfa} for similar objects) that keeps track of whether our insertion points lie in the space $L_1$ or $L_2$. By definition
\begin{equation}\label{loltdef} \begin{split}
		&\epsilon(p_1,p_2)=1 , ~~~~~~{\rm if ~(p_1, p_2)~} \in L_1 \\
		&\epsilon(p_1,p_2)=-1 , ~~~~~~{\rm if ~(p_1, p_2)~} \in L_2 \\
	\end{split}
\end{equation}
It follows from the discussion above that
\begin{equation}\label{onetwo}
	\epsilon(p_1,p_2)=-\epsilon(p_2,p_1)=1
\end{equation}

With this notation in place, it follows that the action of interchange on identical operators is given by
\begin{equation}\label{braidingong} \begin{split}
		& G_{ee, a}^{r_1 r_2}  \rightarrow e^{ \pi i \epsilon(p_1, p_2) (2 h_R-h_a)} G_{ee, a}^{r_1 r_2}\\
		& G_{eo, a}^{r_1 r_2}  \rightarrow  e^{ \pi i  \epsilon(p_1, p_2) (2 h_R-h_a)} G_{eo, a}^{r_1 r_2}\\
		& G_{oe, a}^{r_1 r_2}  \rightarrow - e^{ \pi i  \epsilon(p_1, p_2) (2 h_R-h_a)} G_{oe, a}^{r_1 r_2}\\
		& G_{oo, a}^{r_1 r_2} \rightarrow  -e^{ \pi i  \epsilon(p_1, p_2) (2 h_R-h_a)} G_{oo, a}^{r_1 r_2}\\
	\end{split}
\end{equation}
where $a$ is the irrep to which the two insertions $R$ fuse. If the CG coefficients corresponding to this fusion are symmetric, the block picks up the phase on RHS of \eqref{braidingong} under braiding, on the other hand if the CG coefficients are antisymmetric, the block picks up the phase with a negative sign under braiding(interchange).

Since the full wavefunction must obey Bose/Fermi statistic under interchange, the coeffcient $S$-matrices must transform under interchange as follows:
\begin{equation}\label{braidingongsm} \begin{split}
		& \mathcal{S}_{ee, a}^{r_1 r_2}(p_2,p_1,p_1',p_2') = \pm e^{ -\pi i  \epsilon(p_1, p_2)(2 h_R-h_a)} \mathcal{S}_{ee, a}^{r_1 r_2}(p_1,p_2,p_1',p_2')\\
		& \mathcal{S}_{eo, a}^{r_1 r_2}(p_2,p_1,p_1',p_2') = \pm e^{ -\pi i  \epsilon(p_1, p_2)(2 h_R-h_a)} \mathcal{S}_{eo, a}^{r_1 r_2}(p_1,p_2,p_1',p_2')\\
		& \mathcal{S}_{oe, a}^{r_1 r_2}(p_2,p_1,p_1',p_2') = \mp e^{ -\pi i  \epsilon(p_1, p_2)(2 h_R-h_a)} \mathcal{S}_{oe, a}^{r_1 r_2}(p_1,p_2,p_1',p_2')\\
		&\mathcal{S}_{oo, a}^{r_1 r_2}(p_2,p_1,p_1',p_2') = \mp e^{ -\pi i  \epsilon(p_1, p_2)(2 h_R-h_a)} \mathcal{S}_{oo, a}^{r_1 r_2}(p_1,p_2,p_1',p_2')\\
	\end{split}
\end{equation}

The RHS of the unitarity equation (analogue of to \eqref{rhsunitbos}) gets modified for finite $k$ to be
\begin{equation}\label{rhsunitbosq}
	\sum_{a,r} G^{rr}_{ee,a} \mathcal S^e_{a,id}+G^{rr}_{oo,a} \mathcal S^o_{a,id}
\end{equation}
for identical bosons, and
\begin{equation}\label{rhsunitbosq}
	\sum_{a,r} G^{rr}_{ee,a} \mathcal S^o_{a,id}+G^{rr}_{oo,a} \mathcal S^e_{a,id}
\end{equation}
for identical fermions
where
\begin{equation}\label{ident}
	\begin{split}
		\mathcal S^e_{a,id}(p_1,p_2,p_1',p_2')=\delta(p_1-p_1')\delta(p_2-p_2')+ e^{\pi i (2h_R-h_s)\epsilon(p_1, p_2)}\delta(p_1-p_2')\delta(p_2-p_1')\\
		\mathcal S^o_{a,id}(p_1,p_2,p_1',p_2')=\delta(p_1-p_1')\delta(p_2-p_2')- e^{\pi i (2h_R-h_s)\epsilon(p_1, p_2)}\delta(p_1-p_2')\delta(p_2-p_1')
	\end{split}
\end{equation}
Note that \eqref{ident} agrees with the first and last of \eqref{braidingongsm}.

Putting \eqref{expofsmatrixmtidq} in the unitarity equation with \eqref{rhsunitbosq} on the RHS, we get

\begin{equation}\label{smatrixsuniidbos}\begin{split}
		&\sum_{\rm final ~states} \sum_{r'} ( \mathcal S_{ee,a}^{r_1r'})^* \star  \mathcal S_{ee,a}^{r_2r'} + (\mathcal S_{eo,a}^{r_1r'})^* \star  \mathcal S_{eo,a}^{r_2r'}
		= \mathcal S^e_{a,id} \delta_{r_1, r_2} \\
		&\sum_{\rm final ~states} \sum_{r'} ( \mathcal S_{oe,a}^{r_1r'})^* \star  \mathcal S_{oe,a}^{r_2r'} +{(\mathcal S}_{oo,a}^{r_1r'})^* \star  \mathcal S_{oo,a}^{r_2r'}
		= {\mathcal S}^o_{a,id} \delta_{r_1, r_2} \\
		&\sum_{\rm final ~states} \sum_{r'} ( \mathcal S_{ee,a}^{r_1r'})^* \star  \mathcal S_{oe,a}^{r_2r'} + (\mathcal S_{eo,a}^{r_1r'})^* \star  \mathcal S_{oo,a}^{r_2r'}
		= 0 \\
		&\sum_{\rm final ~states} \sum_{r'} ( \mathcal S_{oe,a}^{r_1r'})^* \star  \mathcal S_{ee,a}^{r_2r'} +{(\mathcal S}_{oo,a}^{r_1r'})^* \star  \mathcal S_{eo,a}^{r_2r'}
		= 0 \\
	\end{split}
\end{equation}
when the initial particles are identical bosons, while it takes the form
\begin{equation}\label{smatrixsuniidbos}\begin{split}
		&\sum_{\rm final ~states} \sum_{r'} ( \mathcal S_{ee,a}^{r_1r'})^* \star  \mathcal S_{ee,a}^{r_2r'} + ({\mathcal S}_{eo,a}^{r_1r'})^* \star  \mathcal S_{eo,a}^{r_2r'}
		= \mathcal S^o_{a,id} \delta_{r_1, r_2} \\
		&\sum_{\rm final ~states} \sum_{r'} ( \mathcal S_{oe,a}^{r_1r'})^* \star  \mathcal S_{oe,a}^{r_2r'} + ({\mathcal S}_{oo,a}^{r_1r'})^* \star  \mathcal S_{oo,a}^{r_2r'}
		= \mathcal S^e_{a,id} \delta_{r_1, r_2} \\
		&\sum_{\rm final ~states} \sum_{r'} ( \mathcal S_{ee,a}^{r_1r'})^* \star  \mathcal S_{oe,a}^{r_2r'} + (\mathcal S_{eo,a}^{r_1r'})^* \star  \mathcal S_{oo,a}^{r_2r'}
		= 0 \\
		&\sum_{\rm final ~states} \sum_{r'} ( \mathcal S_{oe,a}^{r_1r'})^* \star  \mathcal S_{ee,a}^{r_2r'} +{(\mathcal S}_{oo,a}^{r_1r'})^* \star  \mathcal S_{eo,a}^{r_2r'}
		= 0 \\
	\end{split}
\end{equation}
when they are identical fermions.

Finally, when the two initial particles are also identical to the two final particles (so that, in particular, $R=R'$), the precise form of the $S$-matrix (including the identity piece, see Appendix \ref{mstn}) is given by the following replacement of
\eqref{compsmat}
\begin{equation}\label{compsmatid}
	\begin{split}
		&{\mathcal S}^{rr'}_{ee,a}= \delta_{rr'} \mathcal S^e_{a,id}
		+ i \tau^{rr'}_{ee, a}\\
		&{\mathcal S}^{rr'}_{eo,a}= i \tau^{rr'}_{eo, a}\\
		&{\mathcal S}^{rr'}_{oe,a}=  i \tau^{rr'}_{oe, a}\\
		&{\mathcal S}^{rr'}_{oo,a}= \delta_{rr'} \mathcal S^o_{a,id}
		+ i \tau^{rr'}_{oo, a}\\
	\end{split}
\end{equation}

As we have described in great detail earlier in this section (and as was the case in subsection \ref{ipg}), the crossing rules are determined completely by the compounding and transformation properties of conformal blocks, and so are unaffected by the
somewhat inelegant analysis of this subsection. For this reason the discussion of this subsection (which we have included in this paper only for completeness) will play no further role in this paper.

\section{Conventions and framing}\label{conventions}

Through this paper all Wilson lines are defined in the `vertical-framing' convention, defined in \cite{Witten:1989wf}, in which the framing vector is constant and everywhere emerges out of the paper, towards the reader. This framing convention differs from the `zero-self-linking number' framing convention used by Witten in his earlier paper, \cite{Witten:1988hf}, which does not generalize naturally to situations involving bulk interaction-vertices.

In the vertical-framing convention, it turns out that Wilson lines that are topologically the same (in the sense of their shape as one-dimensional curves) are sometimes different as quantum operators because they differ in the topology of their framing field.\footnote{In particular the self-linking number of the Wilson loop and the
	loop translated by the framing vector field may differ between two Wilson lines that are continuously connected in the sense of shape. We will see an example of this below.} We thus need to understand how Wilson lines transform under a twist in framing. This can be figured out as follows.
When cut in a transverse manner, a Wilson line in the representation $R$ is seen as a primary operator
$\phi_R(z)$ of weight $h_R$ ($h_R$ is the weight of the primary operator in the WZW theory associated with the Chern-Simons theory in question). Under a rotation $z \rightarrow e^{i \alpha} z$,
$\phi_R(z)$ maps to the operator $e^{i h_R \alpha} \phi(e^{i \alpha} z)$. The factor $e^{i h_R \alpha}$
tells us that the effect of a $2 \pi$ counterclockwise rotation of the framing vector at $\phi$
\footnote{Completely explicitly, this $2 \pi$ rotation can be affected as follows. Cut the Wilson line. Hold it so that one of the free ends is coming out towards you. Then rotate this end counterclockwise by $2 \pi$. After performing this rotation,
	reglue it to the other end of the Wilson line. Note that if you perform the same operation at the other cut end of the Wilson line you produce the same final twist. The key point here is that
	one is instructed to rotate in the counterclockwise direction when looking at the cut end in question.}  is to multiply the Wilson line by $e^{ 2 \pi i h_R}$. With this rule at hand,
it is not difficult to convince oneself of the identities depicted in Fig 9 of \cite{Witten:1989wf} (which relate a Wilson line with an `overpass' or `underpass' to a phase times a topologically identical Wilson line without the overpass).

Recall that in any block in which representations $R_a$ and $R_b$ couple to $R_c$  we have
\begin{equation}\label{abccoupl}
\phi_{R_a}(z_1) \phi_{R_b}(z_2) \sim \frac{ \phi_{R_c}(z_2)}{(z_1-z_2)^{h_a+h_b-h_c} }
\end{equation}
It follows from the discussion above that under the anticlockwise rotation $z_i \rightarrow e^{i \alpha} z_i$, this block picks up the phase $e^{-i \alpha(h_a+h_b -h_c)}$. This fact is used in the
derivation of the Skein relations reviewed in Appendix \ref{cgs}.

It follows, in particular, that the expectation value of a circular Wilson loop in vertical framing is
not completely invariant under continuous changes in the topology of the loop -- a change that generates an overpass or an underpass changes the expectation value of the loop. This is consistent with the topological nature of the theory in the following manner. Consider a circular Wilson loop
with vertical framing and no over or underpasses. If we now create an overpass (or underpass) in such a loop, while changing the framing vector field in a continuous manner, then the resultant framing vector field ( after we have created the over or under pass) cannot be vertically framed; the maneuver needed to bring the framing field back to vertical results in a phase for the Wilson line, as explained above.

However a simple circular Wilson loop -- with no over or under passes -- has exactly the same expectation value as in the `zero self linking number framing convention' adopted by Witten in
his original paper \cite{Witten:1988hf}. That this is the case follows from the fact that a simple Wilson loop and its displacement along the vertical framing vector, clearly have zero self-linking
number. \footnote{However the same is not true for an unknotted Wilson loop that has an over or under pass with vertical framing.}. As in the case of the zero self linking number convention, a simple circular Wilson loop
in representation $R$ evaluates to the quantum dimension $D_R$, and in particular, is real and positive. The value of this Wilson loop is also invariant under reflections.

In addition to the framing of Wilson lines one needs to carefully define the framing of all interaction vertices. For this purpose, we also adopt Witten's conventions in the paper \cite{Witten:1989wf}. All our interaction vertices are defined by the convention displayed in
Fig. 12(a) of that paper. Using three strips of paper stapled together at the vertex as an aid to imagination, it is also not difficult to verify the identities depicted in Fig. 15 of \cite{Witten:1989wf} which relates an interaction vertex to a phase times the same vertex with a double twist between two of its legs \footnote{The manipulation here involves turning the interaction vertex inside out.}.

Throughout this paper we also adopt the conventions listed in Fig. 16 of \cite{Witten:1989wf}. These
conventions relate the interaction vertices with two different cyclical orderings at the interaction vertex (these cyclical orderings are well defined because we are forced to keep track of the framings at the interactions). We believe that the conventions of Fig. 16 of \cite{Witten:1989wf}
are automatically satisfied if we use exactly the same Clebsch-Gordan coefficients (and same regulation scheme)  to define the Wilson lines that appear on LHS and RHS of the identities depicted in Fig. 16 of \cite{Witten:1989wf}. In the special case that two of the representations $a, b$ or $c$ in Fig 16 of \cite{Witten:1989wf} are equal, the fact that we insist on using the same Clebsch-Gordan convention to couple $a$, $b$ and $c$, in that order, on the two sides of these identities, means that the vertex on the RHS, viewed locally (i.e., viewed in the cyclical order that the Wilson line legs join together at the interaction vertex) equals either plus one or minus one times the local vertex that appears on the LHS, depending on whether the coupling in question is symmetric or antisymmetric between the identical legs.
\footnote{See the parenthetic remark at the end of section 2.2 in \cite{Witten:1989wf}.}

A state (or ket) in Chern-Simons theory can be defined by a path integral with a specified tangle of Wilson lines (and specified end points). The dual state (or bra) corresponding to the same
state is given as follows. One flattens out the part of the boundary which hosts all Wilson line end points (somewhat analogeous to  Fig \ref{prodfig}, except that the flattened out
region now contains {\it all} Wilson line end points). One then reflects the
diagram about the flat surface, and simultaneously complex conjugates all explicit numbers  (e.g., Clebsch-Gordon coefficients) that appear in the definition of the diagram. One of the consequences of the reflection
is that colour lines switch their direction of flow. It follows that the reflection of a vertex that couples representations $R_a$, $R_b$ and $R_c$ in a particular cyclical ordering now couples representations $R_a^*$, $R_b^*$ and
$R_c^*$ in the reverse cyclical ordering. \footnote{The fact that reflection results in the complex conjugation of representations that appear in an interaction vertex also forces us to complex conjugate the Clebsch-Gordan coefficients at the vertex.} \footnote{A consistency check of this definition of the bra goes as follows. According to our definitions, if  $\langle a|b\rangle$ is computed by a given closed Wilson line (or Wilson line tangle) then $\langle b|a\rangle$ must be given by the mirror reflected Wilson line (or Wilson line tangle). It follows as a consequence that the expectation value of a  Wilson line and its mirror reflection must be complex conjugates of each other. This is indeed the case. One can verify it in a
simple example as follows. Consider a circular Wilson line  with one `overpass' (of the form depicted in Fig. 9(a) or the top left diagram in 9(c) of \cite{Witten:1989wf}) at the bottom of the loop. The reflection of this Wilson line has an `underpass', of the form depicted in the bottom left diagram
of Fig. 9(c) \cite{Witten:1989wf}. As explained in Fig. 9(c) of \cite{Witten:1989wf}, the phase factors associated with these two `passes' are indeed complex conjugates of each other, as predicted.}

The discussion in the paragraph above suggests a natural relationship between vector space of three-point couplings of the representations $R_a, R_b$ and $R_c$ and the space of three-point couplings of the representations $R_a^*$, $R_b^*$ and $R_c^*$. If we denote the three-point coupling with given CG coefficient $C$ and a particular cyclical ordering of lines at the interaction vertex as the vector $|s\rangle$ ($|s \rangle$ is a vector in the space of 3 point couplings of $R_a$, $R_b$ and $R_c$) then we define the dual vector $\langle s|$ to be the three-point coupling of $R_a^*$, $R_b^*$ and $R_c^*$ with the CG coefficient $C^*$ and the opposite cyclical ordering (the reversal of cyclical ordering is a consequence of the reflection in the paragraph above). The inner product $\langle t|s \rangle$, on the space of three-point couplings (i.e., on the space of equivalence classes of
CG coefficients) is defined by gluing of the ket $s$ with the bra $t$, i.e., by Witten's quantity
$K(a, b, c, \epsilon^i, {\tilde \epsilon}_j)$. The fact that this inner product is positive definite
(i.e., that $ \langle s |s \rangle$ is positive, and vanishes only when $|s\rangle$ vanishes)
then follows from the discussion of the previous paragraph.

Using the facts and conventions
reviewed in this Appendix, it is easy to see that the quantity $K(a, b, c, \epsilon^i, {\tilde \epsilon}_j)$ is left invariant under any re-ordering
of the Wilson lines at the interaction vertices that appear on either of the vertices of the LHS of Fig \ref{threeptorthogtw}, provided the corresponding (mirror reflected) reordering is also performed
on the other vertex. It follows that $K(a, b, c, \epsilon^i, {\tilde \epsilon}_j)$ defines
an inner product on the space of equivalence classes of CG coefficients themselves, without the
need to specify an ordering in interaction vertices. We see that this inner product is closely analogous to its classical counterpart listed on the LHS of \eqref{cgorthog}.

\section{The multi-valuedness of blocks and the channel expansion of the $S$-matrix} \label{multisheet}

In \eqref{sumtopn} we have expressed the $S$-matrix as a sum over products of component $S$-matrices and conformal blocks $G_i$, where the blocks $G_i$ also have a particular prescribed cut structure (i.e., are evaluated on a given specified sheet). The choice of both the basis for the blocks $G_i$, as well as the sheet on which each $G_i$ is evaluated is made arbitrarily.

The original definition \eqref{sumtop}, however,  makes it clear that the full $S$ (unlike any given conformal block) is  single valued. That
$S$ has no cuts may be seen as follows. Let us start with any particular collection of the end points for the path integral, and move each of the end points infinitesimally. Then every trajectory that contributes to the first
path integral has a nearby trajectory that contributes to the second path integral. As the action for trajectories (in a non singular gauge) is a continuous function of trajectories, it follows that the path integral is a smooth function of
its end points, and in particular has no cuts. Another way of saying the same thing is that the path integral, viewed
as a function of the end points, is the final wave function of the system (atleast in the non relativistic limit). Wave functions (in nonsingular gauges) are always single valued and have no cuts.

Unlike the full answer, the basis functions $G_i$ certainly (and by definition) have cuts across which $G_i$ have discontinuities. Recall that $G_i$ are smooth continuous functions only on branched cover of $S^2$.
The path integral that evaluates the $S$-matrix, on the other hand, is single valued on
$S^2$ not its branched cover.

If we keep the cut structure of $G_i$ fixed, but move from one sheet to another, the block changes according to the formula $G_i'= M_i^jG_j$. (where $G_j$ is the value of the block on the original sheet). It follows that the value of $G_i$ at the bottom of a cut is related to its value at the top of the cut by
the relation
\begin{equation}\label{you}
	G_i^{{\rm bot}} = M_i^jG_j^{{\rm top}}, ~~~~{\rm or } ~~~~~ G^{{\rm bot}} = M G^{{\rm top}}
\end{equation}
We have an equation of the form \eqref{you} for every cut in the conformal block. Given that $S$ is single valued,
it follows that
\begin{equation}\label{vitransf}
	(\mathcal S^i)^{\rm bot} = (\mathcal S^j)^{\rm top} M^{-1}
\end{equation}

In a similar manner, under a change of basis for conformal blocks
\begin{equation}\label{basischange}
	G \rightarrow B G, ~~~~ \mathcal S \rightarrow \mathcal S B^{-1}
\end{equation}
so that $S$ remains invariant. Very roughly speaking, the  ${\mathcal S}$ transform under the monodromy operation  like
antiholomorphic blocks, so that the product ${\mathcal S}^i G_i$ (like full correlation functions in WZW theory with the diagonal modular invariant) is single valued.

A more complicated operation on $G_i$ may be affected by changing the location of its cut (not merely `which sheet' with fixed cut location, as in \eqref{you}. Such an operation results in a position dependant transformation on $G_i$, and a corresponding position dependent transformation on $\mathcal S_i$, designed to ensure that the full $S$-matrix ${\mathcal S}$
stays invariant.

\section{Projector blocks without interactions}
\label{quantexamp}

In this section we present an alternate construction of the blocks $G_{s/a}$ and $G_{{\rm Adj}/I}$ presented in section \ref{fundnew} -- and correspondingly an alternate way of obtaining the crossing relations presented in subsection \ref{cross} -- using a basis of blocks with no bulk interaction vertices. The constructions presented in this Appendix are possible $H_{ {\rm in}}=H_{ {\rm out}}$ in the example under study.

\subsection{Fundamental-fundamental scattering}

\begin{figure}[h]
	\centering
	\includegraphics[scale=.4]{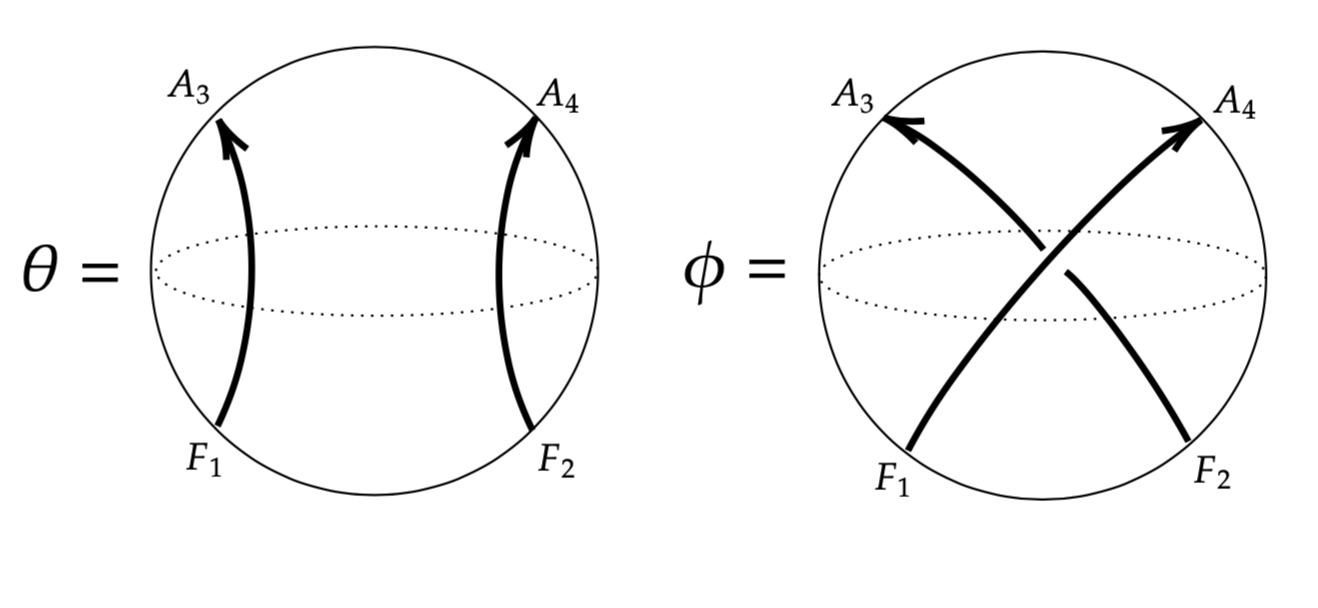}
	\caption{Definition of blocks $\theta$ and $\phi$ }
	\label{thph}
\end{figure}

In this subsection we re-analyse the scattering of two fundamental particles with each other (see subsection \ref{ffs}). In the `classical' version of this problem, a useful basis for the two dimensional space of invariant tensors  was given by the `direct' and `exchange' index contractions defined in \eqref{ginve}. The Chern-Simons
analogue of these index structures are the conformal blocks $\theta$, $\phi$ depicted in Fig. \ref{thph}. As in the rest of this paper, all Wilson line configurations in this section are defined in the vertical framing convention (see Appendix \ref{conventions}.)

\begin{figure}[H]
	\centering
	\includegraphics[scale=.4]{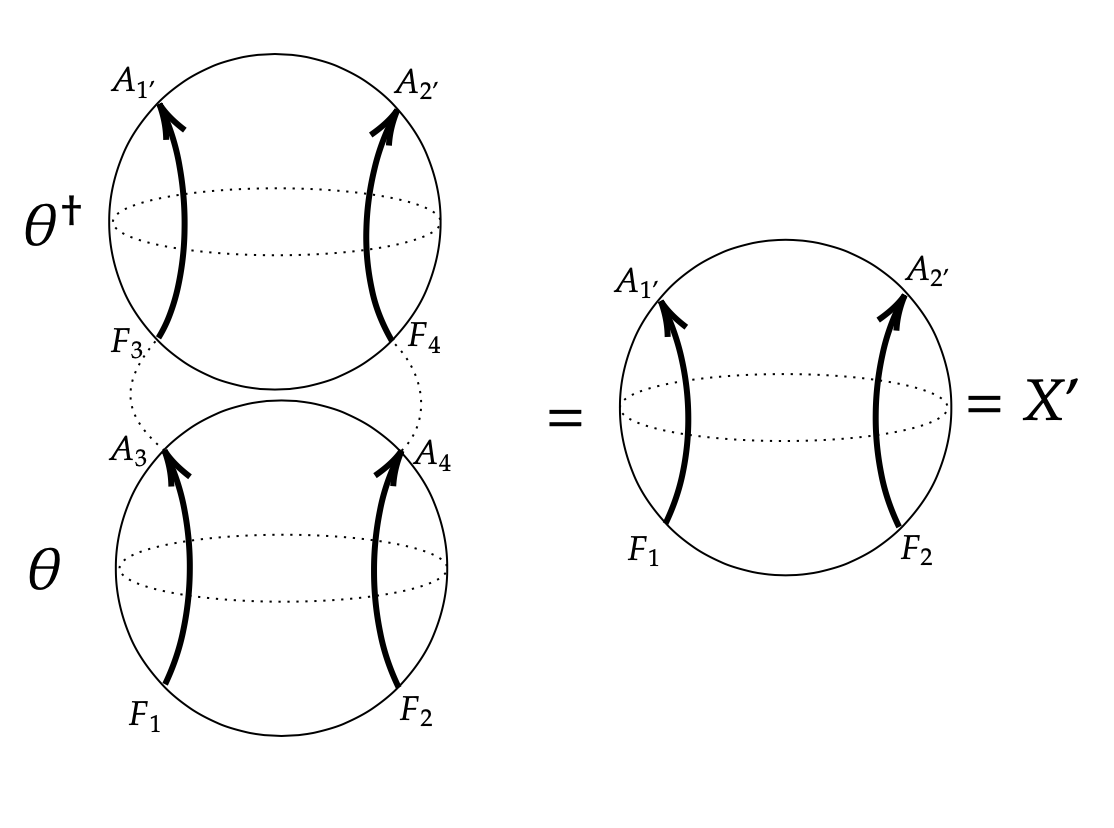}
	\caption{$\theta\times\theta^\dagger$}
	\label{comp1}
\end{figure}

\begin{figure}[H]
	\centering
	\includegraphics[scale=.4]{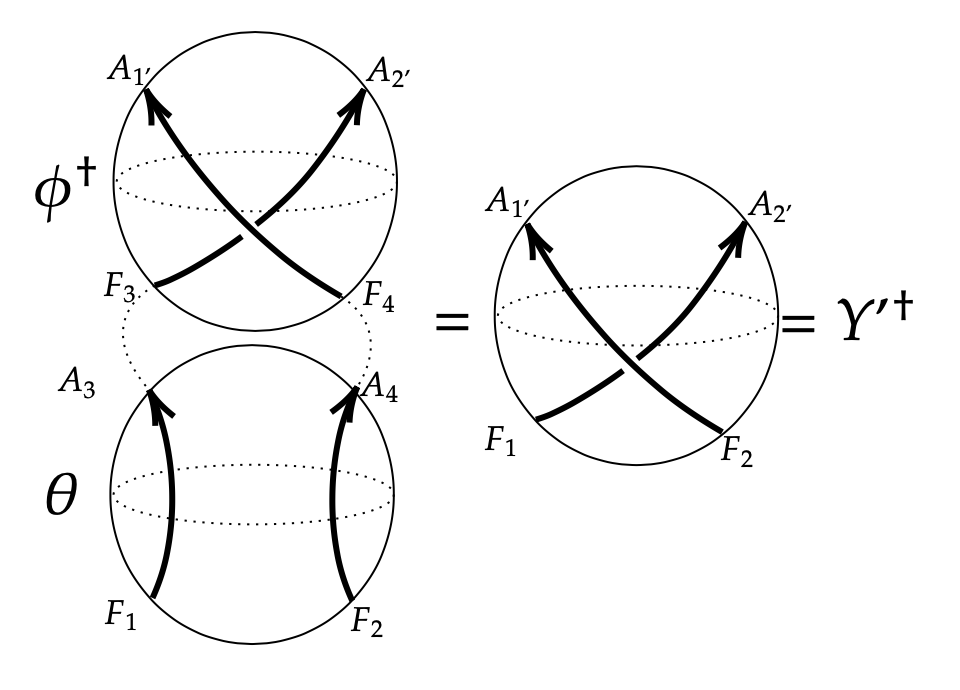}
	\caption{$\theta\times\phi^\dagger$ }
	\label{comp2}
\end{figure}

\begin{figure}[H]
	\centering
	\includegraphics[scale=.4]{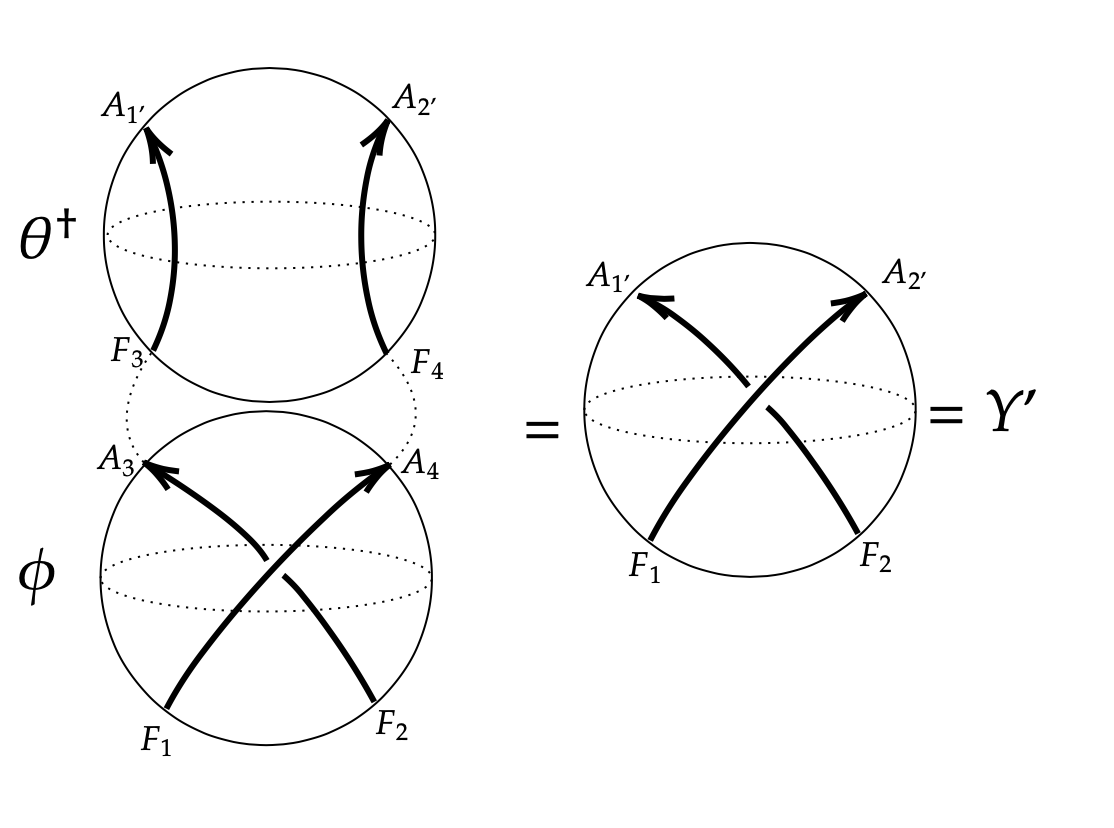}
	\caption{$\phi\times\theta^\dagger$ }
	\label{comp3}
\end{figure}

\begin{figure}[H]
	\centering
	\includegraphics[scale=.4]{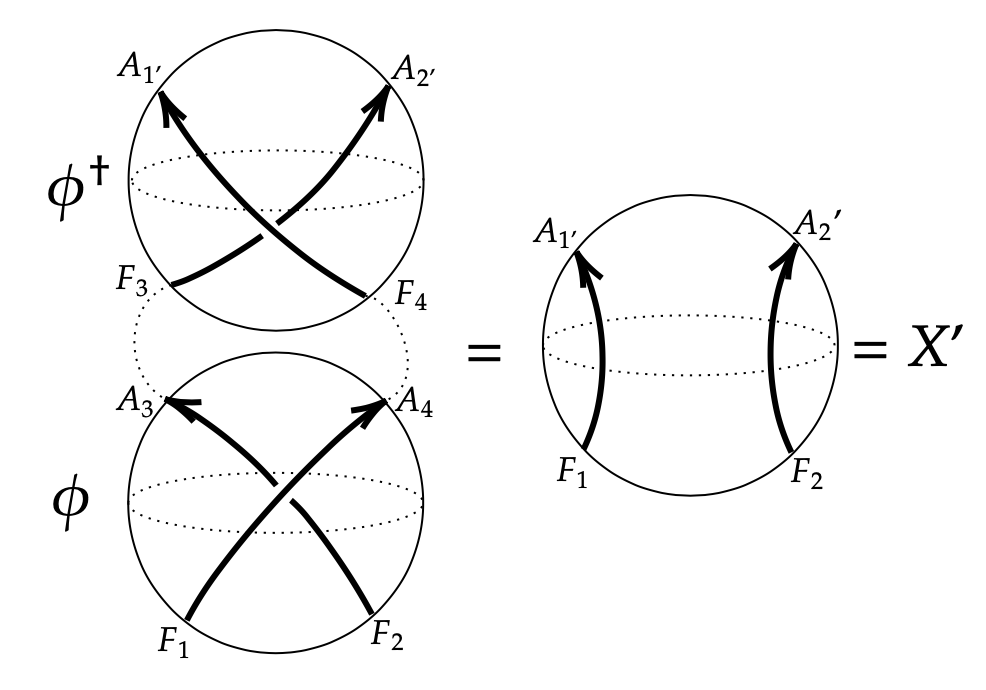}
	\caption{$\phi\times\phi^\dagger$ }
	\label{comp4}
\end{figure}

As discussed in the subsection \ref{cSm}, for imposing unitarity we need to compound these blocks with their Hermitian conjugates. Using compounding relevant to $FF \rightarrow FF$ scattering (see  subsection \ref{cSm}), we find   \\\\
\begin{equation}\label{comprel}
	\begin{split}
		\theta \times \theta^\dagger = X'  \\
		\theta \times \phi^\dagger= Y'^\dagger \\
		\phi \times \theta^ \dagger= Y' \\
		\phi \times \phi^\dagger = X' \\
	\end{split}
\end{equation}
The derivation of \eqref{comprel} is presented in Figs. \ref{comp1}, \ref{comp2}, \ref{comp3}, \ref{comp4}. The blocks $X'$ -- defined in Fig. \ref{comp1} -- is the same as the block $\theta$, but
in the space of blocks defined on $H_{ {\rm in}} \times H_{ {\rm in}}^*$ rather than $H_{ {\rm in}} \times H_{ {\rm out}}^*$.
Had we adopted the notation of \ref{fundnew}, we would have used the symbol ${\hat \theta}$ for
$X'$ (we avoid using this terminology to prevent clutter later in this section). In a similar manner $Y'$, defined in Fig \ref{comp3}, is the block $H_{ {\rm in}} \times H_{ {\rm out}}^*$ that we would have named ${\hat \phi}$, had we been using the terminology of section \ref{fundnew}.

The blocks $\theta^\dagger$ $\phi^\dagger$, $X^{' \dagger}$ and $Y^{'\dagger}$ are defined,
as usual, by reflecting the blocks $\theta$, $\phi$ $X'$ and $Y'$ around the flattened final surface
(which contains all final insertions) \footnote{Equivalently, we could have reflected around the flattened surface that contains all insertions: both procedures give the same answer. It follows that
the dagger operation in the sense of compounding, is the same as the dagger operation in the sense
of states in Witten's Hilbert space \cite{Witten:1988hf}.} . Explicitly, $\theta^\dagger$ is depicted in Fig. \ref{comp1},
$\phi^\dagger$ is depicted in Fig. \ref{comp2}, $Y^{'\dagger}$ in Fig. \ref{comp3} and $X^{'\dagger}$ in Fig. \ref{comp4}.

The RHS of the first, third and fourth of \eqref{comprel} are presented in terms of the basis blocks
$X'$ and $Y'$.
However the second of \eqref{comprel} is presented in terms of $Y^{'\dagger}$. As the
relevant space of blocks is two dimensional \footnote{ Assuming that $k>1$; in the special case $k=1$ the space of blocks is one dimensional.}, it is possible to reexpress $Y^{'\dagger}$ in terms of the blocks $X'$ and $Y'$. This can be achieved by use of the Skein relations. The Skein relations
in the vertical framing convention are explicitly listed in Appendix \ref{cgs}, separately for the $SU(N)_k$, Type I and Type II theories. We pause here to remind the reader how these relations are derived and to understand their transformation under level rank duality.

The Skein relation presented in Fig. 8 of \cite{Witten:1988hf} (recall Fig. 8 of \cite{Witten:1988hf} applies in the vertical framing convention, as employed in this paper) can be rewritten in the
notation of this Appendix as
\begin{equation}\label{skeistruc}
	Y^{'\dagger} - {\rm Tr}(B)  X'+
	{\rm det}(B) Y'=0
\end{equation}
where $B$ is the `half monodromy' matrix of Seiberg and
Moore. The eigenvalues
of the matrix $B$ are given by
\begin{equation}\label{evan}
\left( e^{ \pi i \left( 2 h_F - h_s \right)}, - e^{ \pi i \left( 2 h_F - h_a \right)} \right)
\equiv (\lambda_1, \lambda_2)
\end{equation}
where $h_s$ and $h_a$ are the spins of the primary in the symmetric and antisymmetric representations.
It follows that ${\rm Tr}(B)= \lambda_1+\lambda_2$ and ${\rm det}(B)=\lambda_1 \lambda_2$. Inserting
these equations into \eqref{skeinrelsun} makes that equation explicit, and allows us to solve for
$Y^{'\dagger}$ in terms of $X'$ and $Y'$.
\footnote{  The explicit values of $h_F$, $h_s$ and $h_a$ are listed in \eqref{dimensions},
\eqref{totheor} and \eqref{dimtt} for the case of the $SU(N)_k$, Type I and Type II theories
respectively. }

As we have explained in subsection \ref{pis} and subsection \ref{pb}, the analysis of unitarity
is simplest in the so-called `projector basis' for blocks. In the simple example of the current
section, the fusion of two fundamentals generically (i.e., for $k>1$) produces two representations;
the symmetric and antisymmetric representation. Consequently we wish to move to the basis of
blocks $G_s$ and $G_a$, that respectively play the role of the `projectors' onto the symmetric and antisymmetric representations. $G_s$ and $G_a$ are the bock analogues of the invariant tensors $T_s$ and $T_a$ defined in \eqref{ppscatff}. In section \ref{fundnew} we have already found one representation for these blocks in terms of Wilson lines (see Fig. \ref{nblocks1}). We will now find an alternate representation of the same blocks as linear combinations of $\theta$ and $\phi$.

Let us set
\begin{equation}
	\begin{split}
		G_s= a_1 \theta + a_2 \phi \\
		G_a= b_1 \theta + b_2 \phi
	\end{split}
\end{equation}
and constrain the coefficients $a_1$, $a_2$, $b_1$, $b_2$ by the requirement that the two new blocks be orthogonal to each other (recall that in the `projector' basis, blocks obey \eqref{projopsn})
\begin{equation}
	G_s^\dagger  \times G_a =0
\end{equation}
Using \eqref{comprel} and \eqref{skeistruc}, we find that this  condition constrains the coefficients of the blocks $\phi$ and $\theta$ in $G_s$ and $G_a$ to obey the relations \\
\begin{equation} \label{ratioson}
	\begin{split}
		\frac{a_2}{a_1}= -\lambda_2^{-1}\\
		\frac{b_2}{b_1}= -\lambda_1^{-1}
	\end{split}
\end{equation}
where $(\lambda_1,\lambda_2)$ are the eigenvalues of the $B$ matrix (see \eqref{evan}).

One of the relations \eqref{ratioson} defines the symmetric block $G_s$, while the other defines
the antisymmetric block $G_a$. In order to see which is which (and also to obtain a consistency check on the results \eqref{ratioson}) note that the blocks $G_s$ and $G_a$ must be eigenvectors of the
Seiberg and Moore half monodromy operator $B$; with eigenvalues respectively given by $\lambda_1$ and
$\lambda_2$. Using the fact that the $B$ matrix that has these eigenvalues is implemented by the interchange accomplished by a clockwise rotation by $\pi$ either on the two antifundamentals or on the two fundamentals (see Appendix \ref{conventions}), it follows immediately from the definition of the blocks presented in Fig \ref{thph} and Fig. \ref{comp2} that
\begin{equation}\label{bact}
	\begin{split}
		B |{\phi}\rangle &= |\theta\rangle\\
		B|\theta\rangle &= |\phi^\dagger\rangle={\rm Tr} B |\theta\rangle - {\rm det} B| \phi\rangle \\
	\end{split}
\end{equation}
In going from the second to the third expression in \eqref{bact} we have used the Skein relation
\eqref{skeistruc} with the replacement $Y^{'\dagger} \rightarrow \phi^\dagger$, $Y' \rightarrow \phi$ and $X' \rightarrow \theta$.
\begin{figure}[H]
	\centering
	\includegraphics[scale=.4]{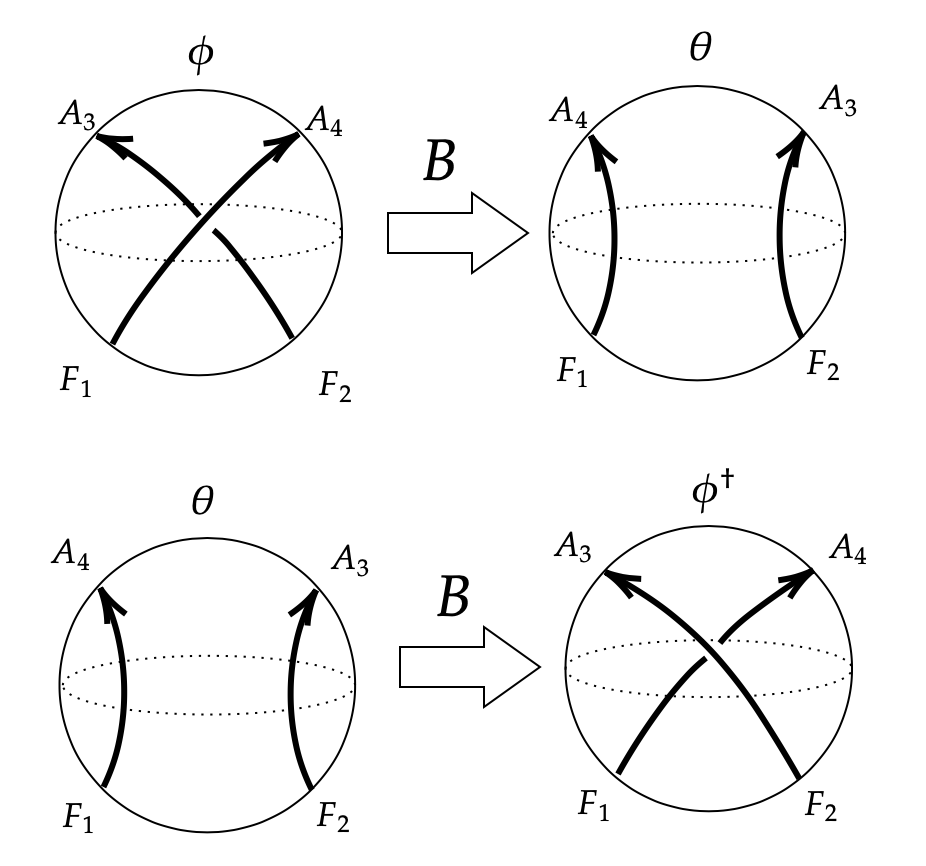}
	\caption{Action of the monodromy operator $B$ }
	\label{monod1}
\end{figure}
It follows from \eqref{bact} that the half monodromy matrix $B$, in the  $\theta-\phi$ basis, takes the explicit form
 \begin{equation} \label{bexpform}
	B = \left(\begin{array}{cc}
		(\lambda_1+\lambda_2)& 1 \\
		- \lambda_1 \lambda_2  & 0
	\end{array}\right)
\end{equation}
It is easily verified that the eigenvalues of the matrix listed in \eqref{bexpform} are $\lambda_1$ and
$\lambda_2$, and that the eigenvectors corresponding to these eigenvalues, respectively, obey the first and second of \eqref{ratioson}.

In order to fix the absolute normalization of the blocks, we now impose the condition
$G_{s/a}^\dagger \times G_{s/a}= {\hat G}_{s/a}^\dagger$. Using \eqref{comprel}, it is straightforward to verify that this requirement fixes
\begin{equation}\label{symasb}
	\begin{split}
		G_s=\frac{-\lambda_2^{-1}\theta+\phi}{\lambda_1^{-1}-\lambda_2^{-1}} \\
		G_a=\frac{-\lambda_1^{-1}\theta+\phi}{\lambda_2^{-1}-\lambda_1^{-1}}
	\end{split}
\end{equation}
Note that in the limit $k \to \infty$, the expressions for the blocks $G_s$, $G_a$ in terms of
$\theta$ and $\phi$ becomes identical to the expressions for the classical invariants
$T_s$ and $T_a$ in terms of $T_d$ and $T_e$ (see \eqref{ppscatff}).

With these choices it is easy to see that
\begin{equation}\label{compsas}
	\begin{split}
		&{\hat G}_s:=G_s \times G_s^\dagger = \frac{-\lambda_2^{-1}X'+Y'}{\lambda_1^{-1}-\lambda_2^{-1}}\\
		&{\hat G}_a':=\frac{-\lambda_1^{-1}X'+Y'}{\lambda_2^{-1}-\lambda_1^{-1}}
	\end{split}
\end{equation}

It is also easy to moreover verify ${\hat G}_s+ {\hat G}_a= X=G_{ {\rm id}}$. Recall that the block $X$ plays the role of the identity
element under the compounding rule relevant for $FF \rightarrow FF$ scattering. It follows that the matrices in \eqref{compsas} do indeed obey
\eqref{projopsn} and \eqref{projcomplete}, as expected on general grounds.

We now expand the $S$-matrix in the form \eqref{smofprese} relevant to fundamental fundamental scattering as \\
\begin{equation}
	S= \mathcal S_s G_s +\mathcal S_a G_a
\end{equation}
As in subsection \ref{upb}, we find
\begin{equation} \label{unitsas}
	S S^\dagger = \mathcal S_s \star \mathcal S_s^* ~G_s \times G_s^\dagger +\mathcal S_a \star \mathcal S_a^* ~G_a \times G_a^\dagger
	= \mathcal S_s \star \mathcal S_s^* ~{\hat G}_s +\mathcal S_a \star \mathcal S_a^* ~{\hat G}_a
\end{equation}
The requirement that our $S$-matrix be unitary, tells us that
\begin{equation}\label{unitspec}
	\mathcal S_s \star \mathcal S_s^* {\hat G}_s +\mathcal S_a \star \mathcal S_a^* {\hat G}_a= \mathcal S_{ {\rm id}} {\hat X}= \mathcal S_{ {\rm id}} ({\hat G}_s+ {\hat G}_a)
\end{equation}
where $\mathcal S_{ {\rm id}}$ is the momentum space identity.
Equating the coefficients of independent blocks in \eqref{unitspec} we conclude that
\begin{equation}
	\mathcal S_s \star \mathcal S_s^*= \mathcal S_a \star \mathcal S_a^*=1
\end{equation}

In summary, the $S$-matrices that are the coefficients of the blocks $G_s$ and $G_a$, obey
the standard, usual, unitarity relations.

\subsection{Fundamental-antifundamental scattering}
\begin{figure}[H]
	\centering
	\includegraphics[scale=.2]{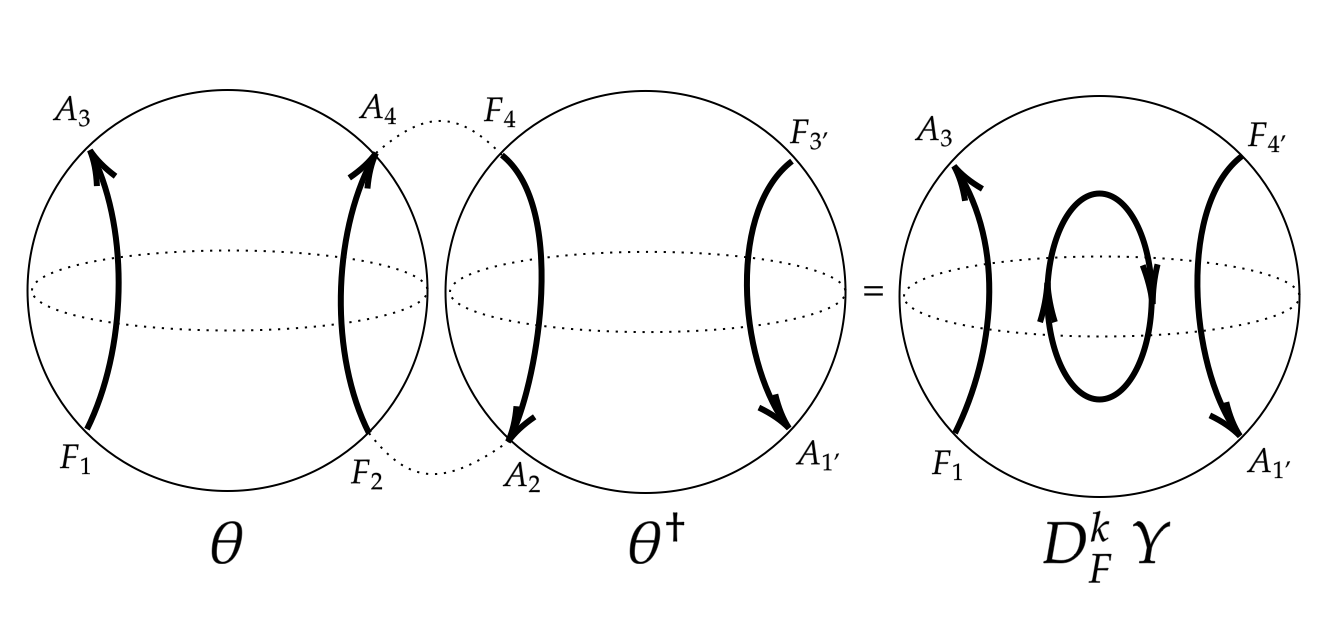}
	\caption{$\theta\times\theta^\dagger$ }
	\label{compf1}
\end{figure}

\begin{figure}[H]
	\centering
 	\includegraphics[scale=.2]{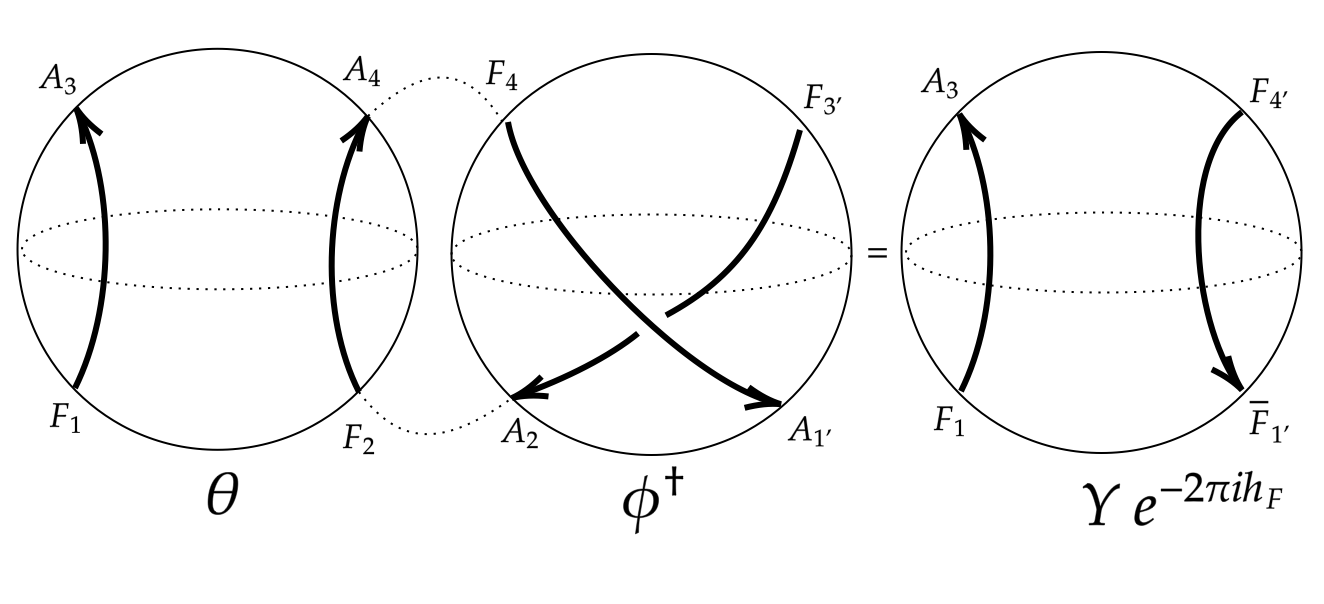}
	\caption{$\theta\times\phi^\dagger$}
	\label{compf2}
\end{figure}

\begin{figure}[H]
	\centering
	\includegraphics[scale=.2]{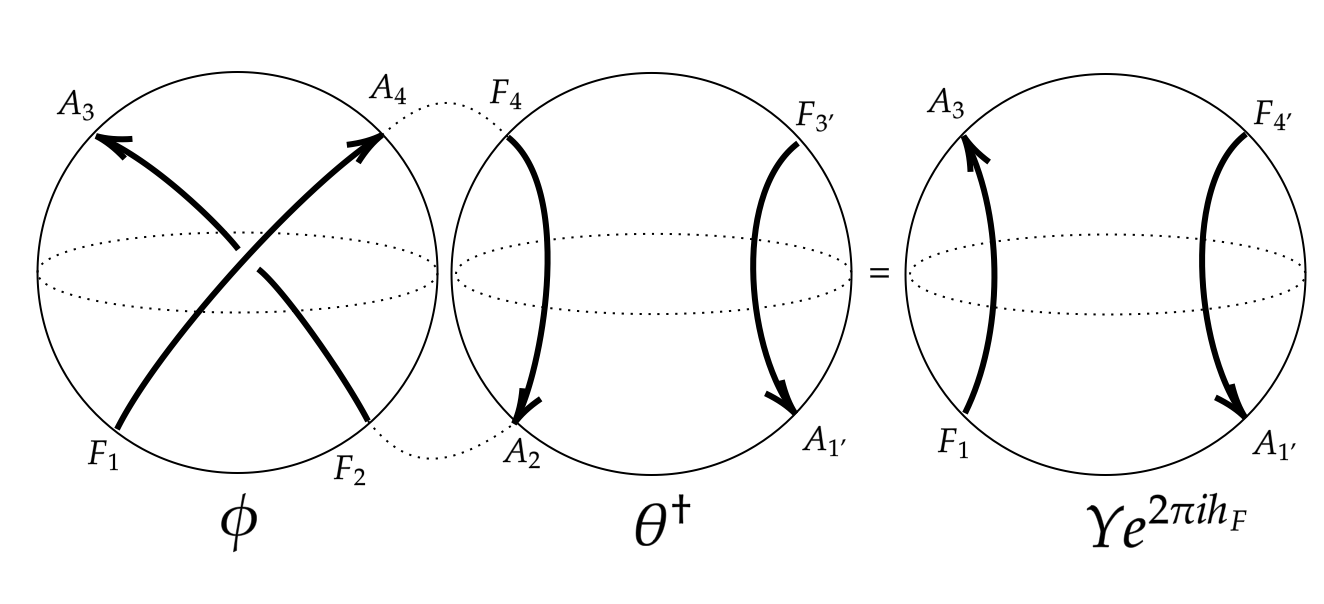}
	\caption{$\phi\times\theta^\dagger$}
	\label{compf3}
\end{figure}

\begin{figure}[H]
	\centering
	\includegraphics[scale=.3]{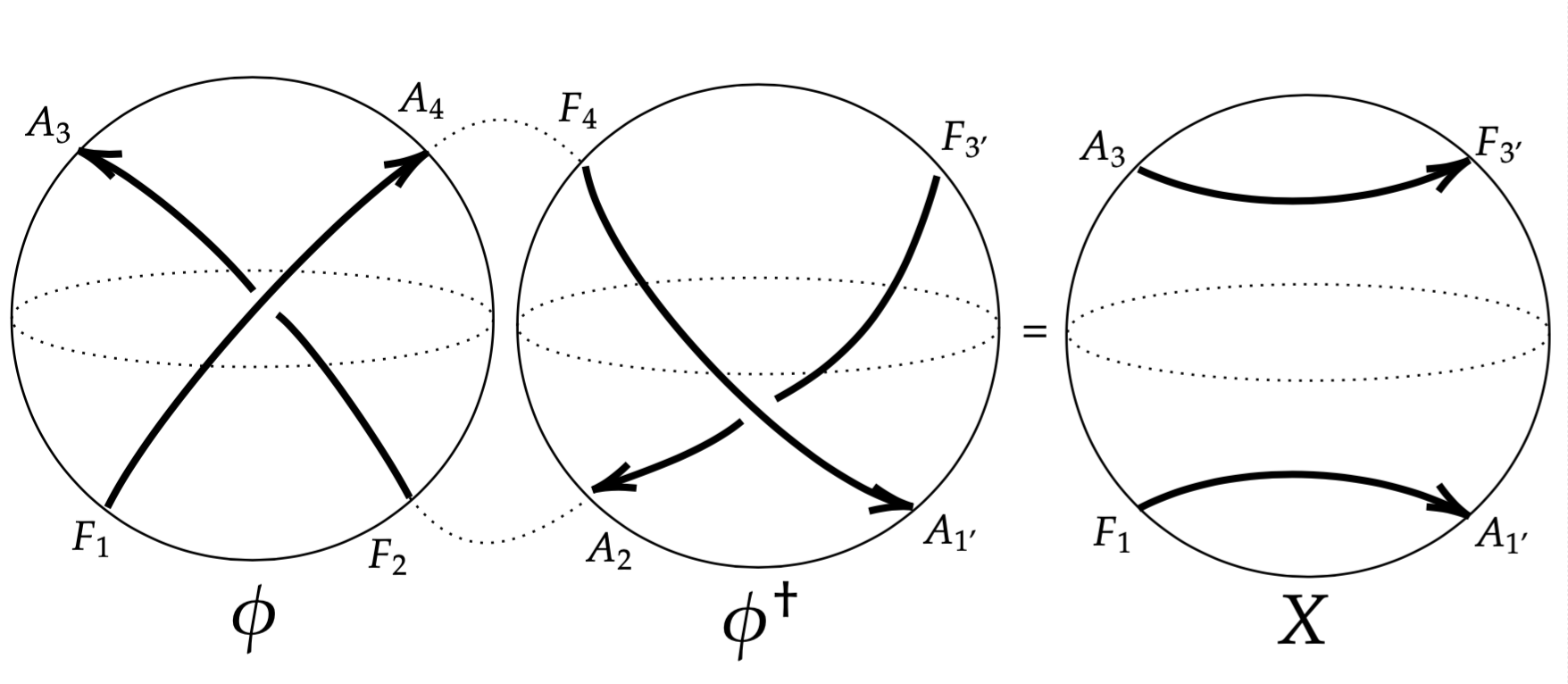}
	\caption{$\phi\times\phi^\dagger$}
	\label{compf4}
\end{figure}

We now turn to the study of the same scattering process in the cross channel, i.e., the study  FA-FA scattering. Once again we will determine the projector blocks in this basis as a linear combination of $\theta $ and $\phi$. Note that the rule for the compounding of blocks in this channel is different from the rule presented in the previous subsection.  In the current context we compound with time running from left to right.
As follows from Figs. \ref{compf1}, \ref{compf2}, \ref{compf3},  and \ref{compf4}, the  compounding rules relevant to unitarity in this channel are

\begin{equation}
	\theta \times \theta^\dagger=  D^k_F Y~~,\theta \times \phi^\dagger= e^{-2\pi i h_F}Y~~,
	\phi \times \theta^\dagger =Ye^{2\pi i h_F}~~, \phi \times \phi^\dagger= X.
\end{equation}
The phases that appear in this equation come from the phases in Fig. \ref{compf2} and Fig. \ref{comp3}, and are the phases associated with overpasses and underpasses in Wilson loop in the vertical framing convention (see Appendix \ref{conventions} for a discussion).

The basis blocks $X$ and $Y$ (for initial to initial star scattering in the fundamental - antifundamental channel) are defined (in terms of their Wilson line representation) in Figs. \ref{compf1}, \ref{compf2}, \ref{compf3},  \ref{compf4}.
\footnote{Note that the expression for the circular Wilson loop may be obtained directly
from the Skein relations in the zero self linking number framing convention, and so is the same for the $SU(N)_k$, Type I and Type II theories (as the Skein relations are independent of $U(1)$ charge in the zero self linking number convention(See \eqref{skeinsfz})).}

As in the previous subsection, we now move to the `projector' type basis for blocks, i.e., to the basis consisting of $G_I$ and $G_{ {\rm Adj}}$. We have already found one representation for these blocks in terms of Wilson lines in Fig. \ref{nblocks2}; here we will find an alternate representation as a linear combination of the basis blocks $\theta$ and $\phi$.

\begin{figure}[H]
	\centering
	\includegraphics[scale=.3]{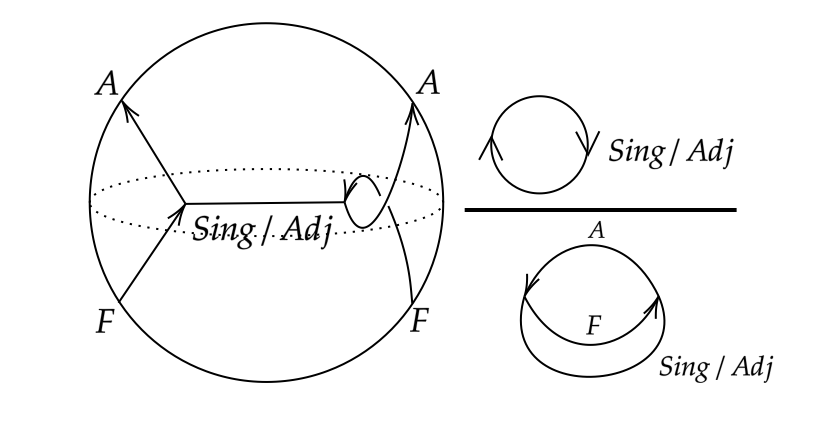}
	\caption{Singlet and adjoint blocks with a twist}
	\label{flipoft}
\end{figure}

Before proceeding with this determination, however,
we need to deal with an important subtlety. If we compare the blocks $\theta$ and $\phi$ defined in
Fig. \ref{thph} with the symmetric and adjoint blocks defined in Fig. \ref{nblocks2}, we see that these sets of blocks have different locations for the Wilson line end points. While the antifundamental insertions in Fig. \ref{thph} are both at the top
of their diagrams, the antifundamental insertion in Fig. \ref{nblocks2} is on the top of the diagram on the left, but at the bottom of the diagram on the right. Before we can find an alternate representation of $G_{{\rm Adj}/I}$ in terms of $\theta$ and $\phi$, we need to move the insertions that appear in this block to the location of the insertions in $\theta$ and $\phi$. As the blocks $G_{{\rm Adj}/I}$ are multivalued,
this maneuver involves making a choice. At the beginning of subsection \ref{cross} we we were faced with the same choice for essentially the same reason. In that section we had to move the insertions in $G_{{\rm Adj}/I}$ to the locations of the insertions of $G_{s/a}$ so that we could take the inner product of these two blocks: but this is precisely the same issue we are faced with here, because $\theta$ and $\phi$ have insertions at the same locations as the blocks $G_{s/a}$. In that section we chose to continue the blocks $G_{{\rm Adj}/I}$ to the locations of the insertions of $G_{s/a}$ (or, equivalently, of $\theta$ and $\phi$) following the arrow between
the first and second diagrams on the second line of Fig. \ref{defadjsing}. In order to match
the results of subsection \ref{cross} we make the same choice here. In the rest of this subsection we will, in other words,
find the representation of the block depicted in Fig \ref{flipoft}
in terms of $\theta$ and $\phi$.

Let $G_{{\rm Adj}/I}$, depicted in Fig.
\ref{flipoft}, be given by the general linear combination of $\theta$ and $\phi$,
\begin{equation} \label{lctheph}
	\begin{split}
		G_I= a'_1 \theta +a'_2 \phi\\
		G_{ {\rm Adj}}=b'_1 \theta +b'_2 \phi\\
	\end{split}
\end{equation}
The orthogonality condition
\begin{equation}\label{monblocks}
	G_I^\dagger \times G_{ {\rm Adj}} =0, ~~~	G_{ {\rm Adj}}^\dagger \times G_{I} =0
\end{equation}
fixes the ratio of coefficients to be
\begin{equation} \label{relcoef}
	\begin{split}
		\frac{a'_2}{a'_1}=0\\
		\frac{b'_1}{b'_2}=\frac{-e^{2\pi i h_F}}{D^k_F}
	\end{split}
\end{equation}

\begin{figure}[h]
	\centering
	\includegraphics[scale=.2]{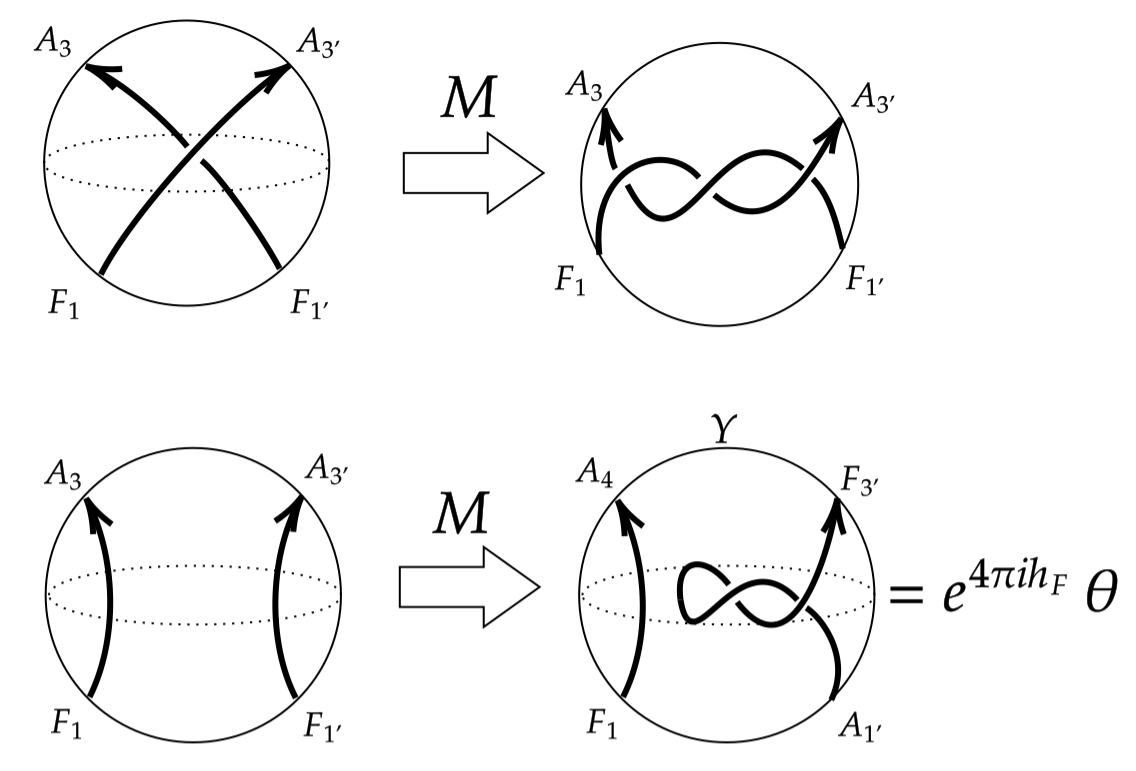}
	\caption{Action of the monodromy operator $M$ on states $\phi$ and $\theta$ respectively}
	\label{monod2}
\end{figure}

As in the previous subsection, we can learn which of the two solutions \eqref{relcoef} is proportional to $G_{{\rm Adj}/I}$ by computing the monodromy of the blocks \eqref{monblocks} with \eqref{relcoef} inserted (the requirement that the blocks defined by \eqref{relcoef} each have a definite monodromy is also a check of the result \eqref{relcoef}). In order to perform this computation, we note that the action of the monodromy operation $M$ (a $2 \pi$ clockwise rotation of the $A$ and $F$ insertions on the right $\phi$ and $\theta$) is given as in Fig. \ref{monod2}.
 As demonstrated in Fig. \ref{monod2}, the action of the relevant monodromy operation on $\theta$
 yields $\theta$ multiplied by a phase (we have used the `overcrossing' and `undercrossing' rules described in Appendix \ref{conventions} to determine this phase). The action of the monodromy operation on $\phi$ is a bit more complicated; we use the Skein relations \eqref{skeistruc} to reexpress the RHS of the first line of fig. \ref{monod2} in terms of $\theta$ and $\phi$. Our final results are
\begin{equation} \label{maction}
\begin{split}
&M \theta= e^{4\pi i h_F}\theta\\
&M \phi=e^{4\pi i h_F}\frac{\lambda_1+\lambda_2}{\lambda_1 \lambda_2}\theta-\frac{1}{\lambda_1 \lambda_2}\phi
\end{split}
\end{equation}
where $\lambda_1$ and $\lambda_2$ are given in \eqref{evan}.
Using \eqref{maction}, it easy to verify that the first and second lines of \eqref{relcoef} yield eigenvectors of the matrix $M$,
with respective eigenvalues
given by $e^{2\pi i(2h_{F})}$ and $e^{2\pi i (2h_{F} -h_{ {\rm Adj}})}$.
It follows that the first of \eqref{relcoef} is proportional to $G_{I}$, while the second is proportional to $G_{ {\rm Adj}}$.

As in the previous subsection we now choose normalizations of $G_I$ and $G_{ {\rm Adj}}$ to ensure that
\begin{equation}\label{compsing}
\begin{split}
&G_I^\dagger \times G_I= {\hat G_I},\\
&G_{ {\rm Adj}}^\dagger \times G_{ {\rm Adj}}=
{\hat G}_{ {\rm Adj}}
\end{split}
\end{equation}
Here $G_I$ and $G_{ {\rm Adj}}$ are the
linear combinations of $\theta$ and $\phi$ \eqref{lctheph} that respectively obey the first and second of \eqref{relcoef}, while
${\hat G}_I$ and ${\hat G}_{ {\rm Adj}}$ are effectively the same $G_I$ and
$G_{ {\rm Adj}}$, with a $\pi$ rotation counterclockwise performed on the two rightmost operators.
\footnote{The reason we should expect the result of the compounding to yield ${\hat G}_I$ and ${\hat G}_{ {\rm Adj}}$ rather than
the `original' blocks ${ G}_I$ and ${G}_{ {\rm Adj}}$ is the following. In this Appendix we are attempting to find an alternate representation of the blocks depicted in Fig. \ref{nblocks2}, after the twist depicted in Fig. \ref{flipoft}. As our compounding rule (which glues operators on the right) is invariant under a rotation of the glued operators, the result of the compounding should be exactly as in subsection
\ref{cross}, and so should yield
blocks of Fig. \ref{nblocks2} without the twist depicted in Fig. \ref{flipoft}. As the twist in Fig. \ref{flipoft} was accomplished by a clockwise rotation of $\pi$, the result of our compounding rule should be the blocks $G_{{\rm Adj}/I}$ constructed in this section subject to a counterclockwise rotation by $\pi$, i.e., the blocks ${\hat G}_{I/{\rm Adj}}$. }
In order to process \eqref{compsing}
we note that the RHS of \eqref{compsing} is given by precisely the linear combinations that define $G_{I/{\rm Adj}}$ except that the rightmost two insertions in each of $\phi$ and $\theta$ are rotated counterclockwise by $\pi$.
Plugging the linear combinations
\eqref{relcoef} into \eqref{compsing} and equating
coefficients of the independent blocks, we find that
\begin{equation} \label{adjsingprojnint}
	\begin{split}
		G_I=e^{2\pi i h_F}\theta/D^k_F \\
		G_{ {\rm Adj}}=\phi-e^{2\pi i h_F}\theta/D^k_F
	\end{split}
\end{equation}

With these normalizations
\begin{equation}\label{compsingn}
	\begin{split}
		&G_I':=G_I \times G_I^\dagger= Y/D^k_F,\\
		&G_{ {\rm Adj}}':=G_{ {\rm Adj}} \times G_{ {\rm Adj}}^\dagger=X-Y/D^k_F.
	\end{split}
\end{equation}
where the blocks $X$ and $Y$ were defined, for instance, in Figs.
\ref{compf3} and Fig. \ref{comp4}.

It follows immediately from
\eqref{compsingn} that
\begin{equation}\label{decompunit}
	{\hat G}_I + {\hat G}_{ {\rm Adj}} = X
\end{equation}
Notice that $X$ is the identity block under compounding in the `horizontal' direction. It follows that the blocks $G_I'$ and $G_{ {\rm Adj}}'$ obey \eqref{projopsn} and \eqref{projcomplete} as expected.
\footnote{Note, on the other hand, that \begin{equation}\label{decompunitthf}
	{G}_I + {G}_{ {\rm Adj}} = \phi
	\end{equation}
This is exactly as we should expect, as $\phi$ equals the horizontal identity block $X$ after a clockwise rotation of $\pi$ on the two rightmost operators, i.e., after the twist performed in
\eqref{flipoft}. In other words
\eqref{decompunitthf} is precisely
the equation \eqref{idap} after
the twist of Fig. \ref{flipoft}, as we should have expected.}

\subsection{Crossing}

Using the explicit expressions for $G_s, G_a, G_I, G_A$ in terms of $\theta$ and $\phi$, it is easy to find one set of these blocks in terms of the other; we find
\begin{equation}\label{blockstrnf}
	\begin{split}
	&G_s=\frac{G_I e^{-2\pi i h_F}\lfloor N+1 \rfloor_q  + G_{ {\rm Adj}}e^{\pi i(h_{ {\rm Adj}}-2h_F)}}{\lfloor 2 \rfloor_{q} } \\
	&G_a=\frac{G_I e^{-2\pi i h_F}\lfloor N-1 \rfloor_{q} - G_{ {\rm Adj}} e^{\pi i(h_{ {\rm Adj}}-2h_F)}}{\lfloor 2\rfloor_{q}}
	\end{split}
\end{equation}
in precise agreement with \eqref{blockstrnfmain}.

\subsection{Direct comparison of blocks}

In this Appendix so far we have found an alternate representation of the blocks $G_{s/a}$ and $G_{I/{\rm Adj}}$ by determining linear combinations of the blocks $\theta$ and $\phi$ that satisfy all the properties (monodromy and compounding rule) that we demanded of these blocks in subsection
\ref{cross}. In this subsection  we verify, for instance \eqref{adjsingprojnint}, by taking the inner product of both sides of this equation with $\langle \theta |$ and $\langle \phi|$ and checking we get the same answer on the LHS and RHS. \footnote{All inner products are taken using Witten's sewing prescription for inner products in the space of four point blocks.}

Using the representations of $\theta$ and $\phi$ given in Fig. \ref{thph}, it is easy to verify that
\begin{equation}\label{thetaphiip} \begin{split}
&\langle \theta | \theta \rangle = \left( D_F^k \right)^2 \\
&\langle \phi | \phi \rangle = \left( D_F^k \right)^2 \\
&\langle \phi | \theta \rangle = e^{- 2 \pi i h_F}  D_F^k\\
&\langle \theta | \phi \rangle=e^{2 \pi i h_F}  D_F^k\\
\end{split}
\end{equation}
where the phases in the last two of \eqref{thetaphiip} are obtained using the over and under crossing
rules of \cite{Witten:1989wf} (see Appendix \ref{conventions}).

Taking the inner product of \eqref{adjsingprojnint} with $\langle \theta|$ and $\langle \phi|$,
we conclude that it must be that
\begin{equation}\label{GGtp} \begin{split}
		&\langle \theta | G_I\rangle =  e^{2 \pi i h_F} D_F^k  \\
		&\langle \phi | G_I \rangle = 1\\
		&\langle \phi | G_{ {\rm Adj}} \rangle = \left(D_F^k\right) ^2 -1\\
		&\langle \theta | G_{ {\rm Adj}}  \rangle=0\\
	\end{split}
\end{equation}

\begin{figure}[h]
	\centering
	\includegraphics[scale=.2]{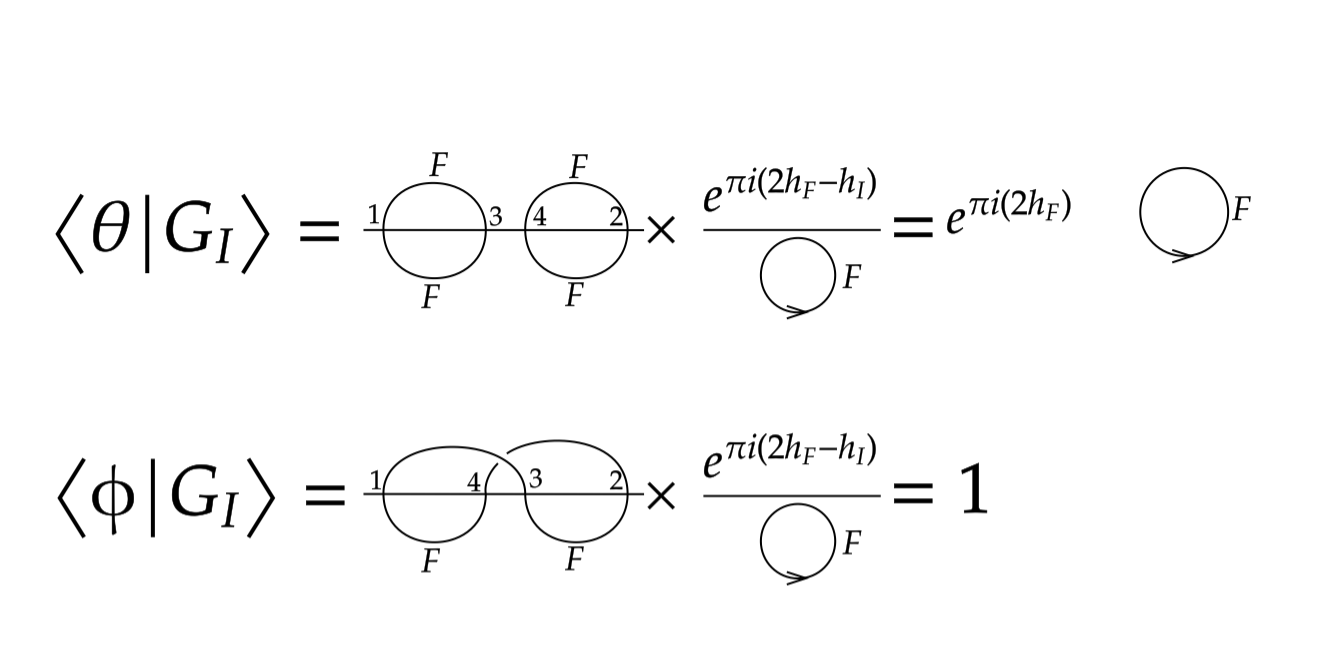}
	\caption{Inner product of singlet with $\theta$ and $\phi$}
	\label{thphi}
\end{figure}

It is not difficult to  independently verify that the representations of $G_{I/{\rm Adj}}$ given in Fig. \ref{flipoft} obey the equation \eqref{GGtp}. In the case of the block $G_I$ we find the equation given in Fig. \ref{thphi}
(the phase on the RHS of Fig. \ref{thphi} is a consequence of the twist of the interaction lines
on the right in Fig. \ref{flipoft} )

\begin{figure}[h]
	\centering
	\includegraphics[scale=.2]{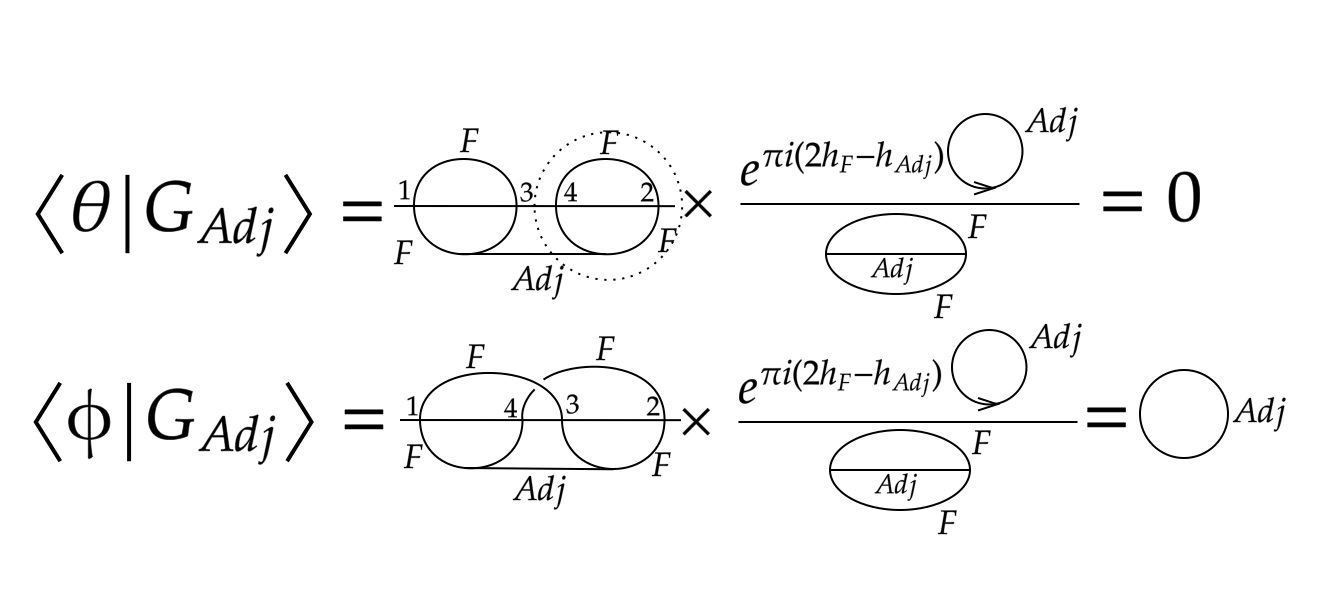}
	\caption{Inner product of adjoint with $\theta$ and $\phi$}
	\label{thphad}
\end{figure}

In the case of the block $G_{ {\rm Adj}}$ we find the equation given in Fig \ref{thphad}.
The fact that the RHS of the first line of \ref{thphad} vanishes follows from cutting the figure along the dotted circle and noting that the cut has single insertion in the adjoint representation
\footnote{Of course any block with a single nontrivial primary insertion vanishes.}. In the second line of Fig.
\ref{thphad}, the leftmost diagram cancels the diagram in the denominator upto a phase (obtained by
rearranging the lines at, say, the left interaction vertex -- see Appendix \ref{conventions}). This
phase (see Appendix \ref{conventions}) exactly cancels the explicit phase in this equation
\footnote{As above, the explicit phase is a consequence of the twist in Fig. \ref{flipoft}.}.
The final answer is the quantum dimension of the adjoint, which exactly equals $(D_F^k)^2-1$ (reflecting the fact that the adjoint is obtained by taking the product of fundamental and antifundamental, and then removing the singlet, which has quantum dimension unity).

In a similar manner we have also checked that the inner product of the LHS and RHS of \eqref{symasb}
match each other.

\section{Groups, Casimirs, scaling dimensions and Skein relations} \label{cgs}

We will be interested in three gauge groups. These are the $SU(N)_k$ theory, the so-called Type I i.e., the  $U(N)_{k, k+N}$ theory and the
so-called Type II i.e., the $U(N)_{k, k}$ theory.

In each case the formula for the dimension of a representation $R$ is given by
\begin{equation}\label{dimrepR}
	h_r= \frac{c_2(R)}{\kappa} + \frac{Q^2}{2N(k_{eff})}
\end{equation}
Here $c_2(R)$ is the quadratic Casimir of the non abelian part of the gauge group (i.e., of $SU(N)$).
In the case that the gauge group is $SU(N)$ $Q$ is zero for all representations.  is the $U(1)$ charge of the representation of the gauge group in a normalization in which the fundamental has charge 1. $k_{eff} = k$ for the Type II theory, and $k_{eff}=k+N$ in the Type theory.

The formula for the $SU(N)$ Casimir is given as follows. Let $y_i$ denote the number of boxes in the $i^{th}$ row of the Young Tableaux. Define
\begin{equation}\label{Youngsum}
	X(Y) = \sum_i y_i (y_i - 2i)\ .
\end{equation}
and let $|Y|$ denote the total number of boxes in the Young Tableaux. Then
\begin{equation}
	2 C_2(Y) = X(Y) + |Y|(N+1) - \frac{|Y|^2}{N}
\end{equation}
For the fundamental, symmetric, antisymmetric and adjoint, respectively, we find $X=-1, 0, -4, -(N-1)^2+1$ and $|Y|$ equals $1, 2, 2, N$.
It is now easy to compute the dimensions of the representations of interest.

In the case of the $SU(N)$ theory we find
\begin{equation}\label{dimensions}
	h_{F}= \frac{N-\frac{1}{N}}{2\kappa}, ~~~h_s=
	\frac{N+1-\frac{2}{N}}{\kappa}, ~~~~h_a=
	\frac{N-1-\frac{2}{N}}{\kappa}, ~~~~
	h_{ {\rm Adj}}=\frac{N}{\kappa}
\end{equation}

In the case of the Type I theory we find
\begin{equation}\label{totheor}
	h_{F}= \frac{N}{2\kappa}, ~~~h_s=
	\frac{N+1}{\kappa}, ~~~~h_a=
	\frac{N-1}{\kappa}, ~~~~
	h_{ {\rm Adj}}=\frac{N}{\kappa}
\end{equation}

In the case of the Type II theory we find
\begin{equation}\label{dimtt}
	h_{F}= \frac{N+\frac{1}{|k|}}{2\kappa}, ~~~h_s=
	\frac{N+1+\frac{2}{|k|}}{\kappa}, ~~~~h_a=
	\frac{N-1+\frac{2}{|k|}}{\kappa}, ~~~~
	h_{ {\rm Adj}}=\frac{N}{\kappa}
\end{equation}

Recall that the $SU(N)$ and Type II theories are dual to each other, and the Type I  theories are also dual to each other.  Relatedly (and carefully taking all signs into account) we
see that
\begin{equation}\label{dimendiff}
	\begin{split}
		& h^{SU}_F(N, k) - h_{F}^{II}(|k|, -{\rm sgn}(k) N)= \frac{{\rm sgn}(k)}{2}, ~~~ h^{I}_F(N, k) - h_{F}^{I}(|k|, -{\rm sgn}(k) N)= \frac{{\rm sgn}(k)}{2},\\
		& h^{SU}_s(N, k) - h_a^{II}(|k|, -{\rm sgn}(k) N)= {\rm sgn}(k), ~~~ h^{I}_s(N, k) - h_a^{I}(|k|, -{\rm sgn}(k) N)= {\rm sgn}(k),\\
		& h^{SU}_a(N, k) - h_s^{II}(|k|, -{\rm sgn}(k) N)= {\rm sgn}(k), ~~~ h^{I}_a(N, k) - h_s^{I}(|k|, -{\rm sgn}(k) N)= {\rm sgn}(k),\\
		& h^{SU}_{ {\rm Adj}}(N, k) - h_{ {\rm Adj}}^{II}(|k|, -{\rm sgn}(k) N)= {\rm sgn}(k), ~~~ h^{I}_{ {\rm Adj}}(N, k) - h_{ {\rm Adj}}^{I}(|k|, -{\rm sgn}(k) N)= {\rm sgn}(k)
	\end{split}
\end{equation}
\eqref{dimendiff} may be regarded as a check on the relations \eqref{dimensions}, \eqref{totheor}, \eqref{dimtt}.

As we have explained in the main text, the
anyonic phase for scattering of two fundamentals in the symmetric and antisymmetric channels, and of a fundamental and an antifundamental in the
adjoint and singlet channels are, given by
$2 \pi \nu$ where
\begin{equation} \begin{split} \label{allnew}
		&\nu_s= 2 h_{F} - h_s\\
		&\nu_a= 2 h_{F} - h_a\\
		&\nu_A= 2 h_{F} - h_{ {\rm Adj}}\\
		&\nu_I= 2 h_{F} \\
	\end{split}
\end{equation}

Using the explicit results for holomorphic dimensions presented above we find that in the Type I theory
\begin{equation} \begin{split} \label{nuone}
		&\nu_s= -\frac{1}{\kappa} \\
		&\nu_a=  \frac{1}{\kappa}\\
		&\nu_A= 0\\
		&\nu_I=  \lambda  \\
	\end{split}
\end{equation}
in the Type II theory
\begin{equation} \begin{split} \label{nutwo}
		&\nu_s= -\frac{1}{\kappa} \\
		&\nu_a=  \frac{1}{\kappa}\\
		&\nu_{ {\rm Adj}}= \frac{1} {|k|\kappa} \\
		&\nu_I= \lambda + \frac{1} {|k|\kappa} \\
	\end{split}
\end{equation}
and in the $SU(N)_k$ theory
\begin{equation} \begin{split} \label{nusu}
		&\nu_s= -\frac{1}{\kappa} \\
		&\nu_a=  \frac{1}{\kappa}\\
		&\nu_{ {\rm Adj}}= -\frac{1} {N\kappa} \\
		&\nu_I= \lambda - \frac{1} {N\kappa} \\
	\end{split}
\end{equation}

Let us now turn to a study of the Skein relations in these theories. Let us first work in vertical framing. If we use the Witten's notation in Fig. 8 of \cite{Witten:1988hf} we find that for all three theories,
\begin{equation}\label{skeinrel} \begin{split}
&\frac{\beta}{\gamma} =- e^{2\pi i h_{f}} \left( e^{ -\pi i h_s} - e^{- \pi i h_a} \right)\\
&\frac{\alpha}{\gamma} = -e^{4\pi i h_{F} -\pi i h_s - \pi i h_a} \\
\end{split}
\end{equation}
Specifically for the type I theory we find
\begin{equation}\label{skeinreltypeI} \begin{split}
		&\frac{\beta}{\gamma} = q^{\frac{1}{2} } - q^{- \frac{1}{2}}\\
		&\frac{\alpha}{\gamma} = -1 \\
		&{\rm one ~choice}: \alpha=-1, ~~~\beta= q^{\frac{1}{2} } - q^{ -\frac{1}{2}}, ~~~\gamma = 1\\
	\end{split}
\end{equation}
For the SU(N) theory
\begin{equation}\label{skeinrelsun} \begin{split}
		&\frac{\beta}{\gamma} = q^{\frac{1}{2 N}}   \left(  q^{\frac{1}{2} } - q^{- \frac{1}{2}} \right) \\
		&\frac{\alpha}{\gamma} = -q^{\frac{1}{ N}}   \\
		&{\rm one ~choice}: \alpha=-q^{\frac{1}{2 N}} , ~~~\beta= q^{\frac{1}{2} } - q^{ -\frac{1}{2}}, ~~~\gamma = q^{-\frac{1}{2N}}\\
	\end{split}
\end{equation}
For the Type II theory
\begin{equation}\label{skeinrelsun} \begin{split}
		&\frac{\beta}{\gamma} = q^{-\frac{1}{2 |k|}}   \left(  q^{\frac{1}{2} } - q^{- \frac{1}{2}} \right) \\
		&\frac{\alpha}{\gamma} = -q^{-\frac{1}{ |k|}}   \\
		&{\rm one ~choice}: \alpha=-q^{\frac{-1}{2 |k|}} , ~~~\beta= q^{\frac{1}{2} } - q^{ -\frac{1}{2}}, ~~~\gamma = q^{\frac{1}{2|k|}}\\
	\end{split}
\end{equation}
The Skein relations for the case of zero self linking take the following form:
\begin{equation}\label{skeinsfz}
	\begin{split}
	&\frac{\alpha}{\gamma}=-e^{8\pi i h_F-\pi i (h_s+h_a)}\\
	&\frac{\beta}{\gamma}=-e^{4\pi i h_F}\left(e^{-\pi i h_s}-e^{-\pi i h_a}\right)
	\end{split}
\end{equation}
In \eqref{dimrepR}, U(1) charge for fundamental representation is $1$, and for symmetric and antisymmetric representations is $2$. Therefore, it is easy to see that in \eqref{skeinsfz}, $8h_F-h_s-h_a$ and $4h_F-h_{s/a}$ are independent of the term $\frac{Q^2}{2 N k_{eff}}$ and hence does not depend on level of $U(1)$ part of the Chern-Simons action.

\bibliography{ref}\bibliographystyle{JHEP}

\end{document}